\documentclass{jfm}
\usepackage{graphicx}
\usepackage{epstopdf, epsfig}
\usepackage{xcolor}
\usepackage{amsmath}
\usepackage{amssymb}
\usepackage{bm}
\usepackage[font=footnotesize]{caption}
\usepackage{subcaption}
\usepackage{graphicx}
\usepackage{multirow}
\usepackage{dsfont}
\usepackage{enumerate}
\usepackage[normalem]{ulem}
\usepackage{tikz}
\usepackage{pifont}

\usepackage[colorlinks,citecolor=blue,linkcolor=blue,urlcolor=blue]{hyperref} 

\def\dd{{\, \rm{d}}}

\usetikzlibrary{plotmarks}
\newcommand\marksymbol[2]{\tikz[#2,scale=1.6]\pgfuseplotmark{#1};}

\DeclareRobustCommand\solid{\tikz[baseline=-0.8ex]\draw[thick] (0,0)--(0.5,0);}
\DeclareRobustCommand\dashed{\tikz[baseline=-0.8ex]\draw[thick,dashed] (0,0)--(0.54,0);}
\DeclareRobustCommand\dashdotted{\tikz[baseline=-0.8ex]\draw[thick,dash dot] (0,0)--(0.54,0);}
\DeclareRobustCommand\dotted{\tikz[baseline=-0.8ex]\draw[thick,dotted] (0,0)--(0.54,0);}

\newcommand{\anna}[1]{{ #1}}
\graphicspath{ {.} }

\title{Linear instability and resonance effects in large-scale opposition flow control}

\author{Anna Guseva\aff{1,2}
  \corresp{\email{A.Guseva@leeds.ac.uk}} \and
 Javier Jim\'{e}nez\aff{2}}

\affiliation{
\aff{1}  Department of Applied Mathematics, University of Leeds, LS2 9JT, UK
\aff{2}The School of Aeronautical and Space Engineering, Universidad Polit{\'e}cnica de Madrid,
Plaza Cardenal Cisneros 3, 28040 Madrid, Spain
}

\begin{document}
\maketitle

\begin{abstract}
Opposition flow control is a robust strategy that has been proved effective in turbulent wall-bounded flows. Its conventional setup consists of measuring wall-normal velocity in the buffer layer and opposing it at the wall. This work explores the possibility of implementing this strategy with a detection plane in the logarithmic layer, where control could be feasible experimentally. We apply control on a channel flow at $Re_\tau = 932$, only on the eddies with relatively large wavelengths ($\lambda / h > 0.1$). Similarly to the buffer layer opposition control, our control strategy results in a virtual-wall effect for the wall-normal velocity, creating a minimum in its intensity. However, it also induces a large response in the streamwise velocity and Reynolds stresses near the wall, with a substantial drag increase.  When the phase of the control lags with respect to the detection plane, spanwise-homogeneous rollers are observed near the channel wall. We show that they are a result of a linear instability. In contrast, when the control leads with respect to the detection plane, this instability is inactive and oblique waves are observed. Their wall-normal profiles can be predicted linearly as a response of the turbulent channel flow to a forcing with the advection velocity of the detection plane. The linearity, governing the flow, opens a possibility to affect large scales of the flow in a controlled manner, when enhanced turbulence intensity or mixing is desired.
\end{abstract}

\section{Introduction}\label{sec:intro}
One of the important aspects of fluid dynamics research from a practical point of view is the control of the near-wall turbulence in wall-bounded flows. Industrial devices where such flows appear can benefit significantly from reduction in friction, or, when necessary, increase in turbulent mixing. In the last three decades, significant effort has been made to understand the mechanisms of control in canonical turbulent flows (including flows in pipes, channels, and boundary layers). One of the most successful control strategies is to interfere with the near-wall turbulent cycle, suppressing the formation of streamwise vortices close to the wall and their interaction with streaks of streamwise velocity. This strategy can be implemented via modification of the wall surface by riblets \citep{garcia2011drag}, active modification of the near-wall flow by blowing and suction (opposition control, \citet{choi1994active}),  or near-wall spanwise oscillations \citep[][among others]{quadrio2004critical}. Despite the theoretical progress, practical implementation of these strategies is scarce. In the case of riblets, their technical maintenance is difficult and their relative efficiency reduces with increasing Reynolds number \citep{spalart2011drag}. Spanwise wall oscillations give promising  40\% of drag reduction \citep{quadrio2004critical},  but there is evidence that secondary circulation at the side walls impedes reaching this value in experiments \citep{straub2017turbulent}. Opposition flow control consists of measuring wall-normal (or spanwise) velocity at the detection plane $y_d$ and opposing it at the wall, and gives up to 20\% friction drag reduction  \citep{choi1994active}. This robust control method creates a ``virtual wall" effect, manifested by  a minimum in the turbulent intensity profile of the controlled velocity component. The virtual wall expels small quasi-streamwise vortices away from the wall \citep{jimenez1994structure} and reduces the vertical transport of streamwise momentum near the wall, diminishing drag \citep{hammond1998observed}. \citet{kim2000linear} identified a possible linear physical mechanism of this reduction, relating the suppression of the spanwise variation of velocity in opposition control to weakening of the linear coupling between wall-normal velocity and vorticity near the wall. The control strategy proposed by \citet{choi1994active} soon became a benchmark for optimal flow control strategies \citep{bewley2001dns}, as well as other physics-motivated methods employing wall-based sensors of shear stress or vorticity fluxes \citep{lee1997application,koumoutsakos1999vorticity}. Early experiments approached its implementation by blocking sweep and ejection events near the wall with wall-normal jets \citep{rebbeck2001opposition,rebbeck2006wind}, but these were limited to just one spatially localized pair of a detector and an actuator.

The principal difficulties in the practical implementation of the opposition flow control  are the actuation times and the need of flow reconstruction. Consider, for example, the classic setup of \citet{choi1994active}. It requires observations of velocity field in the buffer layer ($y^+ \approx 10$) and actuation at the wall on the same scales. Here `$+$' denotes wall units, defined in terms of kinematic viscosity $\nu$ and friction velocity $u_\tau$. The characteristic energetic length scales at this height are $\lambda^+_x \approx 1000$ (streamwise), $\lambda^+_z \approx 100$ (spanwise). 
The passing time of these eddies is of the order of milliseconds and they are too fast to be detected and opposed in experiments due to the resolution restrictions of measuring sensors and actuators. Also, a grid of sensors and actuators with spacing less than a millimeter between them renders the control scheme impractical. This draws our attention away from the buffer layer to the logarithmic layer control, where turbulent structures, with lifetimes of the  order of seconds, could indeed be detected and controlled. Recent work of \citet{ibrahim2019selective} showed that complete removal of large scales in the logarithmic layer results in a positive, outward shift of the mean velocity profile, equivalent to drag reduction.  Also, recent Monte Carlo experiment of \citet{pastor2020wall} suggests that a single
actuator, localised in space and located above $y^+ = 50$, could reduce drag by $3-4$\%, by opposing vertical motions near the wall.
\begin{figure}
\centering
\begin{subfigure}[t]{0.45\textwidth}
	\centering
	\caption{} 
    \label{fig:v100_snapshot}
  	\includegraphics[width=\textwidth]{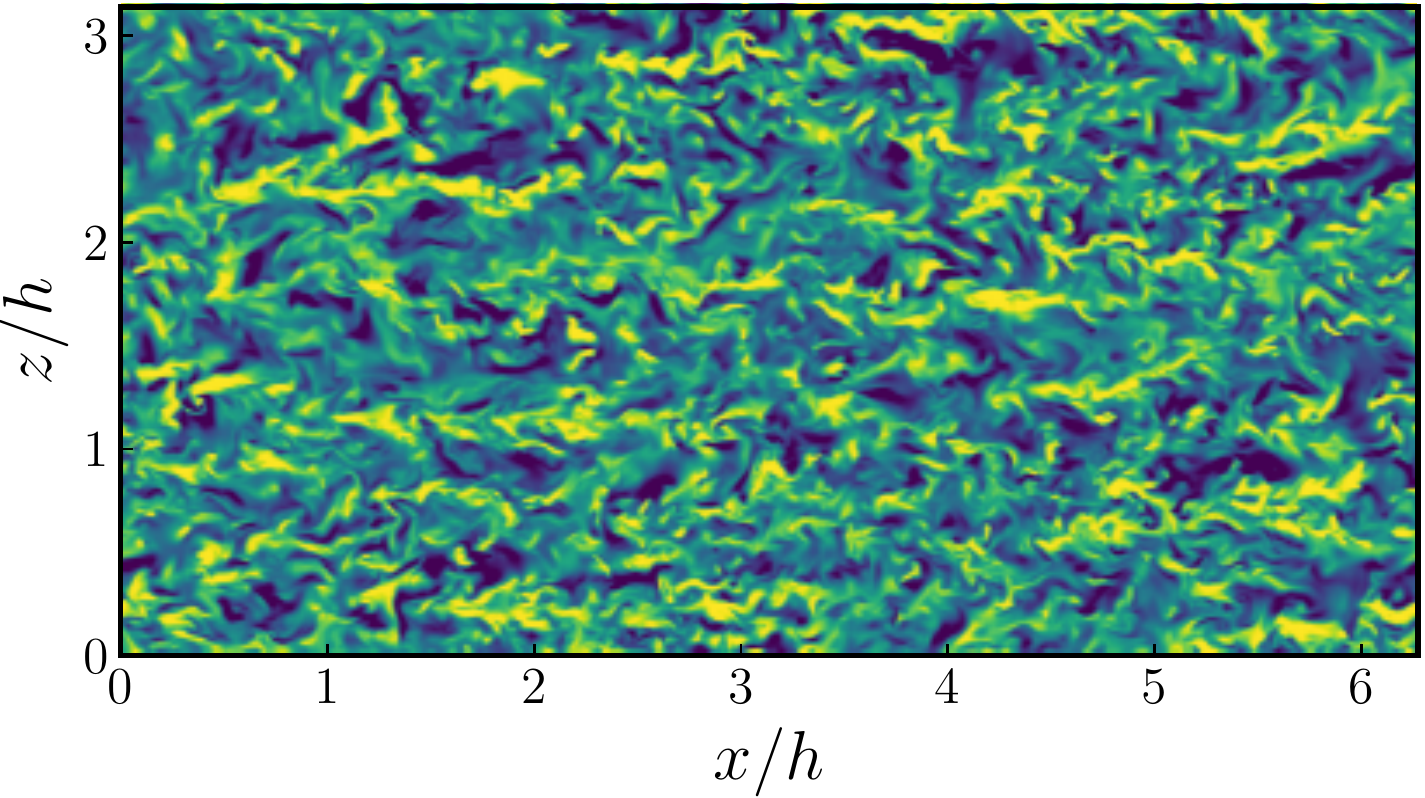}
  \end{subfigure}\hspace{1em}
  \begin{subfigure}[t]{0.45\textwidth}
     	\caption{}
    \label{fig:v100_filtered}
  	\centering
   	\includegraphics[width=\textwidth]{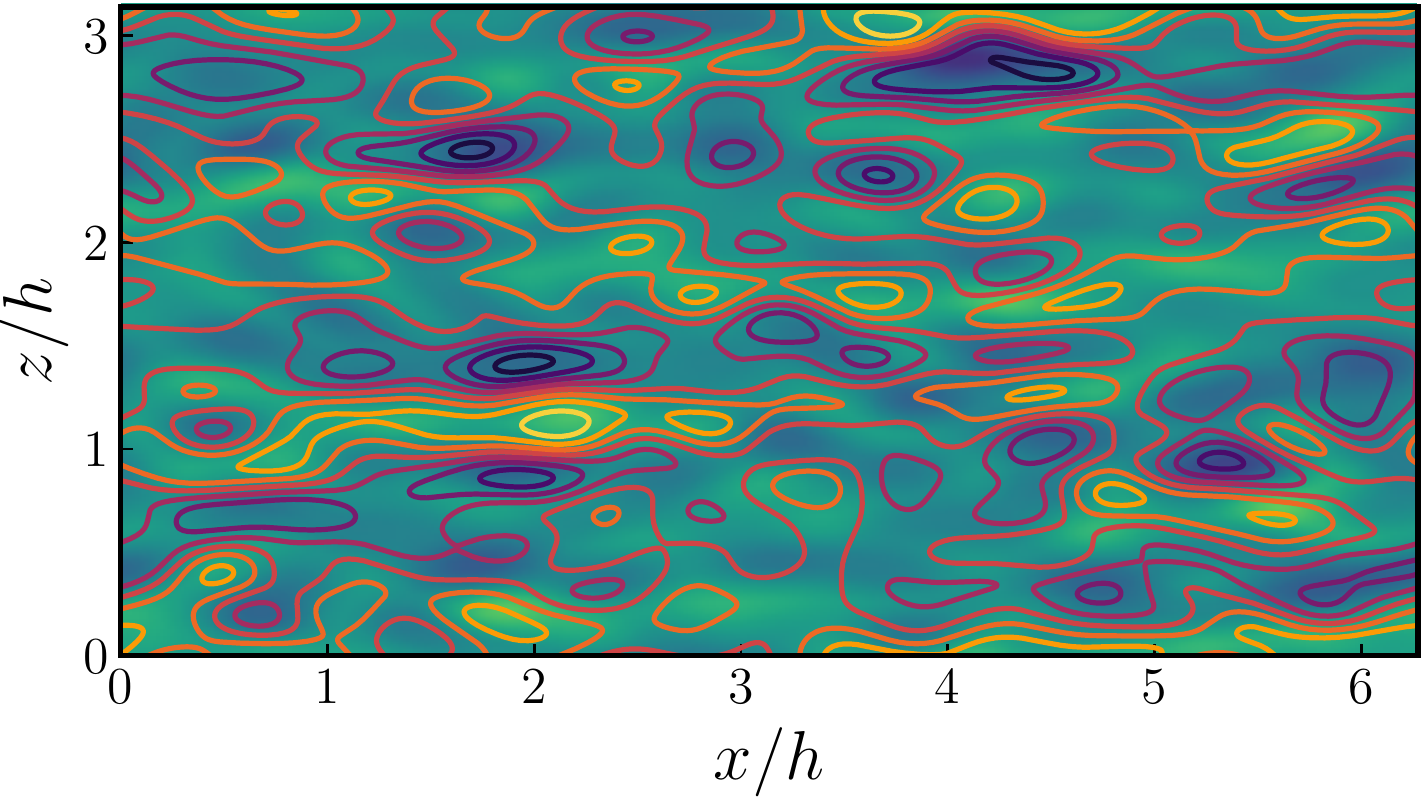}
  \end{subfigure}
  \caption{(a) Instantaneous snapshot of wall-normal velocity at $y/h = 0.1$. (b) Wall-normal velocity from ($a$) filtered with low-pass Gaussian filter on the controlled wavelengths (shaded) with LSE reconstruction from \cite{encinar2019logarithmic} on top (contours). $Re_{\tau} = 932$.}
  \label{fig:v_intro}
\end{figure}
The second difficulty in implementing opposition flow control is that, unlike in experiments or numerical simulations,  the information about the flow above the wall is usually not available in the real world. This constrains our knowledge about velocity field above the wall to flow reconstructions based only on the wall measurements \citep{oehler2018linear,encinar2019logarithmic}.
The main problem of such flow reconstructions is the lack of fidelity far from the wall. The wall is impacted mostly by attached eddies, and their size grows with the distance from the wall. The linear stochastic estimation (LSE) of \citet{encinar2019logarithmic} showed that the farther from the wall, the less information about small-scale flow structure is accessible.  Only scales of wall-normal velocity comparable to the channel height $h$ can be reconstructed with less than 50\% error in the logarithmic layer.
As an illustration of this problem, 
figure~\ref{fig:v100_snapshot} shows a typical instantaneous snapshot of the wall-normal velocity  at $y/h=0.1$, where small structures exist alongside large ones. In figure~\ref{fig:v100_filtered}, that snapshot has been filtered with a Gaussian low-pass filter at the  length scales $\lambda_x/h >1$, $\lambda_z/h > 0.6$. On top of it, the contours of the LSE reconstruction done with the algorithm of \citet{encinar2019logarithmic} are added, and they coincide with the large velocity structures reasonably well.

Despite the need of a control strategy for the structures in the logarithmic layer of wall-bounded turbulent shear flows, its implementation is non-trivial. It has been long known that the performance of opposition control deteriorates if detection plane is lifted above an optimal location of $y_d^+ = 15$  \citep{choi1994active,hammond1998observed}, which can result even in drag increase. \citet{hammond1998observed} related this effect to the inability of the control to establish a virtual wall by allowing high momentum fluid to be drawn in the region between the detection plane and the wall.  This effect can be understood by considering linear mechanisms supporting turbulence in the flow, since a significant part of the dynamics originates from the interaction of  turbulent fluctuations with the mean shear through transient growth \citep{butler1992three,del2006linear}. For a channel flow subject to opposition control, \citet{lim2004singular} showed that the transient growth weakened when $y_d$ was located in the buffer layer close to the wall, and significantly increased if $y_d$ was chosen too far away from it. \citet{chung2011effectiveness} showed that this effect could be mitigated partially by decreasing the amplitude of actuation. \citet{lee2015opposition} later introduced an upstream spatial shift between detection and actuation which improved control performance.   On the other hand, modifying the boundary conditions at the wall can also affect the stability properties of the flow. Introducing wall transpiration permits momentum exchange in the  $y$ direction, often destabilizing otherwise linearly stable flows. An early study of drag increase in turbulent flows over porous surfaces  relates the appearance of large spanwise rollers to an inviscid Kelvin-Helmholtz instability  \citep{jimenez2001turbulent}. The same effect destroys the drag-reducing behaviour of riblets when their characteristic size and spacing are more than $10 - 15$ wall units \citep{garcia2011drag}. \cite{toedtli2020origin} reported a presence of spanwise rollers and linear instability in a channel flow with opposition control of the buffer layer for certain parameters of actuation.

Furthermore, \citet{jimenez2001turbulent} opens a discussion about another mechanism of drag increase in turbulent flows with transpiring walls, that can be active even if the flow is linearly stable. If the flow is forced at a frequency close to the real part of one of its linear eigenvalues, the response of the  system can be quite large. 
The ``response - forcing" framework was generalized for turbulent pipe flow by the resolvent analysis of \citet{mckeon2010critical}, who proposed to decompose the velocity field into a series of optimal response and forcing modes with different frequencies, and to rank them by their importance. Later, resolvent analysis was adapted to the study of opposition control by \citet{luhar2014opposition}, who found that opposition control with detection plane at $y_d^+=10$ suppresses slow response modes localized near the wall, but amplifies faster detached modes. To counteract the latter,  \citet{luhar2014opposition} proposes to employ a phase lag $\phi$ between sensor and actuator. It has been shown by \citet{luhar2014opposition} that negative phases, equivalent to shifting control downstream with respect to ``classic" opposition, result in some improvement of performance. On the contrary, shifting control positively in phase (i.e. upstream) results in unwanted increase of drag.  \citet{toedtli2019predicting} confirmed the capability of the resolvent model to predict friction drag in direct numerical simulations (DNS), showing that optimal negative phase  $\phi = -\pi/4$ allows to slightly shift detection plane up to $y_d^+ = 24$. The results of \citet{luhar2014opposition, toedtli2019predicting} are also in agreement with the conclusion of \citet{pastor2020wall} that locating a localized sensor upstream improves control performance. This improvement is probably produced by cancelling an additional streamwise lag between the control and the detection plane, introduced by the downstream advection of velocity structures by the flow. Advection velocity of the large flow scales is approximately equal to the mean velocity at their wall-normal location \citep{jimenez2018coherent}.

The motivation of this work is to extend the opposition flow control to the large scales of the flow with detection plane in the log layer, i.e. to the scales that can be both observed and controlled. We analyze the effect of the control on the large scales with direct numerical simulation in fully turbulent channel flow ($Re_\tau = 930$). We explore the possibility to affect the eddies of relatively large wavelengths ($\lambda/h > 0.1$) by acting from the wall, and thus to alter the friction created by their presence.  As a side note, we do not attempt to perform linear optimal flow control. There exists a substantial body of work  on linear optimal control  for flows that are close to transition to turbulence \citep{bewley1998optimal}, but application of this theory to fully turbulent flows is not straightforward. The aim of the optimal linear control is to return the flow back to a (low-drag) unstable state. The turbulent mean profile is the result of nonlinear interactions of turbulent flow fluctuations, and is a high-drag state. Linearization around it will not necessarily yield the same results as in near-transitional flows. Nevertheless, \citet{oehler2020linear} applied this technique to the turbulent mean profile in a channel and found that the best performance is achieved when actuator and sensor planes are both located at $y/h=0.3$. While this location is feasible for measurement, it is not very practical for actuation, which is most easily implemented at the wall.  

This paper is structured as follows. We begin by describing  the computational setup and the numerical methods used in the DNS and the linear stability analysis in section~\ref{sec:setup}.  Section~\ref{sec:dnsres} presents the DNS results of the flow affected by the large-scale control. Section~\ref{sec:linstab_invisc} shows  how the presence of the control affects linear stability of the simplified channel flow without viscosity, including the theoretical implications of imposing the control. A more realistic linear model including turbulent viscosity is analysed in section~\ref{sec:listab_vics}. In section~\ref{sec:linstab_and_dns} we exploit linearised flow dynamics to explain part of the DNS results from section~\ref{sec:dnsres}. To clarify the rest, we employ amplified responses of the linearized flow to the control in section~\ref{sec:res_forced}. Finally, section~\ref{sec:discussion} presents discussion of the results and conclusions.

\section{Numerical experiments}\label{sec:setup}
\subsection{Direct numerical simulations}\label{sec:setup_DNS}

 To assess the possibility of large-scale flow control, we simulate turbulent flow in a  channel  with DNS.  In the following,  $u, v, w$ ($\omega_x, \omega_y, \omega_z$) denote the velocity (vorticity) components in the streamwise ($x$), wall-normal ($y$) and spanwise ($z$) directions, respectively. Our numerical scheme is similar to the one of \citet{kim1987turbulence}. We solve equations for the Laplacian of the wall-normal velocity $\nabla^2 v$  and for the wall-normal vorticity $\omega_y$, which are coupled in the nonlinear terms. The advantage of this formulation is that the pressure is eliminated from the equations, and no boundary conditions for pressure are needed. The computational box is periodic in the wall-parallel directions, and this periodicity allows to represent solutions in the form of Fourier harmonics in $x$ and $z$,
\begin{equation}
\begin{aligned}\label{eq:alp_bet}
v (x,y,z,t) &= \sum_{k_x, k_z} \hat{v}(t, y, k_x, k_z) e^{\mathrm{i} (k_x x + k_z z )}, \\ \omega_y (x,y,z,t) &= \sum_{k_x, k_z} \hat{\omega}_y(t, y, k_x, k_z) e^{\mathrm{i} (k_x x + k_x z )},\\
\end{aligned}
\end{equation}
where $\hat{v}$, $ \hat{\omega}_y$ represent complex Fourier coefficients of particular Fourier modes, and $t$ represents time. Wavenumbers $k_{x_n} = 2 \pi n/L_{x}$, $k_{z_m} = 2 \pi m/L_z$ are proportional to integer multiples $n,m$, and inversely proportional to the length and width of the computational domain $L_x$, $L_z$.
 In the wall-normal direction, $y$, the equations are discretized with compact finite differences. Unlike in \citet{kim1987turbulence}, the flow is integrated in time with 4th order Runge-Kutta scheme. For more details on the numerical method, see \cite{flores2006effect}.  
 
In the code formulation, the flow mass flux is kept constant and the pressure gradient is allowed to vary. This way, if a control is applied, the total shear stress $\tau_w = - \overline{u' v'} + \nu (\partial u /\partial y)$, the friction velocity $u_\tau = \sqrt{\tau_w /\rho}$ and the friction Reynolds number $Re_{\tau} = u_{\tau} h / \nu $ vary too ($\rho$ is the fluid density). This becomes important later for the definition of the friction factor $C_f = \tau_w / (0.5 \rho U^2_b)$, which is used to assess control performance. $U_b = (1/2h) \int_{-h}^{h} U dy$ denotes the bulk velocity. Since $u_\tau$ changes, the normalization of the flow in wall units also changes.  The  majority of our DNS results are non-dimensionalized with the respective $u_\tau$ of each case, unless stated otherwise. The uncontrolled flow parameters are identified as $u_{\tau 0}$, $Re_{\tau 0}$, $C_{f 0}$, etc.

Further details of the computational setup can be found in table~\ref{tab:sim_param}. The friction  Reynolds number of the uncontrolled base flow $Re_{\tau 0} = 932$ is relatively large, allowing to gather enough statistics in the logarithmic layer. 
The size of the computational domain is $2 \pi h \times \pi h$, which is large enough to accommodate the structures prevalent at the target location for control $y_d = 0.1 h$ \citep{flores2010hierarchy,lozano2014effect}. The longest wavelengths $\lambda_{x,z} = 2\pi/|k_{x,z}|$ that our simulations can accommodate are $(2\pi, \pi)$ in the $x$ and $z$ directions, respectively, with $k_{x,z}$ denoting a wavenumber pair.
The two left columns of table~\ref{tab:sim_param} give the information about the mesh in collocation space, and the coarsest mesh resolutions in the three directions, indicating that the baseline simulations are well resolved.
\begin{table}
\begin{center}
    \begin{tabular}{l l p{1cm} l l}
       \multicolumn{2}{l}{Channel flow parameters}   & &  \multicolumn{2}{l}{Control parameters} \\ \hline
     $Re_{\tau 0}$ & $932$ & & Control gain $|A|$ &   $[0,1]$\\
     $L_x/h \times L_z/h$ &  $2\pi \times \pi$ & & Streamwise shift $x_0 / (\pi h)$ &  $[-1,1]$ \\
     $N_x$, $N_y$, $N_z$ & $ 512\times 385 \times 512$ & & Controlled wavelengths $\lambda_{x,z}/h$& $[\pi/3,\infty]$, $[\pi/5,\infty]$\\
      $\Delta x^+$, $\Delta z^+$,$\Delta y_{max}^+$ &  $11.5$, $5.7$, $7.7$ && Controlled wavenumbers $k_{x,z} h$& $[0,6]$, $[0,10]$\\
      & & & Detection plane height $y_d/ h$ & $0.1$ \\
    \end{tabular}
    \caption{Parameters of the DNS and control. $L_x$, $L_z$ are streamwise and spanwise sizes of the computational domain, respectively, compared to the half-height of the channel $h$. $N_x$, $N_y$ and $N_z$ are the numbers of grid points in physical space in each direction, $\Delta x^+$, $\Delta z^+$, $\Delta y_{max}^+$ are the spatial resolutions in wall units before de-aliasing. The phase of the complex coefficient $A$ is related to streamwise shift $x_0$ in~\eqref{eq:xlag}. The $(0,0)$ mode was not controlled ($E_{vv}^{00} =0$ from continuity).}
    \label{tab:sim_param}
    \end{center}
\end{table}

 We implement a variation of the opposition control setup that affects only large scales of the flow.  At each time step of the simulation, the wall-normal velocity $v$ is  recorded at the detection plane $y_d/h  \approx 0.1$, which corresponds to $y^+_d \approx 100$ for the base uncontrolled flow. Although here we focus on the effects of full opposition and do not employ LSE,  we use the conclusions from LSE analysis of \citet{encinar2019logarithmic} to guide our choice of controlled length scales. Their flow reconstructions suggest that only the largest structures of $v$ with wavelengths of $\pi/3 < \lambda_x/h < \infty$, $\pi/5 < \lambda_z/h < \infty$ can be reconstructed with at least 50\% accuracy at this wall-normal location (figure~\ref{fig:v_intro}b). Thus the measurement and actuation are performed only for Fourier modes with these wavelengths, with the corresponding wavenumbers $k_{x,z}$ given in table~\ref{tab:sim_param}. To avoid direct forcing of the mean flow, the mode with $k_x, k_z = (0,0)$ is omitted in the control.  In the next step, this measurement is used to oppose the vertical velocity. The control law can be written as a boundary condition at the wall
\begin{equation}\label{eq:ctrl_law}
\hat{v}_w (t, k_x, k_z) = - A \hat{v} (t, y_d, k_x, k_z),
\end{equation}
where the control coefficient in general can be a complex number: $A=|A|\exp(i\phi)$. The control gain $|A|$ shows the relationship between the magnitudes of the control input and  output. The gain $|A|$ and the phase $\phi$ of the control coefficient  are parameters that can be optimized separately for each flow mode, as suggested by \citet{luhar2014opposition}. The phase of control can be interpreted as a shift of the Fourier harmonic in the streamwise direction: $\phi = - k_x x_0$. \anna{In an average sense,} positive values of $x_0$ correspond to a rightwards shift along the $x$-axis of the control with respect to the detection,  (i.e. downstream), and negative ones to a leftward shift (i.e. upstream). In summary,
\begin{equation}\label{eq:xlag}
\hat{v}_w (t, k_x, k_z) = -|A| \exp (i \phi_{k_x}) \hat{v} (t, y_d, k_x, k_z) = -|A| \exp (- i k_x x_0) \hat{v} (t, y_d, k_x, k_z),
\end{equation}
where a different phase $\phi_{k_x}$ is assigned to each $k_x$ to make sure that the same shift $x_0$ is applied to all harmonics, and the control wave train moves as a whole backwards or forward in $x$ in respect to the measurement.   See table~\ref{tab:sim_param} for the compilation of control parameters.

Relating streamwise and phase shifts requires some care. For example,  any non-zero phase shift will  automatically become a spanwise shift for modes with $k_x =0, k_z \neq 0$, which is detrimental for control and causes drag increase \citep{chung2003sensitivity}. In our implementation, however, the modes with $k_x =0$, $k_z \neq 0$, are not affected by  phase shifts, since their phases are zero for any $x_0$ in \eqref{eq:xlag}. In addition, an instantaneous phase shift with mixed arguments in $x$ and $z$ can arise  due to the lack of $k_z \to - k_z$ symmetry in the instantaneous DNS flow, which can be removed by setting equal actuation amplitudes for $k_z$ and $-k_z$ modes \citep{toedtli2019predicting}.  As this results in leaving the mixed-argument term uncontrolled,  and the DNS flow is nevertheless    statistically invariant to $k_z\to-k_z$, we did not implement this correction here. However, it could be potentially important for quantitative comparison of the friction behavior between the controlled DNS and linearized flow models \citep{toedtli2019predicting}.

\subsection{Linearized flow}\label{sec:setup_linstab}
For linear analysis we employ the numerical method from \citet{Schmid2012}, augmented with turbulent viscosity \citep{reynolds1972mechanics,del2006linear,pujals2009note}. The linearized Navier--Stokes operator, written in terms of wall-normal velocity $v$ and wall-normal vorticity $\omega_y = \partial u/\partial z - \partial w/\partial x$, transforms into the Orr-Sommerfeld and Squire equations
\begin{align}
& \left[ \left( \frac{\partial}{\partial t} + U \frac{\partial }{\partial x} \right) \nabla^2 -  U^{''} \frac{\partial }{\partial x} - \nu_t (y) \nabla^4  - 2  \nu'_t (y) \nabla^2  \frac{\partial }{\partial y} -   \nu^{''}_t (y) \left( 2 \frac{\partial^2 }{\partial y^2}  - \nabla^2 \right) \right] v = 0 , \label{eq:OrrSomm}\\
&\left[ \frac{\partial}{\partial t} + U \frac{\partial}{\partial x} - \nu_t (y) \nabla^2 - \nu'_t (y)  \frac{\partial }{\partial y}   \right] \omega_y = - U^{'} \frac{\partial v}{\partial z}, \label{eq:Squire}
\end{align}
with the mean turbulent velocity profile $U(y)$ and  boundary conditions~\eqref{eq:ctrl_law}, supplemented by $\partial v / \partial y|_{y=0,2h} =0$ and $\omega_y|_{y=0,2h} =0$.
The primes in \eqref{eq:OrrSomm}, \eqref{eq:Squire} denote wall-normal derivative, and $v$ at the wall is a function of $v$ at the detection plane in~\eqref{eq:ctrl_law}.
 Note that the linearization of the Navier-Stokes equation was done around the uncontrolled flow profile, although the mean velocity profile changes when control is applied. 
The eddy viscosity profile $\nu_t$, suggested by \citet{cess1958survey}, is an analytic function of $y$. The idea behind it is that, for every spatial harmonic, the background turbulence acts directly through Reynolds stresses and indirectly through the turbulent mean profile. Turbulent viscosity, introduced into the viscous term, is merely a closure for the mean Reynolds stresses (see Appendix A). Periodicity of the flow in wall-parallel directions allows to represent solutions in the form of Fourier harmonics~\eqref{eq:alp_bet} with wavenumbers $k_x$ and $k_z$ as the input parameters of the problem. 

It is common to study~\eqref{eq:OrrSomm}, \eqref{eq:Squire} by introducing a forcing term, accounting for turbulent fluctuations, nonlinearities or noise \citep{mckeon2010critical}, and to simplify the notation by introducing  operators $D \equiv \partial / \partial y$, $L_{OS}$ and $L_{SQ}$, the vector of variables $\mathbf{q}$ and $\kappa^2 = k_x^2 + k_z^2$ as
\begin{equation}\label{eq:oss_matrices}
 \mathbf{q} = 
 	\begin{pmatrix}
 	\hat{v}  \\
	\hat{\omega}_y 
 	\end{pmatrix}, 
 \quad
 M = 
 	\begin{pmatrix}
 	\kappa^2 - D^2 & 0 \\
	0 		  & 1 
 	\end{pmatrix},
 \quad
 L = 
 	\begin{pmatrix} 
	L_{OS} 				& 0 \\
	\mathrm{i} k_z \frac{d U}{d y} & L_{SQ} 
	\end{pmatrix}.
\end{equation}
See Appendix A for more details. Introducing the unknown forcing f, we can write (2.4) in matrix form,  
\begin{equation}\label{eq:f_linsys}
M \frac{\partial}{\partial t}  \mathbf{q} = -L  \mathbf{q} + \mathbf{f}.
\end{equation}
Consider solutions of the form $\mathbf{q} = \mathbf{q_r} (y) e^{- \mathrm{i} \omega t}$, implying the forcing $\mathbf{f} = \mathbf{q_f} (y) e^{- \mathrm{i} \omega t}$. Equation~\eqref{eq:f_linsys} can be re-written as
\begin{equation}\label{eq:freq_linsys}
- i \omega M \mathbf{q_r}  = - L  \mathbf{q_r} + \mathbf{q_f} .
\end{equation}
Note that there are two ways of representing the role of the boundary condition~\eqref{eq:ctrl_law}. In the first one, we could absorb the boundary conditions in the forcing $\mathbf{f}$, leaving the original force-less system and its eigenvectors untouched. However, on a closer look,~\eqref{eq:ctrl_law} not only affects the values of $\hat{v}$ near the wall, but also  the shape of the  eigenvectors above the wall. Therefore, here we constrain the eigensolutions of~\eqref{eq:freq_linsys}, as well as its responses  to the forcing \anna{$\mathbf{q_f}$}, to the condition $\hat v|_{y=0,2h} = - A \hat{v}|_{y=y_d,2h-y_d}$ by modifying the operators $M$ and $L$, as shown in Appendix B. 

  If there is no forcing, $\mathbf{q_f}=0$, the problem is converted into a generalized eigenvalue problem $\omega \mathbf{q_r} = M^{-1}(-\mathrm{i}L) \mathbf{q_r}$ and the complex eigenvalues $\omega \in \mathds{C}$ are sought. In agreement with commonly used notation, we will use $c = \omega /k_x = c_r + \mathrm{i} c_i $ as a measure of the stability of the system, $c_r = \omega_r/k_x$ being the phase speed of the disturbance in the $x$-direction, and  $c_i = \omega_i /k_x$ representing its growth rate. The criterion for instability is $c_i>0$.   The numerical method employed here for linear analysis is detailed in \citet[Appendix~A]{Schmid2012}. In short, we consider the generalised eigenvalue problem obtained by setting $\mathbf{q_f}=0$ in~\eqref{eq:freq_linsys}. This problem contains derivatives in $y$ up to a fourth order and is further discretized in $y$ with spectral Chebyshev collocation method, resulting in an $N\times N$ matrix, with $N$ eigenvalues and eigenvectors. Here we used $N=256$, or $N = 512$ for large values of $|A|$, but we also tested our results with higher $N$ to ensure the absence of spurious eigenvalues.  It is straightforward to reduce~\eqref{eq:OrrSomm},~\eqref{eq:Squire} to an inviscid problem by eliminating viscosity $\nu_t$. The new second-order differential equation in $y$ only requires two boundary conditions on $v_{w}$, one for each wall of the channel. The conditions $\partial v / \partial y =0$ and $\omega_y=0$ at the wall no longer hold in an inviscid flow where wall-parallel velocities are not required to be zero, and therefore should be removed from the system.

\textbf{\textit{Response to a forcing.}} Consider the problem~\eqref{eq:freq_linsys} in a more general form,  where the forcing is nonzero  $\mathbf{q_f} \neq 0$, but its shape is not known \textit{a priori}. In a noisy nonlinear system such as turbulent shear flow, this choice is reasonable. If $\omega$ is real (let us denote it $\omega_f$), it appears as an additional parameter in~\eqref{eq:freq_linsys}. The response of the system to this general forcing can be derived as
    \begin{equation}\label{eq:res_op}
    \mathbf{q_r} = (M^{-1} L  - \mathrm{i} \omega_f I)^{-1} M^{-1} \mathbf{q_f} = \mathcal{H} \left\lbrace M^{-1} \mathbf{q_f} \right\rbrace,  
    \end{equation}
    where $\mathcal{H}$ is the resolvent operator of~\eqref{eq:freq_linsys}.
      The spectral norm of the operator $\mathcal{H}$ represents the maximum  amplification of the response to a forcing with frequency $\omega_f$,
\begin{equation}\label{eq:res_norm}
|| \mathcal{H} || \equiv \sup_{\mathbf{q_f} \neq 0  } \frac{|| \mathbf{q_r}||}{||M^{-1} \mathbf{q_f} (y)||}.
\end{equation}
   This norm, weighted by the total energy of the flow, can be computed as the first (largest) singular value, $\sigma_0$, of the singular value decomposition (SVD) of the operator $\mathcal{H} = U \Sigma V^T$. Here the diagonal matrix $\Sigma$ contains the singular values (relative amplitudes of the response), and the matrices $U$ and $V$ are the optimal responses and forcings, respectively, ranked by the amplitude of response. In the following we consider only the flow responses related to the largest singular values, as in the so-called rank-$1$ model introduced by \citet{mckeon2010critical}.  They proposed two amplification mechanisms: the first one, through the shear $U'$ and related to it transient growth, and the second one, through amplification at the critical layer where the phase speed of the forcing $c_f=\omega_f/k_x$ matches the local mean velocity.

\section{Large-scale control in DNS}\label{sec:dnsres}

\subsection{Virtual wall effect for large scales}\label{sec:stats}
\begin{figure}
\centering
  \begin{subfigure}[t]{0.45\textwidth}
  \caption{}\label{fig:meanvelprof}
  \centering
    \includegraphics[width=\textwidth]{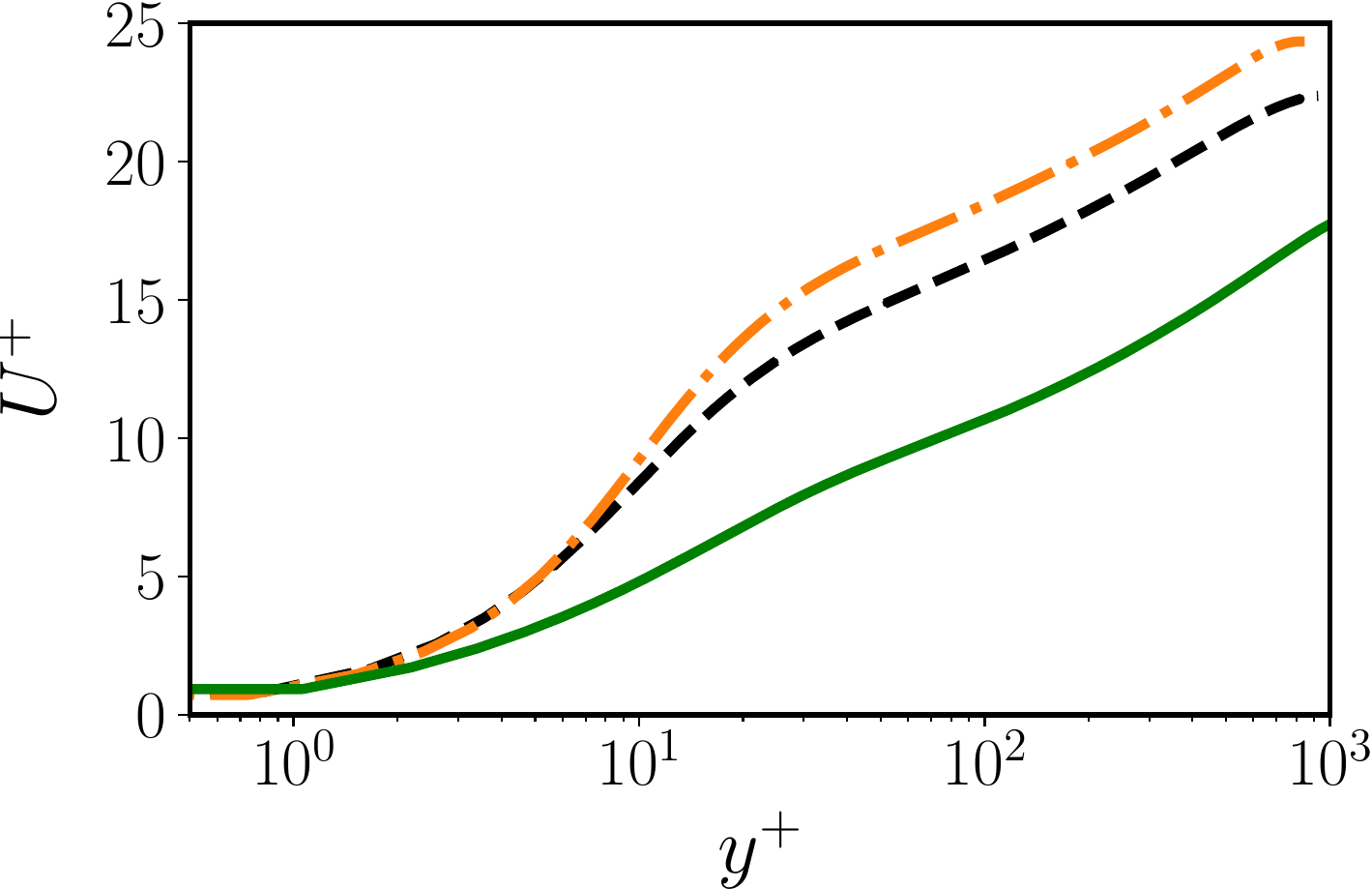}
  \end{subfigure}\hspace{1em}
  \begin{subfigure}[t]{0.45\textwidth}
     \caption{}\label{fig:virtual_wall}
  \centering
   \includegraphics[width=\textwidth]{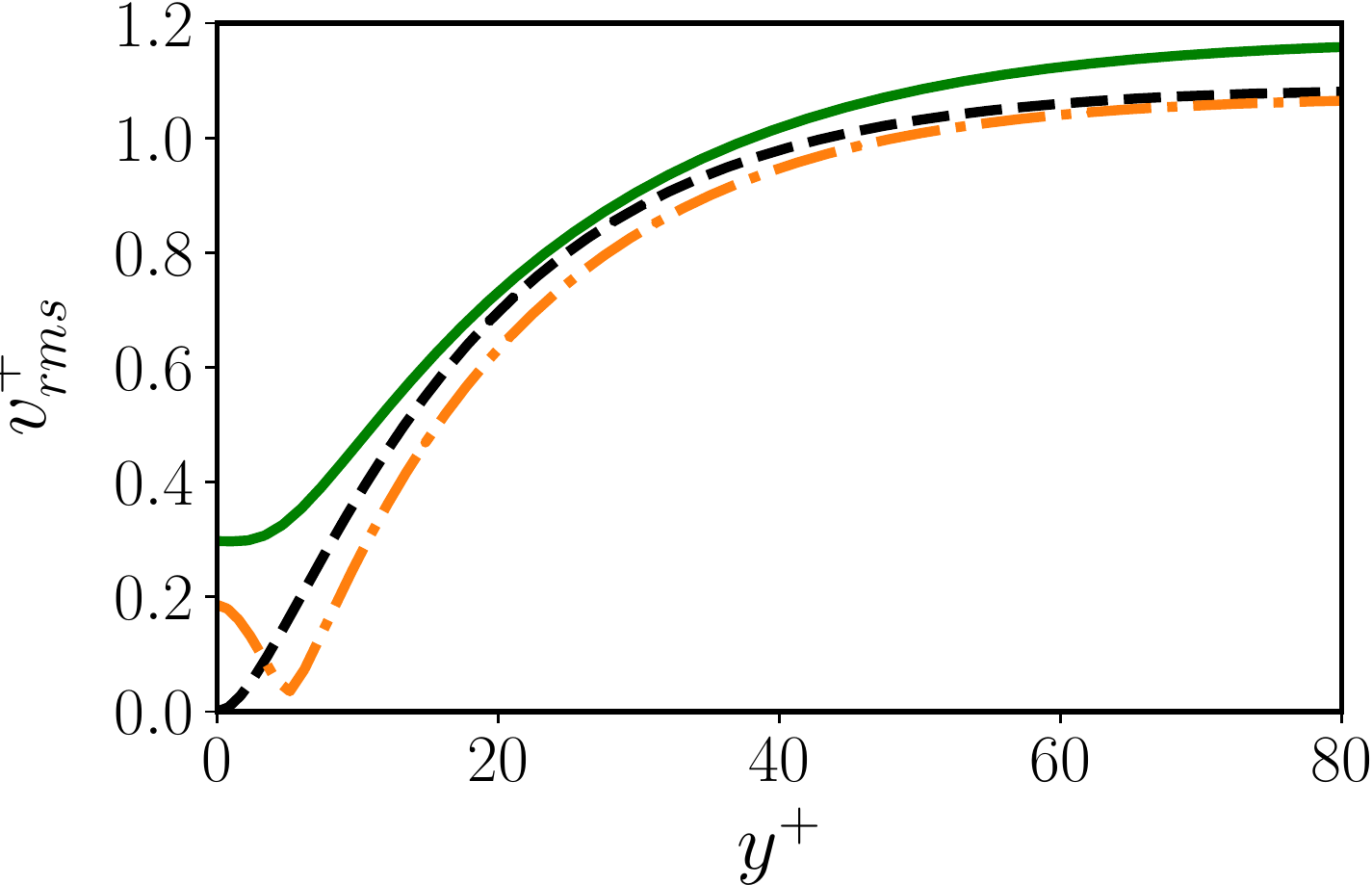}
  \end{subfigure}
    \begin{subfigure}[t]{0.45\textwidth}
        \caption{}\label{fig:vrms_lsonly}
  	\centering
    \includegraphics[width=\textwidth]{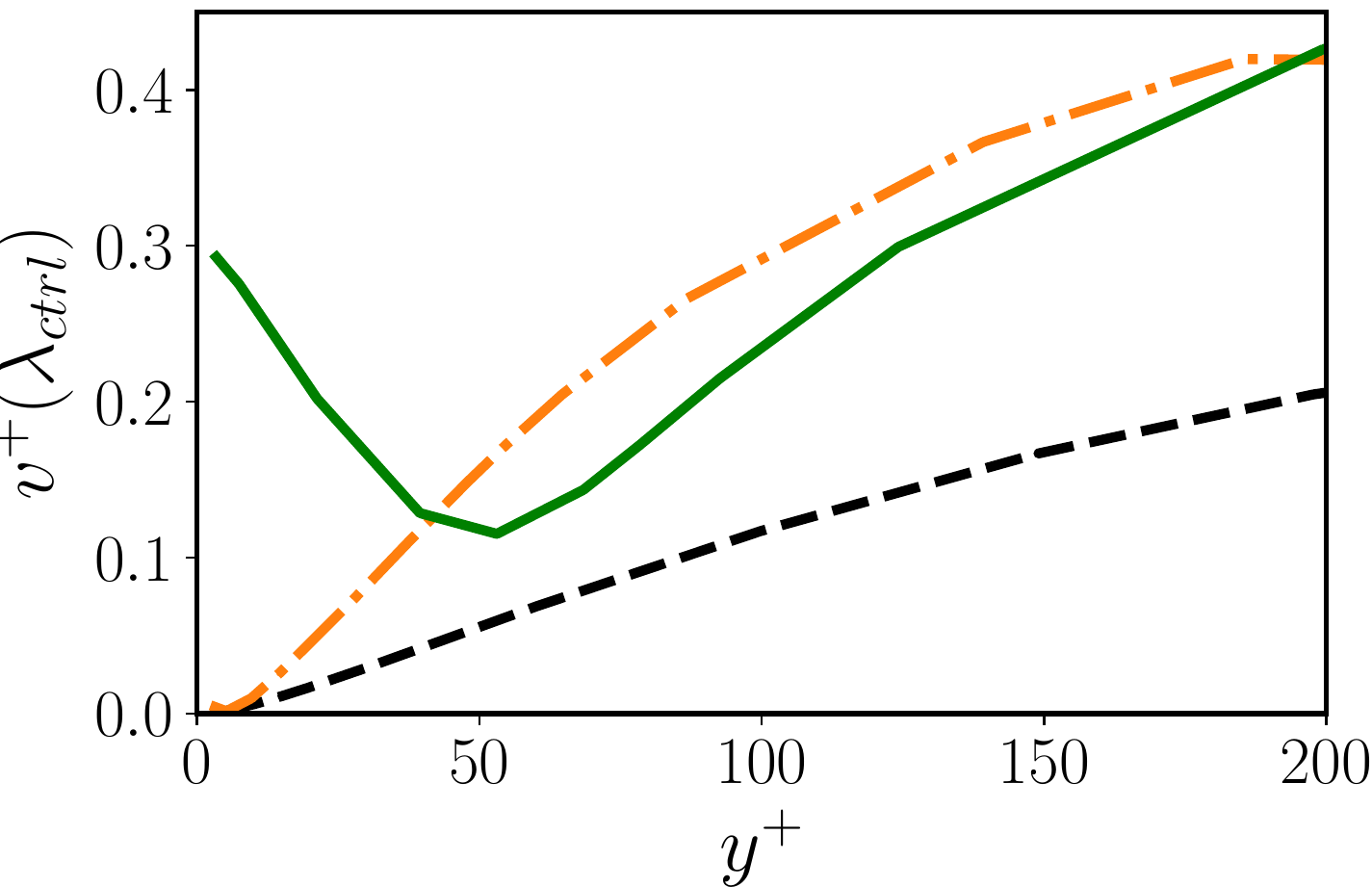}
  \end{subfigure}\hspace{1em}
    \begin{subfigure}[t]{0.45\textwidth}
    \caption{}\label{fig:y0_amp}
    \centering
    \includegraphics[width=\textwidth]{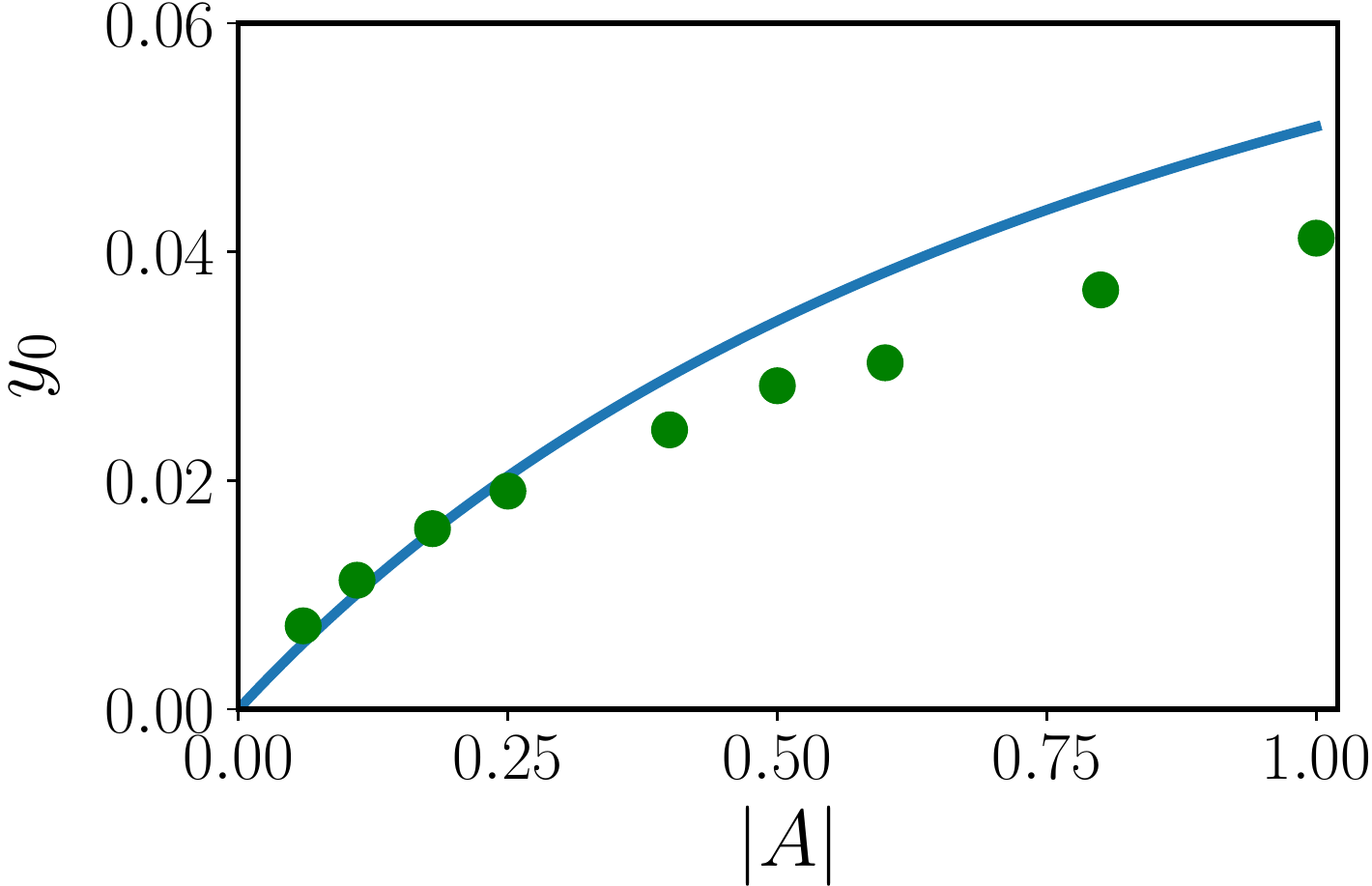}
  \end{subfigure}
  \caption{(a)  Influence of the control on the mean velocity profile $U^+$.  \dashed , without control; \dashdotted , opposition control with all flow scales involved and $y_d^+ = 10$; and \solid , opposition control with only large scales and $y_d/h=0.1$. (b) The rms of wall normal velocity. The virtual wall effect of control with $y^+_d=10$ produces a minimum at $y^+ = 5$, in contrast to the large-scale control at  $y_d/h=0.1$. (c) The contribution to the rms of $v$ from the large scales (see text for more details). (d) Wall-normal coordinate of the local minimum in panel (c) for the large-scale control as a function of the control gain, together with the fit~\eqref{eq:vwall_loc}. Control parameters in panels (a-c): $|A|=1$, $x_0 = 0$.}
  \label{fig:velstat}
\end{figure}
The effect of wall modification is reflected in the turbulent mean profile in the log layer \citep{nikuradse1933stromungsgesetze}. In wall units, $U^+ = \gamma^{-1} \mathrm{log} (y^+) +B$, and the wall modifications such as roughness or control preserve the slope of the logarithmic law, $\gamma^{-1}$, but change the intercept constant of the profile $B$ \citep{townsend1980structure}. The decrease in $B$ is related to an increase in friction factor $C_f$, as in flows above rough walls, while an increase in $B$ is related to drag-decreasing effect \citep{jimenez1994structure}. 
Figure~\ref{fig:meanvelprof} compares the mean velocity profiles of the flow subject to large-scale control, $y_d/h = 0.1$, with the ``classic" opposition flow control, $y_d^+ = 10$, and with the uncontrolled flow. The classic opposition control results, as expected, in a shift of the mean profile upwards with the corresponding decrease in drag, $\Delta C_f = (C_f - C_{f0}) / C_{f0}\approx -0.17$. On the contrary, opposing only large wavelengths results in the shift of the mean velocity profile downwards, and a significant drag increase of $\Delta C_f \approx 50$\%.

A possible physical explanation to the success of opposition flow control in the first case is that opposing flow at the wall creates so-called virtual wall effect for small eddies with sizes about $10-15$ wall units \citep{luchini1991resistance,jimenez1994structure}. These ideas were also proven useful in the application to drag reduction with riblets \citep{garcia2011drag}.  Figure~\ref{fig:virtual_wall} illustrates this effect using the root mean square (rms)   of the wall-normal velocity component,  referred to as $v^+_{rms}$ from now on. In the case of ``classic" opposition control with $y_d^+=10$, the minimum of $v^+_{rms}$ appears  at $y_0^+ = 5$,  midways to the detection plane. There is an expected peak at the wall due to non-zero boundary condition on $v$, but away from the wall the overall intensity of $v$ decreases with respect to the uncontrolled flow. On the contrary, the large scale opposition control strategy  enhances  $v^+_{rms}$ for all $y$. At first sight, there is no visible minimum of $v$ near the wall. To find out whether the chosen control strategy affects large structures, we determine their contribution to  $v^+_{rms}$ from the spectrum of the wall-normal velocity, and denote this quantity as $v^+_{rms} (\lambda_{ctrl})$. It is calculated by adding only the values 
corresponding to controlled wavelengths from table~\ref{tab:sim_param} for each wall-normal location.  Figure~\ref{fig:vrms_lsonly} shows $v^+_{rms}  (\lambda_{ctrl})$ for the  uncontrolled case, for classical opposition control, and for the opposition control of the large wavelengths. Both control strategies increase the contribution to $v^+_{rms}$ from the large structures of $v$, but the large-scale opposition  produces a minimum near the wall for those structures. This minimum is located at $y_0^+ \approx 50$, which corresponds, approximately, to half of the wall-normal distance to the detection plane. The analogous minimum for the classic opposition control is much less pronounced since  there is little energy in large scales of $v$ at $y_d = 10$. 

If $v$ is assumed to vary linearly between the wall and the sensor location, then, given \eqref{eq:ctrl_law}, 
\begin{equation}\label{eq:lin_v_wall}
    \hat{v} /\hat{v} (y_d) = (y/y_d) (1 + A) - A.  
\end{equation}
The location of the virtual wall can be defined as a minimum of $|v/v_d|$, reached at   
\begin{equation}\label{eq:vwall_loc}
y_0 /y_d= \frac{ \Re(A) + (\Re(A))^2 +(\Im(A))^2}{(1 + \Re(A))^2 + (\Im(A))^2},
\end{equation}
where $\Re$ and $\Im$ denote the real and imaginary parts. It reduces to 
\begin{equation}\label{eq:vwall_loc_realA}
  y_0 /y_d= \frac{A}{1 + A} ,
\end{equation}
when $\Im(A)=0$, i.e. when the control coefficient $A$ is purely real.  It follows from~\eqref{eq:xlag}  that this happens when $-k_x x_0 = 2 \pi n$, $n=0, 1, ...$ . One specific case of this condition is when no phase shift is introduced, i.e. $x_0 =0$. In figure~\ref{fig:y0_amp} we test if the assumption of linearity is valid for the large-scale control. The control gain $|A|$ is varied while the phase of control is kept zero, and the location of the minimum in $v^+_{rms} (\lambda_{ctrl})$ is recorded. For small values of $|A| \leq 0.25$ the location of the minimum in $v$ is predicted quite well by \eqref{eq:vwall_loc}, while for larger control gain the trend, while still increasing, no longer exactly follows the linear prediction.

\subsection{Introducing streamwise offset between sensing and actuation}\label{sec:xlag}
\begin{figure}
\centering
  \begin{subfigure}[t]{0.45\textwidth}
      \caption{}\label{fig:cf_mean}
  \centering
    \includegraphics[width=\textwidth]{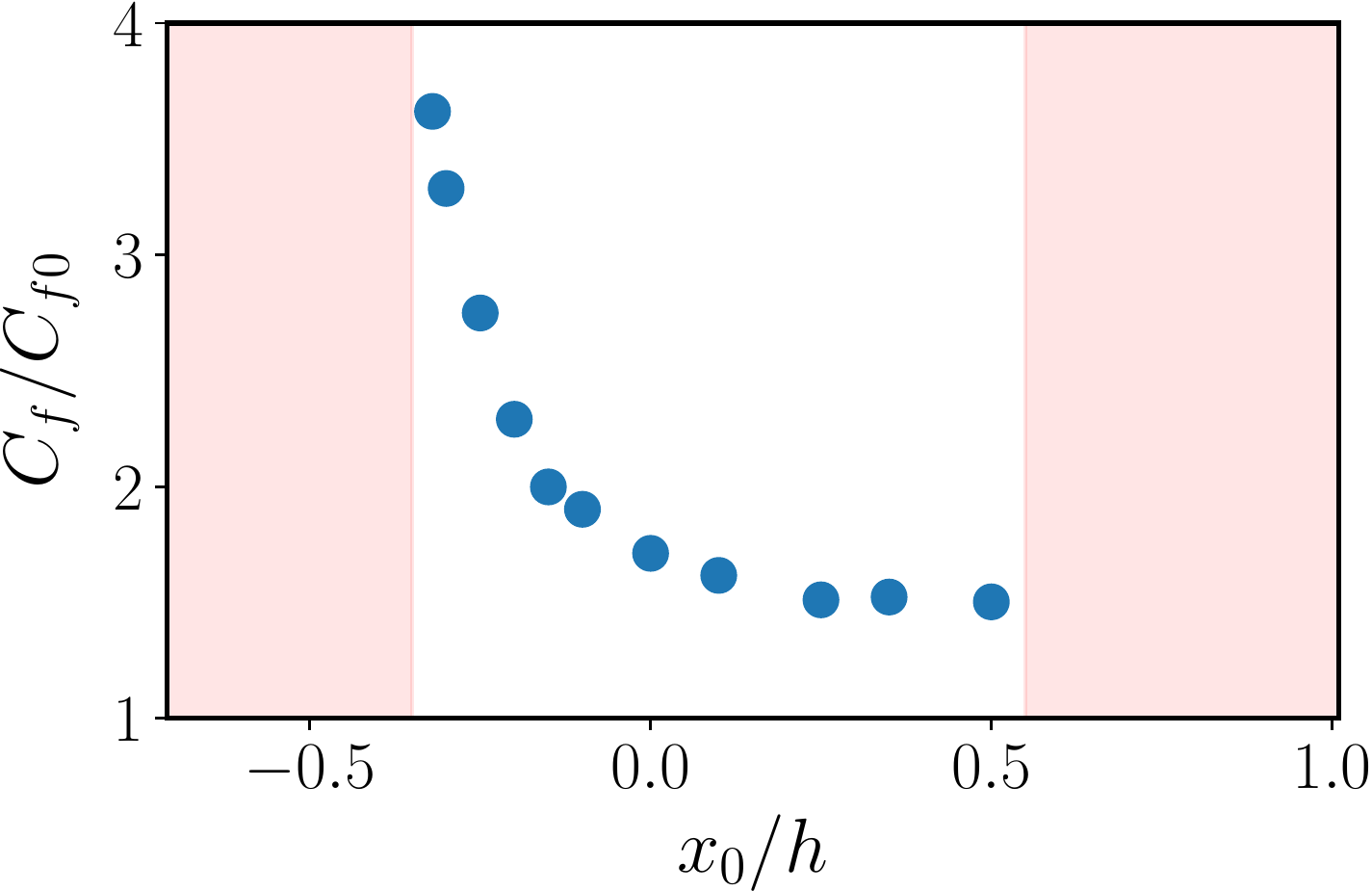}
  \end{subfigure}\hspace{1em}
   \begin{subfigure}[t]{0.45\textwidth}
      	\caption{}\label{fig:cf_amp}
  	\centering
   	\includegraphics[width=\textwidth]{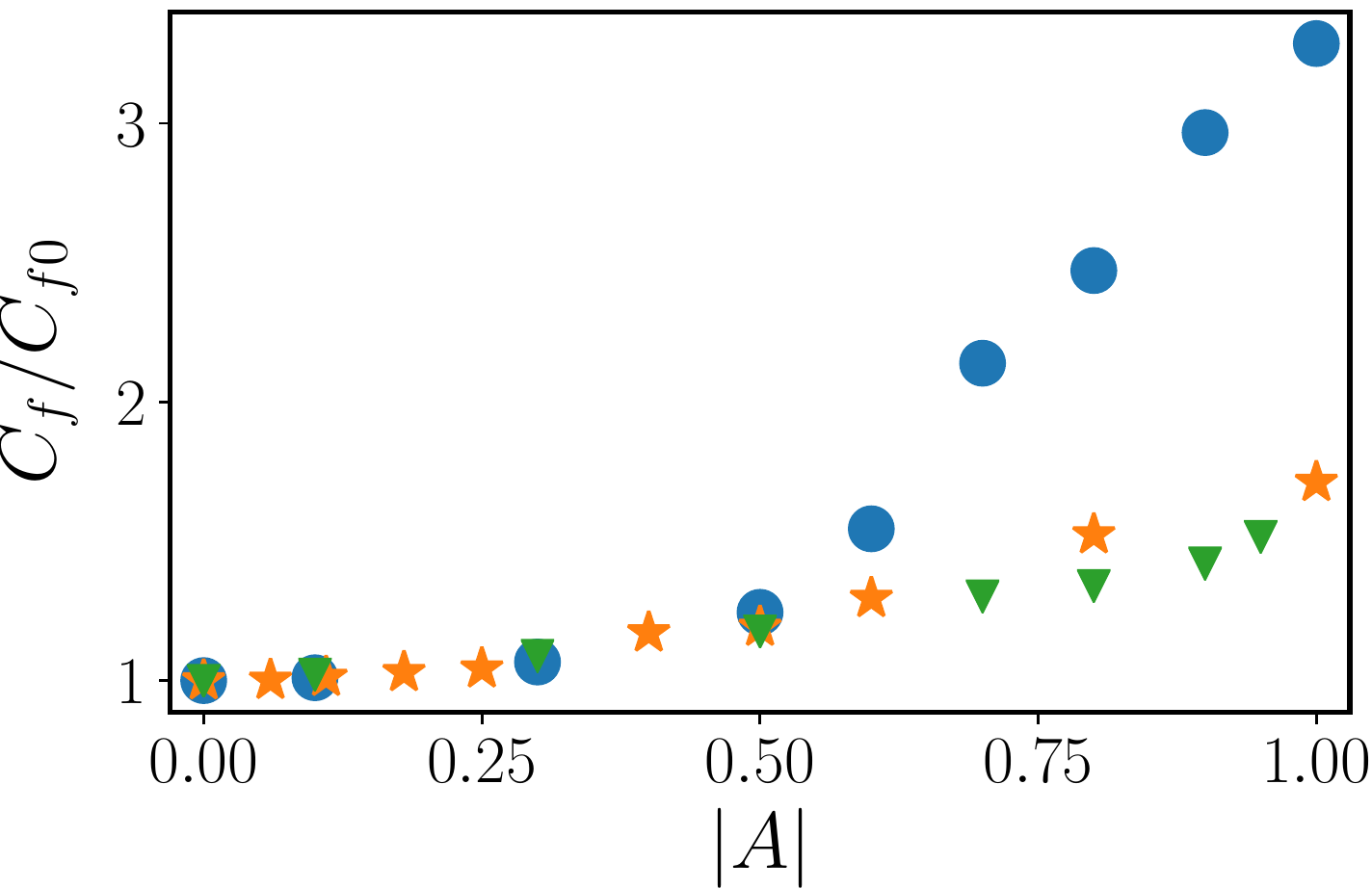}
  \end{subfigure}
  \caption{(a) Friction factor $c_f$ as a function of streamwise shift $x_0$ of the control normalized with friction factor of the uncontrolled flow, $|A| = 1$. Shaded regions denote the areas of the sharp increase in friction where the resolution in the table~\ref{tab:sim_param} is insufficient to converge. (b) Friction factor as a function of the control gain $|A|$. \protect\marksymbol{*}{black}, $x_0 = -0.3$; \protect\marksymbol{star}{black},  $x_0 = 0$;
$\blacktriangledown$, $x_0 = 0.55$.}
  \label{fig:cf}
\end{figure}
In the spirit of~\eqref{eq:xlag} we modify the control by shifting it upstream or downstream, creating an offset between sensing and actuation.
 Figure~\ref{fig:cf_mean} presents results of a control experiment in which $x_0$ is varied in order to search for an optimal control delay. The results of control are relatively insensitive to the downstream shifts: the drag increase stays at the level of 50\% for most of the positive values of $x_0$. Negative, upstream shifts produce much larger drag increase, up to 300\% for $x_0 = -0.3$. Both very large negative ($x_0 \leq -0.35$) and positive ($x_0 \geq 0.55$) streamwise shifts result in control instability. In both cases, the wall-normal velocity grows to the levels at which the resolution of the code is insufficient and the simulations diverge. There exists nevertheless a clear difference between the two limits. While at large negative shifts the instability manifests itself in a gradual increase in drag, for large positive shifts it appears suddenly.  In figure~\ref{fig:cf_amp} we plot the friction factor for the two limits $x_0 = 0.55$ and $x_0 = -0.3$, as well as the regular control with no shift. For all three cases, we observe that for relatively low control gain $|A|\leq 0.5$ the introduction of large-scale opposition only results in a mild increase in drag. The flow almost does not `feel' this control. For higher control gain,  both $x_0=0$ and $x_0=0.55$ saturate around $c_f/c_{f_0} = 1.5$, with the latter performing slightly better. As $|A|\to 0.95$, and $x_0=0.55$,  the friction suddenly increases sharply, and the transition on the right side of figure~\ref{fig:cf_mean} is approached. For $x_0 = -0.3$, which is close to the transition on the left, the drag continuously increases beyond $|A| \approx 0.5$. The large increase in drag obtained with the large-scale control strategy is very different from what is expected from the ``classic" opposition control. Together with the control instability, it shows that large-scale structures in the logarithmic layer in the channel flow are very sensitive to actuation.

\subsection{Spanwise rollers and oblique waves}\label{sec:oblique}

\begin{figure}
\centering
  \begin{subfigure}[t]{0.45\textwidth}
       \caption{}\label{fig:u10p_unc}
     \includegraphics[width=\textwidth]{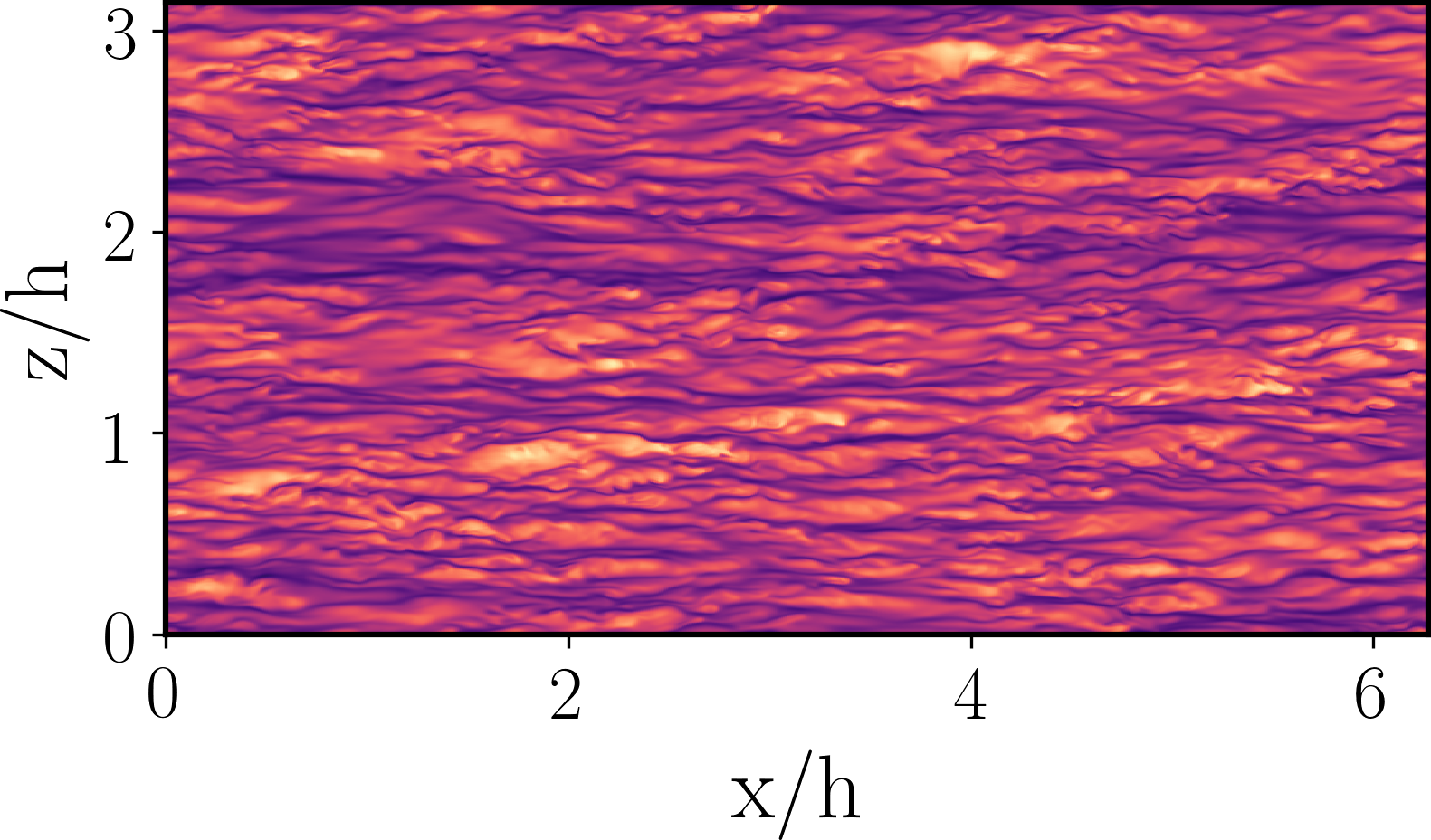} 
  \end{subfigure}\hspace{1em}
  \begin{subfigure}[t]{0.45\textwidth}
       \caption{}\label{fig:u10p_xlag_m03}
  	 \includegraphics[width=\textwidth]{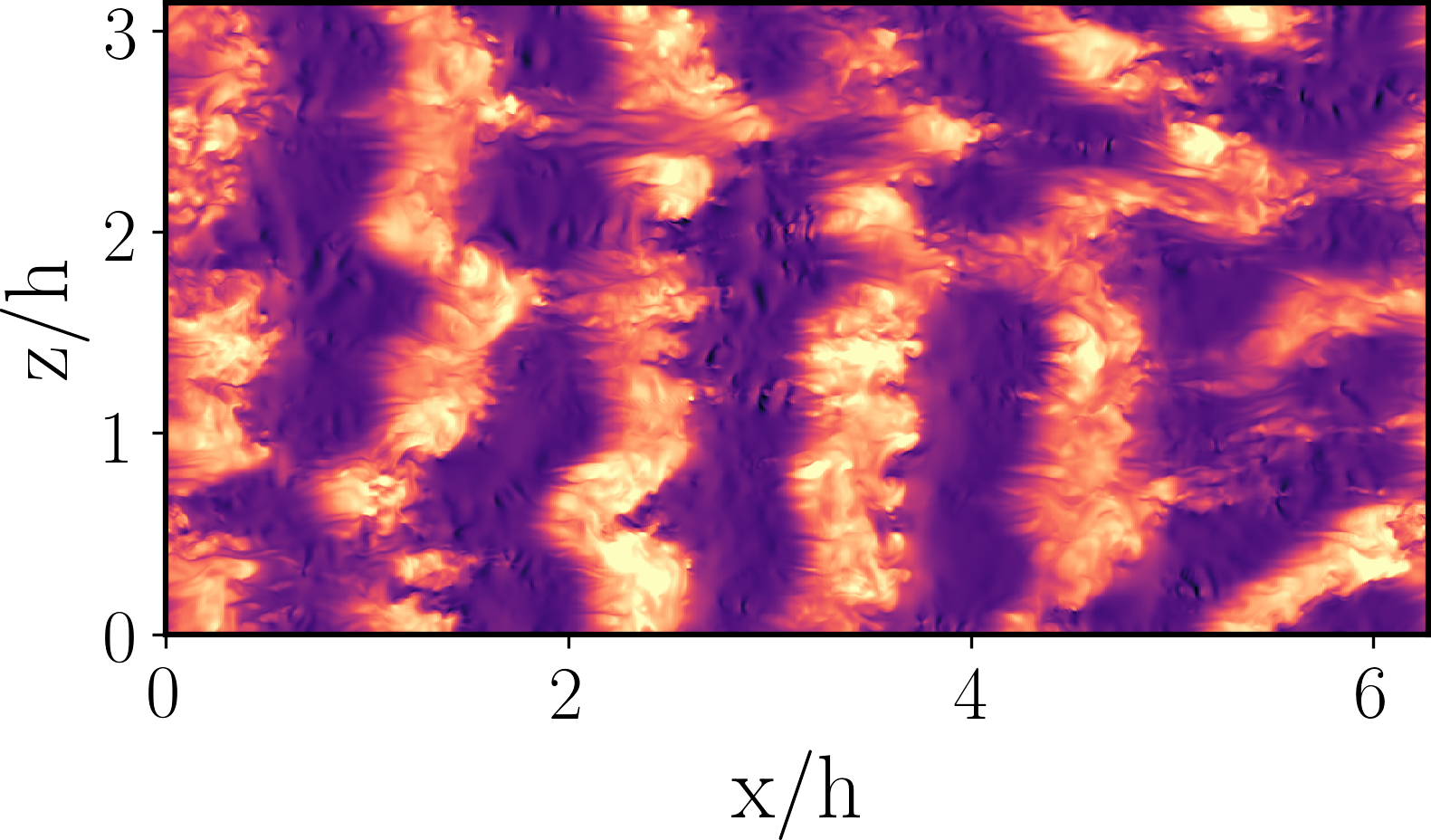} 
  \end{subfigure}\\[1ex]
    \begin{subfigure}[t]{0.6\textwidth}
         \caption{}\label{fig:u10p_full_kxkz}
  	 \includegraphics[width=\textwidth]{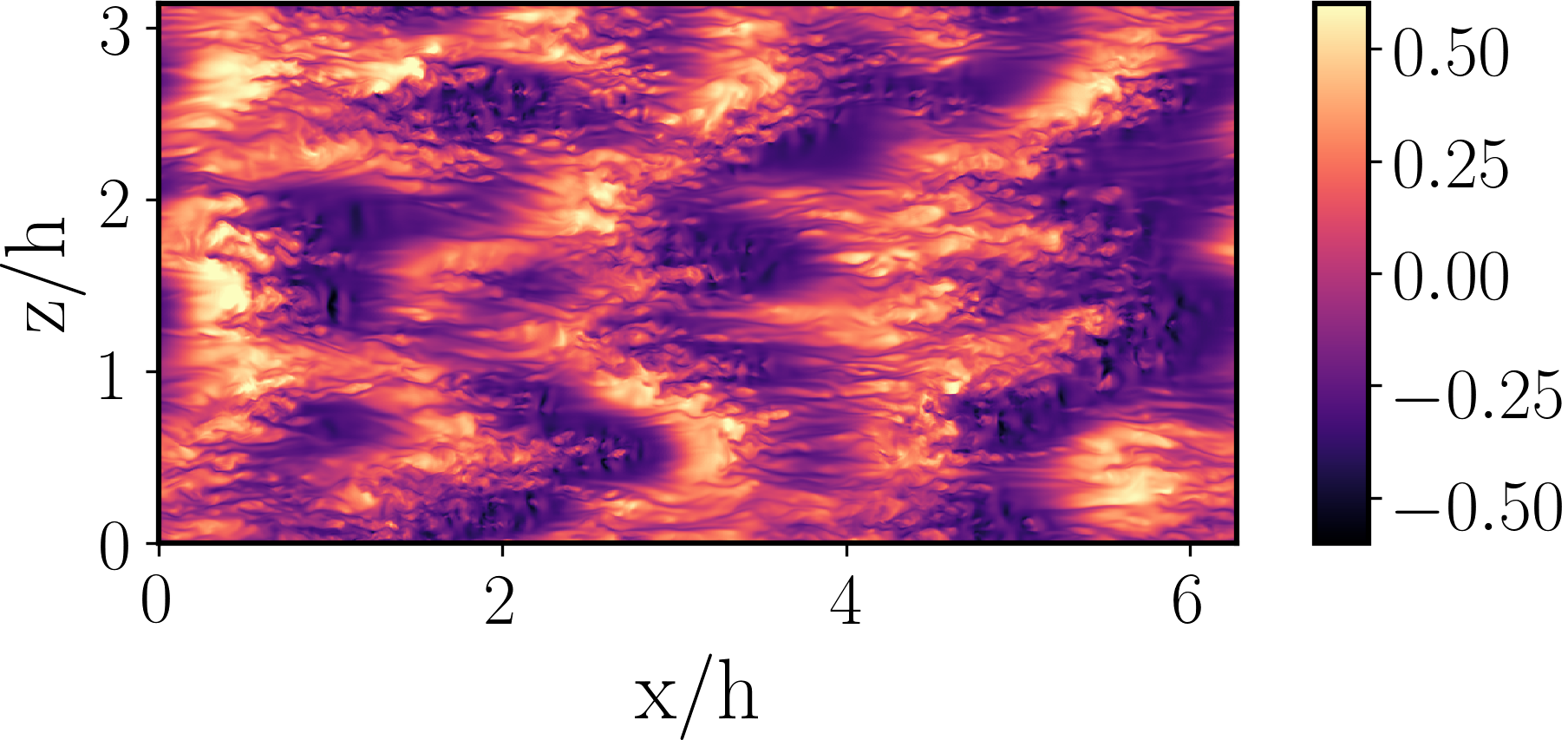} 
  \end{subfigure}\\[1ex]
  \caption{Instantaneous snapshots of streamwise velocity fluctuations near the wall ($y^+ = 10$), normalized with the bulk velocity. (a) Uncontrolled  flow.  (b) Large-scale control with negative phase shift, $x_0 = -0.3$.  (c)  Large-scale control, $x_0 = 0$. The colorbar is the same for all the snapshots. }
   \label{fig:u10p}
\end{figure}

To illustrate the effect of control on the streamwise velocity, we show in figure~\ref{fig:u10p} instantaneous snapshots of the streamwise velocity fluctuations in the buffer layer ($y^+ = 10$). Figure~\ref{fig:u10p_unc} shows a ``normal" snapshot of uncontrolled flow. The flow is populated with buffer-layer low- and high-velocity streaks with footprints of two larger $u$-structures, possibly logarithmic-layer streaks. Figure~\ref{fig:u10p_xlag_m03} also contains a streamwise velocity snapshot, but now for the case with large-scale control and negative streamwise shift $x_0 = -0.3$. This case corresponds to the sharp increase in drag on the left side of figure~\ref{fig:cf_mean}. The streaky structure of the flow is completely lost and, instead, five or six spanwise rollers appear that are almost homogeneous in spanwise direction. Figure~\ref{fig:u10p_full_kxkz} also shows $u$ but for the case of control with no streamwise shift $x_0 = 0$. In this case the streaky structure of the uncontrolled flow is also lost, but spanwise-homogeneous rollers do not appear. Instead, oblique-like waves with inclination in $x-z$ plane occur. As visible from the snapshot, the $(x,z)$ lengths of the waves are approximately $(2h, h)$, which is within the range of controlled wavelengths in table~\ref{tab:sim_param}. 
\begin{figure}
\centering
  \begin{subfigure}[t]{0.4\textwidth}
  	\centering
  	    \caption{}\label{fig:Euu_lx}
    \includegraphics[width=\textwidth]{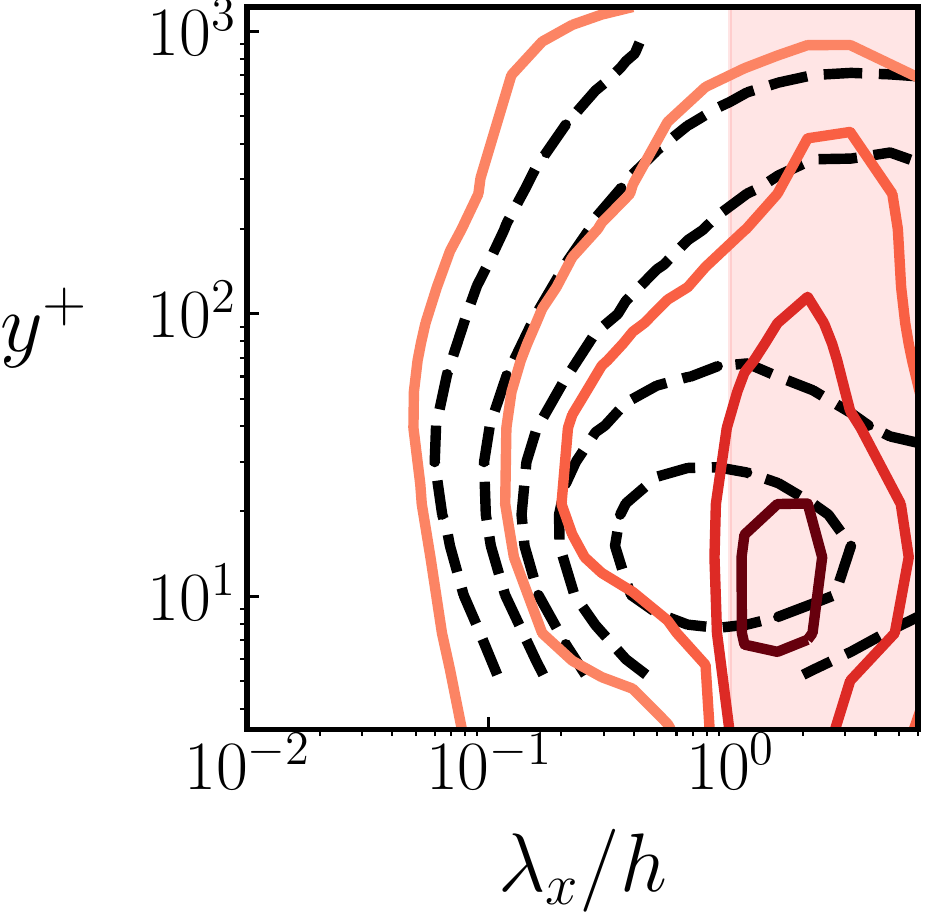} 
  \end{subfigure}\hspace{1em}
  \begin{subfigure}[t]{0.4\textwidth}
      \caption{}\label{fig:Euu_lz}
  	\centering
  	\includegraphics[width=\textwidth]{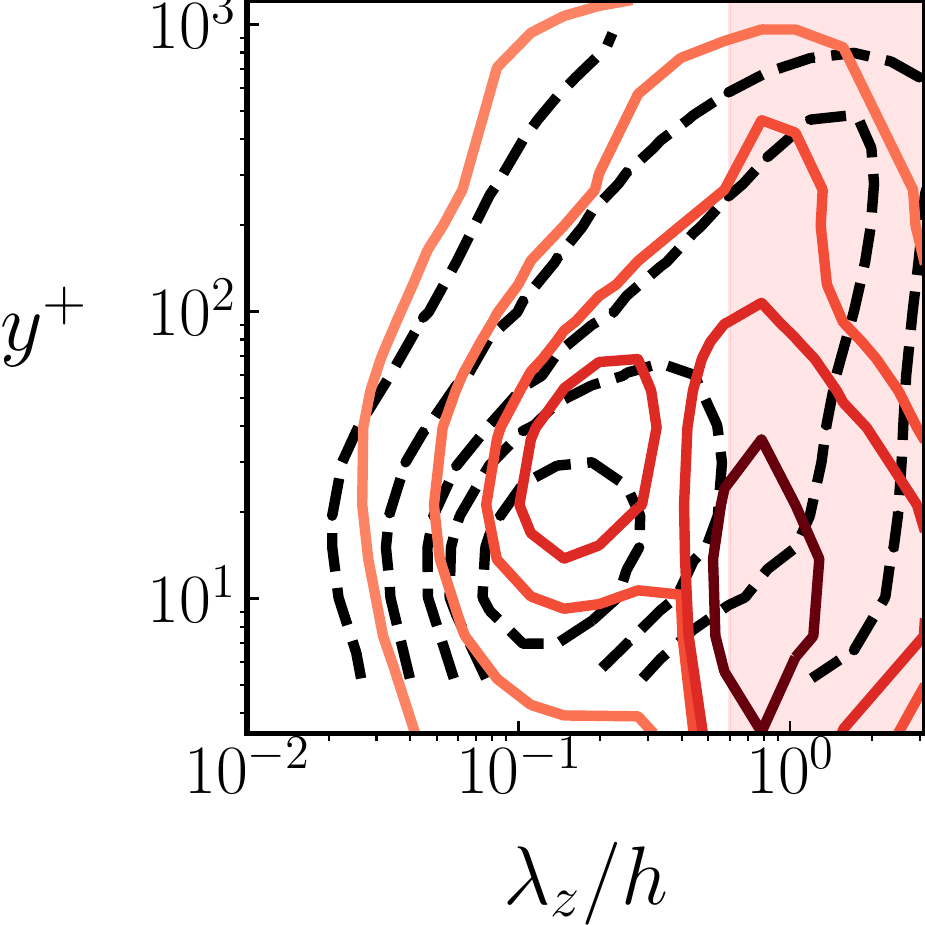} 
  \end{subfigure}\\[1ex]
  \begin{subfigure}[t]{0.4\textwidth}
    \caption{}\label{fig:Evv_lx}
  	\centering
  	\includegraphics[width=\textwidth]{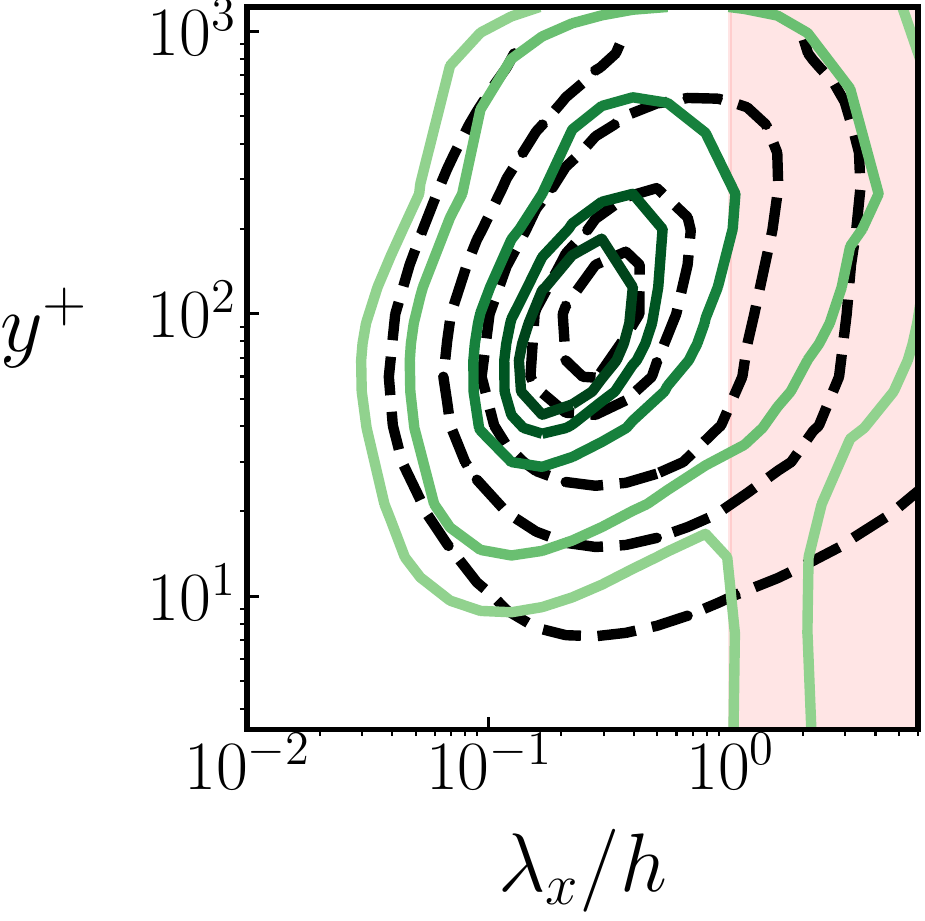}
  \end{subfigure}\hspace{1em}
  \begin{subfigure}[t]{0.4\textwidth}
      \caption{}\label{fig:Euv_lx}
  	\centering
  	\includegraphics[width=\textwidth]{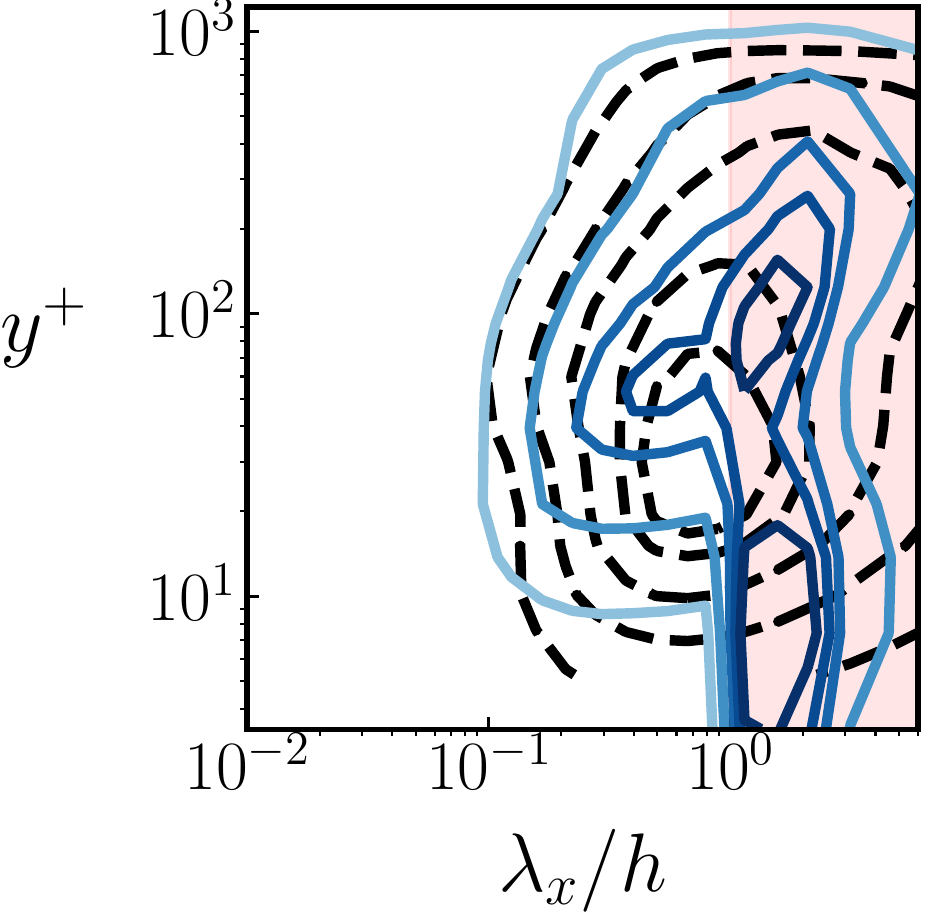}
  \end{subfigure}
\caption{(a,b) Spectrum of $u$ as a function of wall-normal distance and streamwise (a)  and spanwise (b) wavelength. (c) Spectrum of $v$ as function of $y$ and $\lambda_x$. (d) The same for ${-uv}$. The spectrum in (b) was premultiplied with $k_z$, while the rest was premultiplied with $k_x$. All data are normalized with $u_\tau$ and integrated in the third spatial direction. \dashed , the uncontrolled flow, \solid , the controlled flow. The contours contain $97.5$\%, $80$\%, $45$\%, $23$\% and $4$\% of the total spectral mass. Control parameters: $|A| = 1$, $x_0 = 0$, shaded regions denote controlled wave lengths from table~\ref{tab:sim_param}.}
\label{fig:Euu_spec}
\end{figure}

To identify the  length scale of the oblique waves chosen by the flow in figure~\ref{fig:u10p}(c) more precisely, we plot in figures~\ref{fig:Euu_spec}(a,b) the spectrum of the streamwise velocity component, $E_{uu}$,  as a function of $\lambda_x$, $\lambda_z$ and of the distance from the wall.  It is clear from the spectra that applying opposition flow control on the large scales not only creates a significant footprint on the spectrum of the wall-normal velocity component, $E_{vv}$, at the wall, as seen in figure~\ref{fig:Evv_lx}, but also on the spectrum of $u$. This response, although distributed over various length scales, peaks near $\lambda_x/h, \lambda_z/h \approx 2,1$, corresponding to the structures visible in the snapshots. The wall-normal location of the maximum of the streamwise energy component is in the buffer layer at $y^+ \approx 10$. With non-zero $E_{vv}$ at the wall and $E_{uu}$ peaking in the buffer layer, the  Reynolds stress spectrum $E_{uv}$ has also a maximum in that area (figure~\ref{fig:Euu_spec}d). As a result, we observe a significant increase in friction, which is reflected in an increase of the effective friction velocity and of $Re_\tau$ of the controlled flow.  Notice a local minimum in the stress contours at about $y^+ \approx 50$, corresponding to the location of the minimum in rms of $v$ in figure~\ref{fig:vrms_lsonly}. At the same time, some of the energy contours of the controlled flow  are shifted towards the left from the uncontrolled case. This is most visible in the spectrum of the wall-normal component of energy (figure~\ref{fig:Euu_spec}c). 

The  waves observed in Fig.~\ref{fig:u10p} are not found in regular channel flows, suggesting that the applied control strategy can best be understood as a forcing on $v$ at the wall. This forcing can be deleterious in terms of drag reduction, even if it creates a positive virtual-wall effect for large scales (figure~\ref{fig:velstat}c). The increase in drag created in DNS  by large-scale control could signal either the presence of a linear instability, or an amplified linear response of the system to the control. In the next sections we employ the methods of section~\ref{sec:setup_linstab} to show that the oblique waves and spanwise rollers induced by the large-scale control can be explained by the linearized dynamics of the Navier--Stokes equations.

\section{Linear stability of the inviscid channel flow}\label{sec:linstab_invisc}

\subsection{Analysis of the Rayleigh equation}\label{sec:theory_invisc}
We begin by exploring the stability of the inviscid flow subject to control, as it is the most simplified linear model of the channel flow.  The inviscid flow, linearized about the uncontrolled mean profile, is governed by the Rayleigh equation, which is~\eqref{eq:OrrSomm} without viscosity,
\begin{equation}\label{eq:rayleigh}
(U-c) (D^2 - \kappa^2) \hat{v} = U^{''} \hat{v},
\end{equation}
where $c = \omega/k_x=  c_r + \mathrm{i} c_i$. This is a second-order problem with $y\to -y$ symmetry. Any general solution to~\eqref{eq:rayleigh} can be expressed as a linear combination of a symmetric and an antisymmetric solution.  In the uncontrolled flow, the coefficients of~\eqref{eq:rayleigh} are real, and if $\hat{v}$ is an eigenfunction of~\eqref{eq:rayleigh} with eigenvalue $c$, so is its complex conjugate $\hat{v}^*$ with eigenvalue $c^*$. The boundary conditions~\eqref{eq:ctrl_law}, involving the complex coefficient $A$, destroy this property of the flow. Making the change of variables $\hat{v} = \tilde{V} (U-c)$, multiplying by the complex conjugate and integrating over the wall-normal coordinate  \citep[pp. 23-24]{Schmid2012}, we get 
\begin{equation}\label{eq:rayleigh_int}
\int_{0}^{2 h} (U-c)^2 \underbrace{\left( |D \tilde{V}|^2 + \kappa^2 |\tilde{V}|^2 \right)}_{Q \geq 0} \dd y =  \tilde{V}^* (U-c)^2 D \tilde{V} \mid_{0}^{2 h}, 
\end{equation}
where $\tilde{V}^*$ denotes the complex conjugate of $\tilde{V}$. The real and  imaginary parts of~\eqref{eq:rayleigh_int} are
\begin{align}
 \int_{0}^{2h} \left[ (U -c_r)^2 - c_i^2 \right] Q \dd y &= \Re[\tilde{V}^* (U-c)^2 D \tilde{V} \mid_{0}^{2h}],  \label{eq:rayleigh_re} \\
 -2 c_i \int_{0}^{2h} (U - c_r) Q \dd y & =  \Im[\tilde{V}^* (U-c)^2 D \tilde{V} \mid_{0}^{2h}],  \label{eq:rayleigh_imag}
 \end{align}
 respectively.

 In the case of homogeneous boundary conditions, $\hat{v} = \tilde{V} \mid_{0}^{2h} = 0$, the right hand side in~\eqref{eq:rayleigh_imag} is equal to zero. Since $Q$ is non-negative, $(U-c_r)$~must change sign in the interval $[0, 2h]$ for non-trivial solutions. It follows that, 
in the case of impermeable walls, the advection speed of perturbations $c_r$ is bounded by the mean velocity profile 
\begin{equation}\label{eq:cond_on_cr}
     U_{min} < c_r < U_{max}.
\end{equation} 

However, in the case of inhomogeneous boundary conditions, like those introduced by our control, the right hand side in~\eqref{eq:rayleigh_imag} is non-zero and we get   
\begin{equation}\label{eq:cr_constraint}
\int_{0}^{2h} (U - c_r) Q \dd y = - \frac{\Im[\tilde{V}^* (U-c)^2 D \tilde{V} \mid_{0}^{2h}]}{2 c_i} \equiv C_{\Im}
\end{equation}
from~\eqref{eq:rayleigh_imag}.
If $C_{\Im}>0$, then $c_r \leq U_{max}$, or the integral would be negative, but $c_r$ \textit{can be smaller than} $U_{min}$. On the contrary, if  $C_{\Im}<0$, the integral can not be positive, and $c_r \ge U_{min}$. Restriction~\eqref{eq:cond_on_cr} is not applicable here. 
Once we have non-zero vertical velocity at the wall, the lower (or the upper) boundary on the phase speed of perturbations in~\eqref{eq:cond_on_cr} are relaxed.  As we will see below, this will result in $c_r <U_{min}$  or $c_r>U_{max}$ for some parameters of the opposition flow control.

 Assume a symmetric mean velocity profile $U(y)$ such that $U_{0} = U_{2h} = 0$. With~\eqref{eq:ctrl_law} in mind, the boundary conditions on $\tilde{V}$ become
 \begin{equation}\label{eq:vtilde_bc}
    \tilde{V}_0 =  A \tilde{V}_{y_d} \frac{U_{y_d} -c}{c}, \quad \tilde{V}_{2h} = A \tilde{V}_{2h - y_d} \frac{U_{y_d} -c}{c}.
\end{equation}
With the help of~\eqref{eq:vtilde_bc}, the right hand side of \eqref{eq:rayleigh_int} can be re-written as 
\begin{equation}\label{eq:rayleigh_RHS}
\tilde{V}^* (U-c)^2 D \tilde{V} \mid_{0}^{2h}  = c^2 \left( A  \frac{U_{y_d} -c}{c} \right)^*  \left[  \tilde{V}^*_{2h - y_d} [D\tilde{V}]_{2h} - \tilde{V}^*_{y_d}[ D\tilde{V}]_{0} \right].
\end{equation}

It is not straightforward to figure out the sign of the real and complex parts of~\eqref{eq:rayleigh_RHS}, since  $c^2$ and $A^*$ are complex numbers, and $\tilde{V}^*$, $D\tilde{V}$ are complex variables themselves. Some insight can be gained in a special case with $\phi = \pm \pi$, which makes the control coefficient real, $A = A^* = -|A|$. Further simplification comes from considering neutrally stable eigenvalues, $c_i = 0$, related to control.  When $c_i =0$,  the right-hand side of~\eqref{eq:rayleigh_imag} should be zero,  i.e. $\Im[\tilde{V}^* (U-c)^2 D \tilde{V} \mid_{0}^{2h}] =0$, but this does not give us any information about $c_r$. Instead, \eqref{eq:rayleigh_re} becomes the only condition on $c_r$. With the help of \eqref{eq:rayleigh_RHS} this condition reads 
\begin{align}
  \int_{0}^{2h}  (U -c_r)^2 Q \dd y &= - c_r (U_{y_d} - c_r) |A| C_\Re \label{eq:cr_pi_cond}, \\
 \text{where} \quad C_\Re &= \Re\left[  \tilde{V}^*_{2h - y_d} [D\tilde{V}]_{2h} - \tilde{V}^*_{y_d}[ D\tilde{V}]_{0} \right]. \nonumber
\end{align}
 Since the left hand side of~\eqref{eq:cr_pi_cond} is always non-negative, $C_\Re \geq 0$ if $c_r <0$ or $c_r > U_{y_d}$. Expanding \eqref{eq:cr_pi_cond}, we get a quadratic equation for $c_r$,
\begin{equation}\label{eq:cr_quad} 
  c_r^2 \underbrace{\left[ \int_{0}^{2h} Q \dd y - |A| C_\Re \right]}_{I_1} + c_r \underbrace{ \left[ -2  \int_{0}^{2h} U Q \dd y  + U_{y_d} |A| C_\Re \right] }_{I_2} + \underbrace{\int_{0}^{2h} U^2 Q \dd y}_{I_3} = 0,  
\end{equation}
with roots
\begin{equation}\label{eq:cr_roots}
c_{r}^{\pm}  = \frac{-I_2 \pm \sqrt{I_2^2 - 4 I_1 I_3}}{2 I_1}. 
\end{equation}
For the eigenvalues with $c_r^\pm$ to exist, the discriminant of~\eqref{eq:cr_roots} must be non-negative, $I_2^2 - 4 I_1 I_3 \geq 0$. To make further progress in understanding the roots of~\eqref{eq:cr_quad}, we need to know $Q(y)$ and $C_\Re$ for the eigenvectors at each particular $|A|$, and therefore we have to invoke the numerical stability analysis of~\eqref{eq:rayleigh}. The case $c_i = 0$ will be important later on.

\subsection{Inviscid stability: numerical results}\label{sec:num_invisc}

\begin{figure}
\centering
  \begin{subfigure}[t]{0.48\textwidth}
      \caption{}\label{fig:invisc_eig_A1}
  	\centering
    \includegraphics[width=\textwidth]{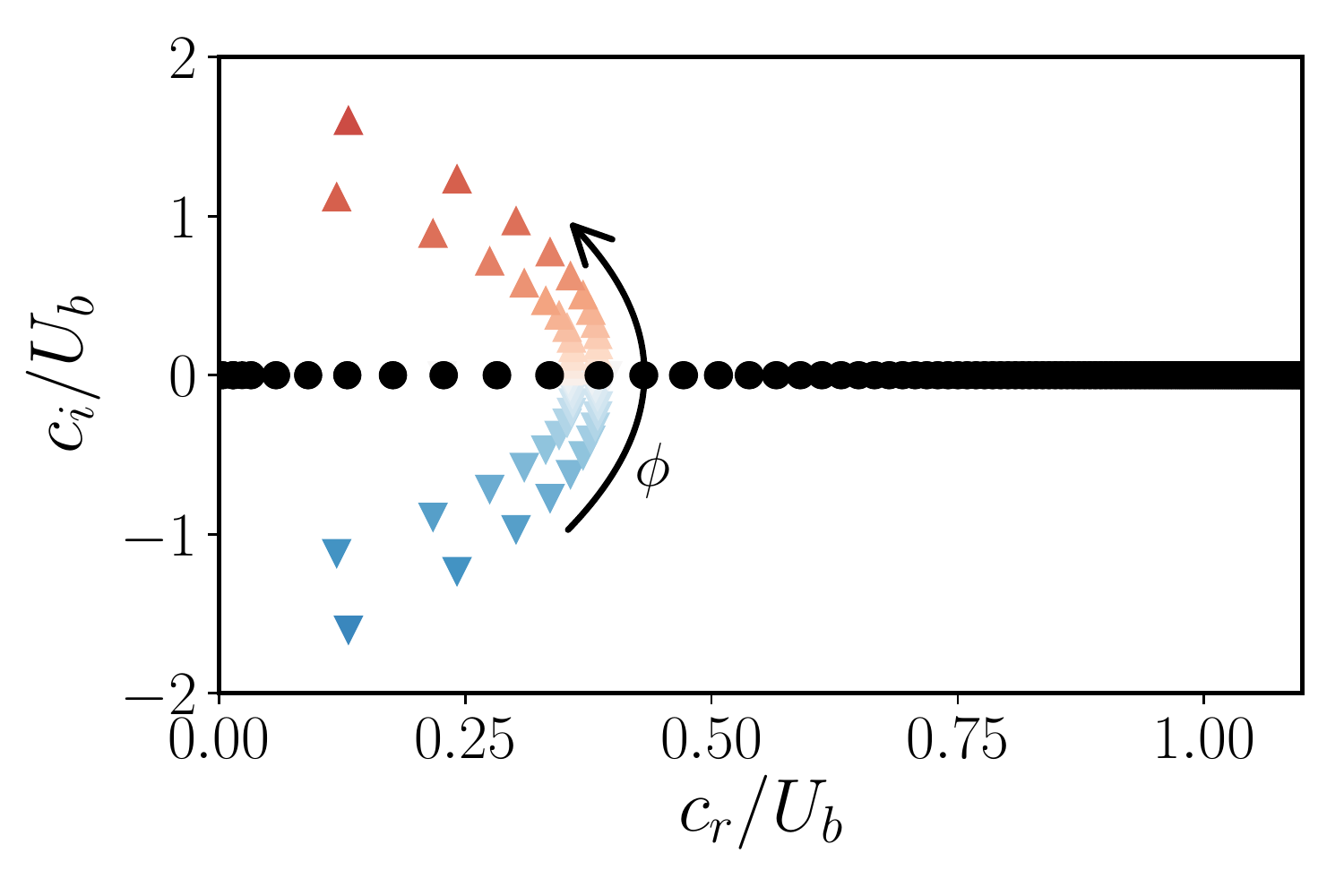} 
  \end{subfigure}\hspace{1em}
    \begin{subfigure}[t]{0.48\textwidth}
      \caption{}\label{fig:invisc_eig_A1_zoom}
  	\centering
  	\includegraphics[width=\textwidth]{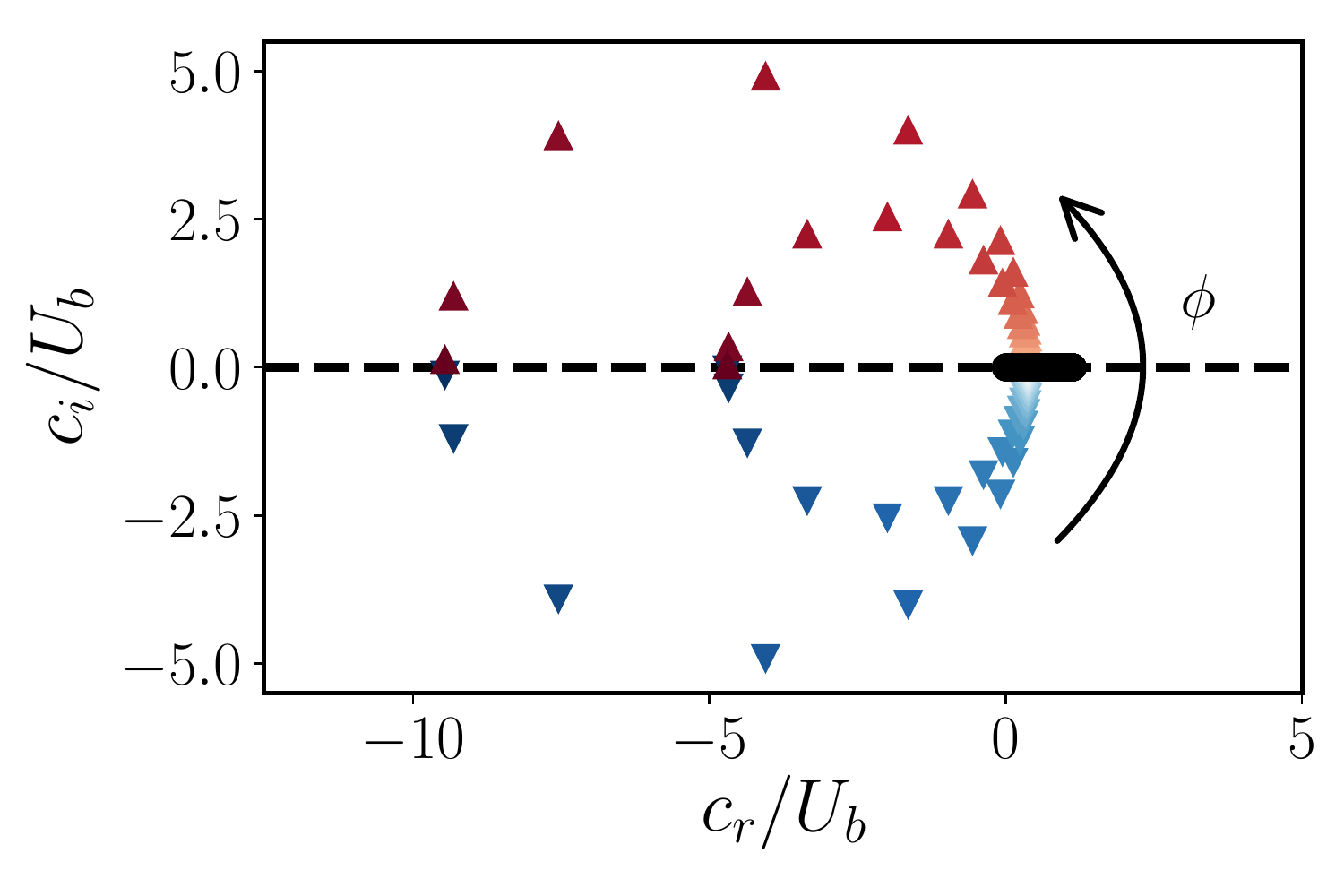} 
  \end{subfigure}\\[1ex]
      \begin{subfigure}[t]{0.48\textwidth}
      \caption{}\label{fig:invisc_eig_A2_zoom}
  	\centering
  	\includegraphics[width=\textwidth]{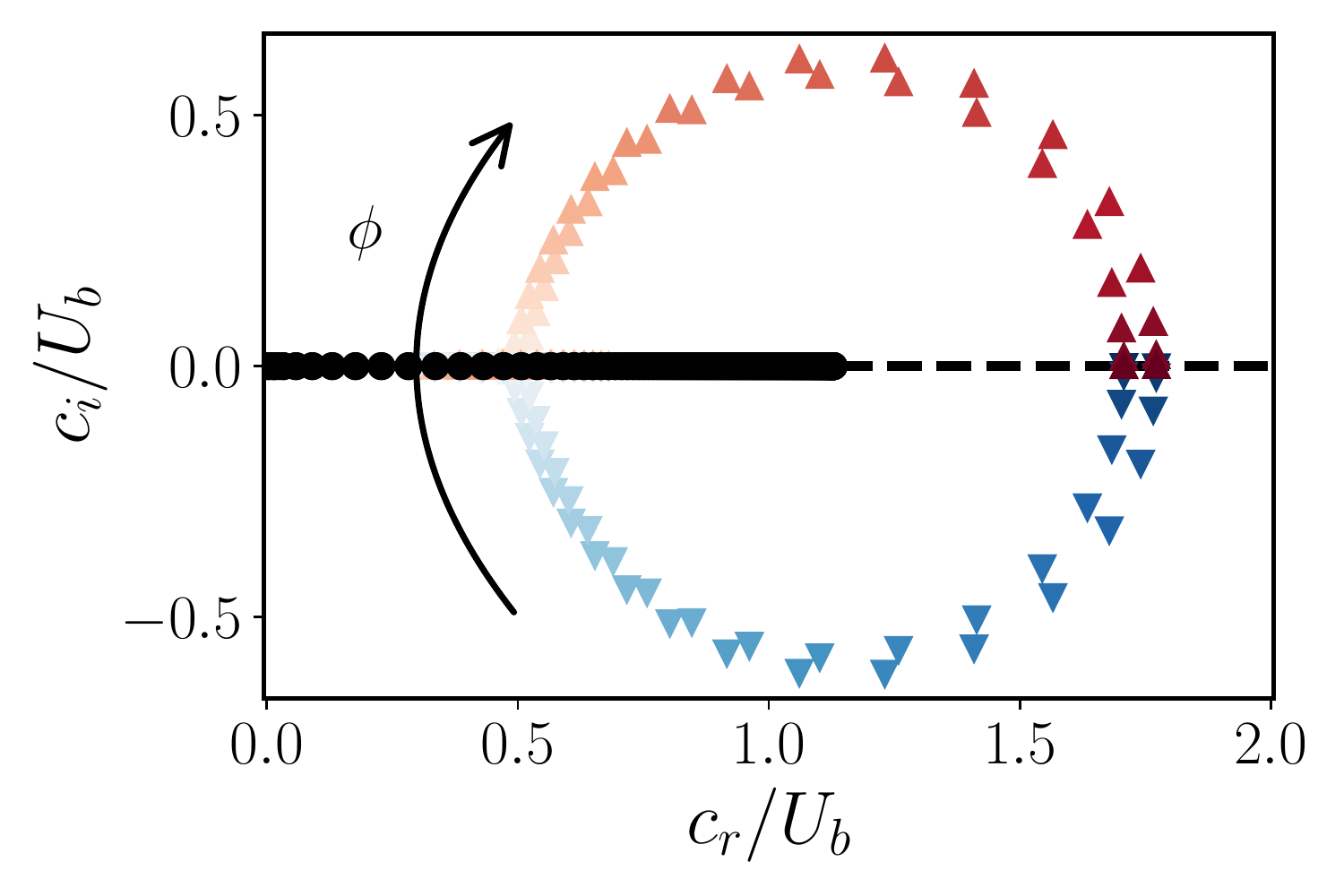} 
  \end{subfigure}\hspace{1em}
    \begin{subfigure}[t]{0.48\textwidth}
     	\caption{}\label{fig:invisc_stab_map}
  	\centering
  	\includegraphics[width=\textwidth]{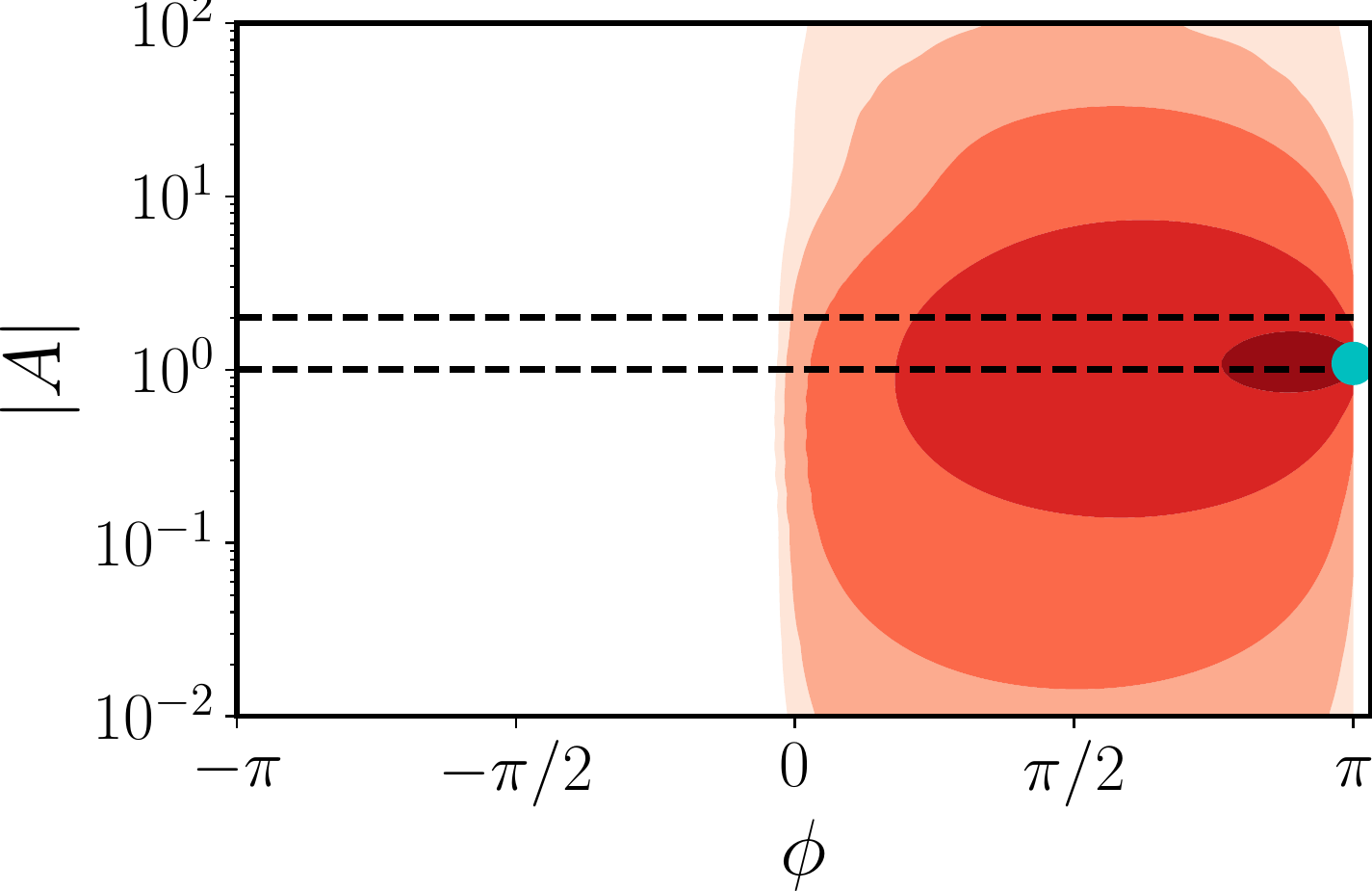}
  \end{subfigure}\hspace{1em}
\caption{ (a) Eigenvalue spectrum of the inviscid problem as a function of control phase $\phi$, $|A| =1$. Negative phases are denoted by blue $\blacktriangledown$, and positive by red $\blacktriangle$. Color intensity shows the increase in  $|\phi|$ as $\phi$ spans the interval $(-\pi, \pi)$. \ding{108} represent the  eigenvalues of the uncontrolled flow.  (b) The same as in (a) but with wider axes, to show the full spectral range. Both axis have the same scale. (c) Eigenvalue spectrum for $|A| =2$. Symbols and colors as in (a).
(d) Inviscid stability map as function of the control gain and phase. Contours, with increasing color intensity: $c_i /U_b = [10^{-4}, 10^{-3},10^{-2},10^{-1},1 ]$. \dashed, $|A| =1$ and $|A|=2$  corresponding to panels (a,b) and (c), respectively. \ding{108} denotes the most unstable $(|A|_f,\phi_f)$ pair of control parameters.  $k_x =1$, $k_z =0$ for all panels.}
\label{fig:inv_eigs}
\end{figure}

We begin  by considering stability of the longest wavelength of the flow, which, in our simulations, is $k_x h =1$, that does not vary in $z$-direction ($k_z=0$), employing the Chebyshev collocation method described in section~\ref{sec:setup_linstab}. Figure~\ref{fig:inv_eigs}(a,b) shows how the control influences stability of this wavemode for $|A| =1$ as a function of the control phase $\phi$.
 Without control, all the eigenvalues of \eqref{eq:rayleigh} are neutrally stable ($c_i =0$) and their phase velocity is restricted to $c_r  \in [U_{min}, U_{max}]$. Pure opposition control with  $\phi = 0$ does not change this property of the flow, which remains neutrally stable. With the introduction of a non-zero control phase, two eigenvalues depart from the real axis, while the rest remain unaffected. Recall that since $A$ becomes complex when $\phi \neq 0$, these eigenvalues are not necessarily complex conjugate pairs. In fact, they appear together  in either the  $c_i >0$ or the $c_i<0$ half-plane. They correspond to a symmetric and an anti-symmetric eigenmode in respect to $y$. In the case of increasing positive phase, $\phi>0$, the eigenvalues move counter-clockwise in the plot, their imaginary part grows, and the flow becomes unstable. If $\phi<0$,  the eigenvalues move clockwise as the phase decreases, symmetrically with respect to the $\phi>0$ case, but have negative imaginary parts and the eigenvectors associated with them are stable.  
Near the threshold of instability, the dependence of the phase velocity seems to be parabolic, but, when a wider range of $\phi$ is considered in figure~\ref{fig:inv_eigs}b, it becomes obvious that the affected eigenvalues execute a quasi-circular motion in the $(c_r, c_i)$ plane. The horizontal axis of this motion lies on the neutral stability line $c_i = 0$. The absolute value of $c_i$ grows with $|\phi|$  for both stable and unstable branches, resulting in large absolute values of growth rate as $|\phi|$ approaches $ \pi$ as well as large negative phase velocities, $c_r<0$. As mentioned in the previous section, these negative phase velocities are unusual compared to a normal channel flow, where $c_r$ is restricted by \eqref{eq:cond_on_cr}, i.e. the minimal mean velocity, \anna{which is} zero at the channel walls. However, the periodicity of the complex control coefficient $A$ requires $A(\pi) = A(-\pi)$, and after reaching a maximum in $c_i$ the absolute value of the growth rate rapidly decreases, until  the stable and unstable branches join at $\phi= \pm \pi$ and $c_i=0$, and flow recovers its neutral stability. This effect is visualised by the overlap of symbols of the two branches on the left in figure~\ref{fig:invisc_eig_A1_zoom}. If we repeat the same numerical experiment with a larger control gain, for example, $|A|=2$, the results are quite different. Figure~\ref{fig:invisc_eig_A2_zoom} illustrates this on the same wave mode  with $k_x h = 1$, $k_z = 0$.  Similarly to $|A|=1$, two unstable eigenvalues exist, with corresponding symmetric and antisymmetric eigenvectors. Positive phases result in unstable flow, and negative phases in stable, with corresponding increase in $|c_i|$ as $|\phi|$ increases. This time, however, the eigenvalues move towards increasing $c_r$ as $|\phi|$ increases, i.e. the unstable eigenvalues move clockwise and their stable counterparts move counter-clockwise until they meet at $\phi =\pm \pi$. 

To further characterize the parametric dependence of the instability, figure~\ref{fig:invisc_stab_map} presents a two-dimensional inviscid stability map of $k_x = 1$, $k_z = 0$ as a function of $|A|$ and $\phi$. The dashed lines in the figure correspond to figures~\ref{fig:inv_eigs}(a-c). As expected from the previous discussion, the unstable region is located in the range of $\phi \in (0, \pi)$. The region with $\phi \in (-\pi, 0)$ is neutrally stable for all values of $|A|$. The isocontours of constant $c_i$ show that the instability growth rate increases with $\phi$, and peaks when $\phi$ is close to $\pi$ for each $|A|$. Note that in contrast with classic opposition control ($\phi = 0$), $\phi = \pm \pi$ is equivalent to ``reinforcement" control where the flow velocity applied at the walls is in-phase with the flow velocity at the detection plane $y_d$. The point $\phi = \pm \pi$ is nevertheless neutrally stable, as we saw in figure~\ref{fig:inv_eigs}(b,c). As the control gain increases, the instability becomes more pronounced, and the maximum growth rate is attained for $|A|_f \approx 1.2$, $\phi_f \approx \pi$. The choice of nomenclature for this pair of control parameters, denoted by a blue circle, will become evident shortly. With further increase in $|A|$ the growth rate begins to decrease, and the flow becomes again almost neutrally stable for $|A| = 10^2$.

So far we have only been concerned with the wave number $k_x h =1$, $k_z =0$, but  wave numbers as high as $k_{x,z} h = 20$ are also affected by the instability and exhibit similar circular motion. For all $k_{x,z}$, the effect of control manifests itself as a pair of eigenvalues with corresponding symmetric and asymmetric eigenvectors. However, the difference between the eigenvalues in the pair becomes smaller as $k_{x,z}$ increases, and their respective eigenvalues become identical. For example, each symbol in figure~\ref{fig:eigspec_flip}, presenting control-related eigenvalues for $k_x h = 6$, $k_z =0$, in reality stands for a pair of almost identical eigenvalues. This simplifies the visual inspection of the data since only one circle of the eigenvalue motion with $\phi$ has to be tracked, and we will use figure~\ref{fig:eigspec_flip} further for convenience.

\subsection{Explaining the flip of the eigenvalue motion}\label{sec:flip_invisc}

\begin{figure}
\centering
      \begin{subfigure}[t]{0.48\textwidth}
     	\caption{}\label{fig:eigspec_flip}
  	\centering
  	\includegraphics[width=\textwidth]{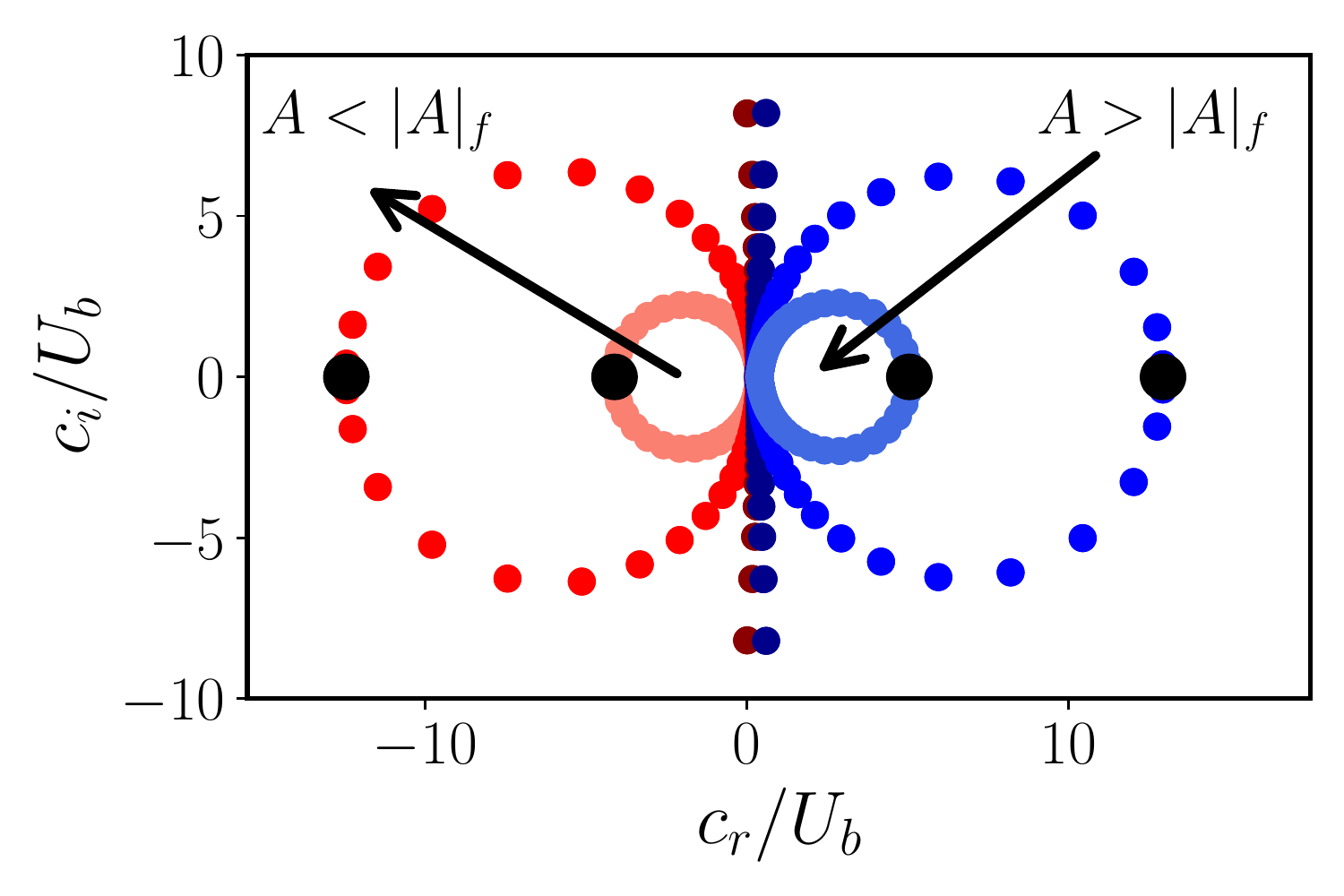}
  \end{subfigure}
        \begin{subfigure}[t]{0.48\textwidth}
      \caption{}\label{fig:theory_amp_I1}
  	\centering
  	\includegraphics[width=\textwidth]{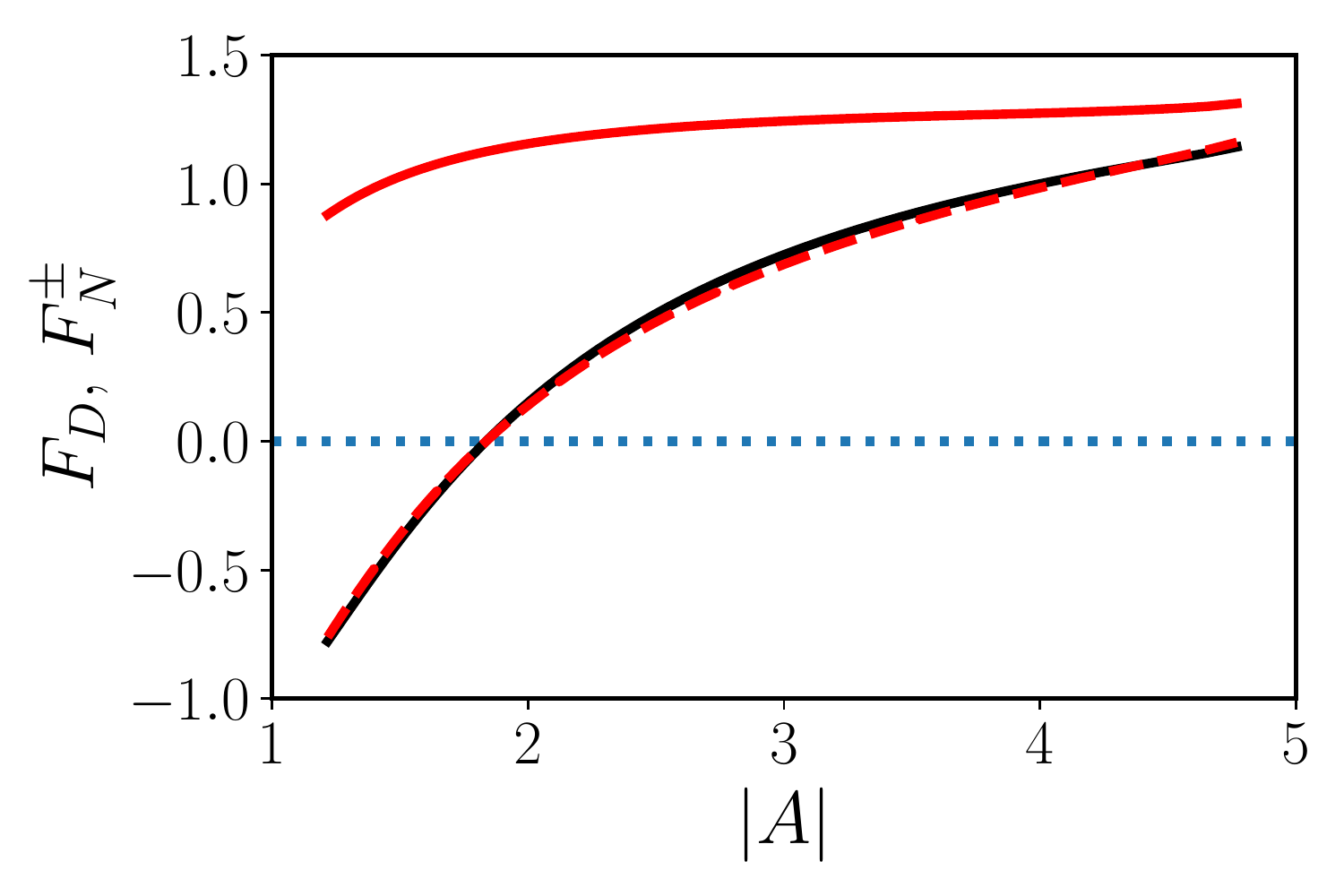}
  \end{subfigure}
    \begin{subfigure}[t]{0.48\textwidth}
      \caption{}\label{fig:theory_amp_cr}
  	\centering
  	\includegraphics[width=\textwidth]{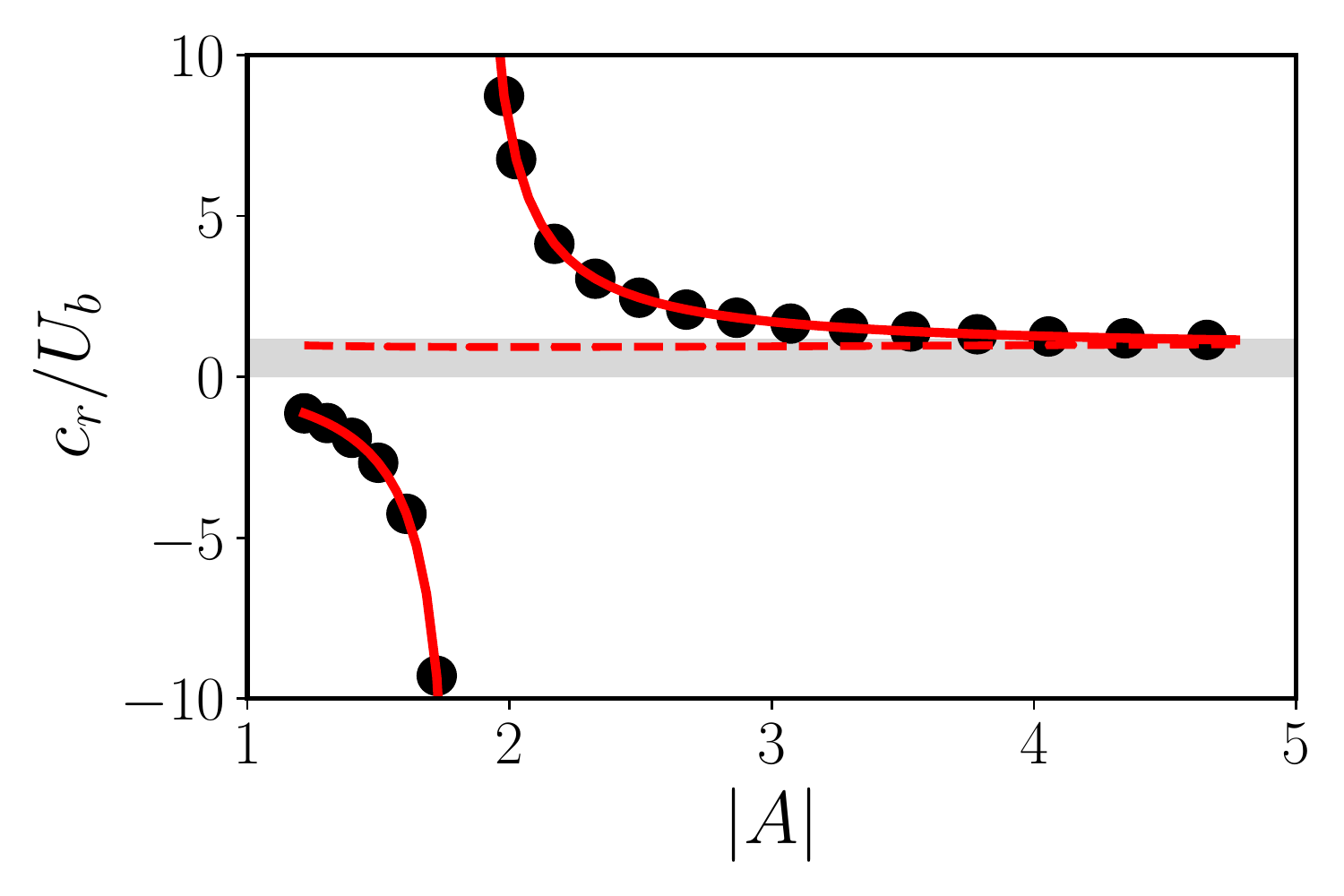}
  \end{subfigure}
  \begin{subfigure}[t]{0.48\textwidth}
      \caption{}\label{fig:Amax_kappa}
  	\centering
  	\includegraphics[width=\textwidth]{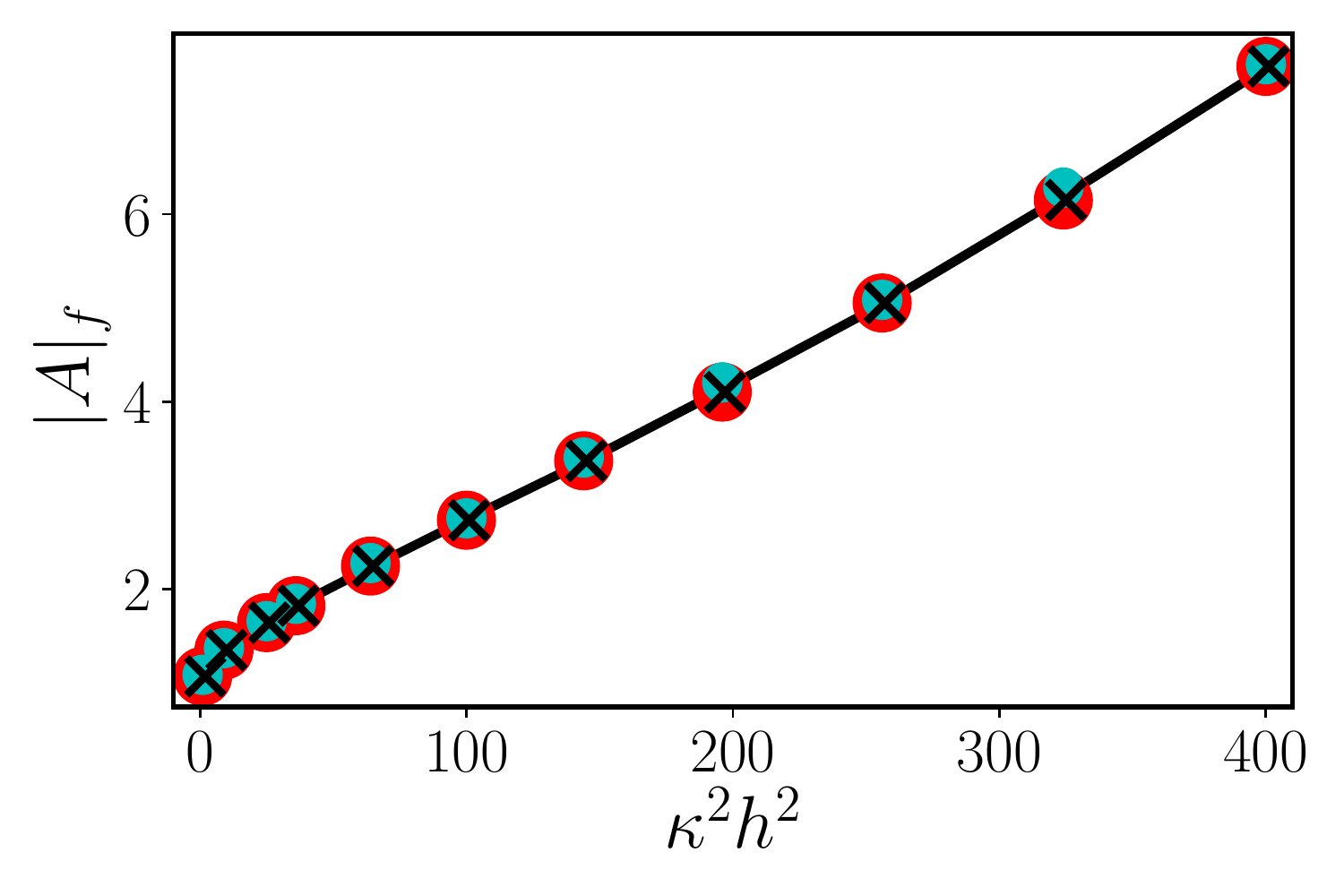}
  \end{subfigure}
\caption{(a) The flip of the eigenvalue motion as the control gain increases from $|A|< |A|_f$  (red) to $|A|> |A|_f$ (blue). $|A| \in [1.6, 1.75, 1.83, 1.84, 1.93, 2.1]$, $k_x  h = 6$, $k_z=0$. Black \ding{108}, eigenvalues with $\phi = \pm \pi$. (b) The numerator and denominator of~\eqref{eq:cr_roots_norm}, calculated from the eigenvectors corresponding to black \ding{108} in (a).  Red: \solid,  $F_N^+$, \dashed, $F_N^-$. Black,  \solid: $F_D$. \dotted~marks zero. (c) \ding{108}, the real part of numerical eigenvalues with the largest $|c_r|$ in panel (a) as a function of $|A|$, together with the analytical expression~\eqref{eq:cr_roots_norm}: \solid, $c_r^+$, \dashed $c_r^-$. Grey rectangle marks the interval of $[U_{min}, U_{max}]$.   (d)  \solid, zeros of the denominator $F_D$ (see panel b), as a function of $\kappa$. Symbols denote the control gain $|A|_f$, resulting in the flip of the eigenvalue motion. Cyan \ding{108}, inviscid flow, $k_x h \in [1,20]$, $k_z = 0$. Red \ding{108}, viscous flow, $k_x h \in [1,20]$, $k_z = 0$.  \ding{53}, viscous flow, $k_x h =1$, $k_z h\in [1,20]$. The data can be approximated with a linear fit $|A|_f = 1.2 + 0.016 \kappa^2 h^2$.
}
\label{fig:flip}
\end{figure}

We observe in figures~\ref{fig:inv_eigs}(b,c) that the direction of eigenvalue motion, as well as the sign of $c_r$ for the fastest-growing modes, depends on $|A|$, and we would like to understand better this dependence. For this purpose, figure~\ref{fig:eigspec_flip} presents the eigenvalue spectrum of $k_x h =6$, $k_z = 0$ for several values of $|A|$. All the eigenvalues near $c_i = 0$, which are very weakly affected by the control and do not exhibit quasi-circular motion, are removed for clarity. Each circle, containing almost identical eigenvalue pairs, is analogous to the ones in figure~\ref{fig:inv_eigs}(b,c), but represents the eigenvalue motion as $\phi $ changes in the interval of $[-\pi, \pi]$ with the same symbol and color. When $|A|\ll 1$, the radius of the circle is very small, and so is $c_i$. With the increase of $|A|$, the growth rates become larger, and equivalently,  the circle that contains them expands. The widening of the circle goes on until $|A| \approx |A|_f$ (for this wavemode, $|A|_f \approx 1.835$). At $|A|_f$, numerically obtained eigenvalues reach extremely large values, and the circle radius tends to infinity at this point. A tiny further increase in $|A|$ makes the eigenvalue motion flip towards the right of the imaginary axis, in the region with $c_r >0$, almost symmetrically. Hence the name $|A|_f$ for this ``critical'' gain, short for $|A|_{flip}$.   After the flip the circle begins to shrink as $|A|$ grows, manifesting the weakening of the instability, and the magnitude of $c_i$ decreases after $|A| = |A|_f$. Eventually the flow comes back to a neutrally stable state.

During the flip, the real eigenvalue at $\phi = \pm \pi$, which is the advection velocity of that neutral mode (black dots in figure~\ref{fig:flip}a), suddenly changes from $c_r < 0$ to $c_r >0$, and the whole circle follows it. Fortunately, we already developed the tools to explain this behavior in   section~\ref{sec:theory_invisc}.  Since we observed that before the flip $c_r < 0$, and after it $c_r \gg U_{y_d}$, it follows that $C_\Re \geq 0$ in \eqref{eq:cr_pi_cond} during the flip.  Recall that $Q = |D \tilde{V}|^2 + \kappa^2 |\tilde{V}|^2 \geq 0$ from \eqref{eq:rayleigh_int}. Then the denominator of~\eqref{eq:cr_roots},
\begin{equation}\label{eq:cr_denom}
    I_1 =  \int_{0}^{2h} Q \dd y - |A| C_\Re = I_q - |A| C_\Re,
\end{equation}
has  contributions from two positive competing terms, and can change sign as $|A|$ grows. When $I_1 = 0$, there is a singularity in~\eqref{eq:cr_roots}, and the eigenvalues $c_r^\pm$ in~\eqref{eq:cr_roots} approach infinity. It is 
instructive therefore to study numerically the denominator $2 I_1$ and the numerator $-I_2 \pm \sqrt{I_2^2 - 4 I_1 I_3}$ of~\eqref{eq:cr_roots}, together with $c_r^\pm$. These quantities are functions of eigenvectors, and do not have a meaningful amplitude, so they must be normalized with a positive-definite quadratic function of $\tilde{V}$. The integral $I_q$ from~\eqref{eq:cr_denom} provides such a norm, and we can normalize both the numerator and the denominator of \eqref{eq:cr_roots}. This gives us an equivalent expression for $c_r^\pm$,
\begin{equation}\label{eq:cr_roots_norm}
c_{r}^{\pm}  = \frac{-\tilde{I}_2 \pm \sqrt{\tilde{I}_2^2 - 4 \tilde{I}_1 \tilde{I}_3} }{2 (1 - |A| \tilde{C}_\Re)} = \frac{F^\pm_N}{F_D}, 
\end{equation}
where $\tilde{}$ reflects our normalization, for example, $\tilde{C}_\Re =  C_\Re/I_q$. Note that $C_\Re/I_q$ itself depends on $|A|$, decreasing through the flip. This dependence  reflects the change in the shape of the eigenvector. While the integral $I_q$ has contributions from the whole channel height, $C_\Re$ represents terms from the boundary. The smaller their ratio is, the more the eigenvector spreads over $y$. 

With this in mind, we go back to the numerical solutions of~\eqref{eq:rayleigh} in search for eigenvalues corresponding to \eqref{eq:cr_roots}. We seek for the eigenvalues represented by black dots in figure~\ref{fig:eigspec_flip}. In order to do this, the phase of control is set to $\phi =\pi$ for each $|A|$, and the full set of real-valued eigenvalues $c$ and eigenvectors $\hat{v}$ is obtained numerically. They are then filtered to get the eigenvalue-eigenvector pair with the largest advection speed subject to conditions $c_r \le 0$ or  $c_r \ge U_{max}$. Otherwise, when $0<c_r < U_{max}$, the eigenvalues with $\phi = \pm \pi$ appear in the region  already populated by the rest of the neutrally stable modes (like those marked with black dots in figure~\ref{fig:inv_eigs}a,c), and it is difficult to identify them. This also implies that the range of control gains is limited by the values of $|A|$ near the flip. 
Using the corresponding eigenvectors, it is straightforward to calculate $C_\Re$, $I_{1,2,3}$ from \eqref{eq:cr_quad}, and $c_r^\pm$ using \eqref{eq:cr_roots_norm}.

Figure~\ref{fig:theory_amp_I1} presents both $F^\pm_N$ and  $F_D$ from~\eqref{eq:cr_roots_norm} as functions of $|A|$ near the flip. The numerator $F^+_N$ is a positive increasing function of $A$, while the denominator $F_D$ changes sign from negative to positive as the gain crosses $|A|_f \approx 1.835$. Thus, $c_r^+$ is negative when $|A|<|A|_f$, and positive afterwards, going through a hyperbolic infinity at $|A|_f$. Figure~\ref{fig:theory_amp_cr} shows an excellent agreement between $c_r^+$ from~\eqref{eq:cr_roots_norm} and the eigenvalues with the largest magnitudes of advection velocity from figure~\ref{fig:eigspec_flip}. This agreement is expected since we used the eigenvectors of~\eqref{eq:rayleigh} to evaluate~\eqref{eq:cr_roots_norm}; however, numerical analysis of~\eqref{eq:rayleigh} did not allow for qualitative explanation of the flow behaviour. With the help of~\eqref{eq:cr_roots_norm}, the inflation, flip, and deflation of the eigenvalue circular motion can now be  inferred from the relation between $F_D$ and $F^+_N$ in figure~\ref{fig:theory_amp_I1}. Unlike $F^+_N$, the second numerator $F_N^-$ closely follows the behavior of the denominator, so the related root $c_r^-$  stays bounded across $|A|_f$ and is of order of unity. This root is in the range of ``regular" eigenvalues of the Rayleigh equation, $U_{min}< c_r^- <U_{max}$, and is unrelated to the eigenvalues that undergo the flip. 
The results in figure~\ref{fig:flip}(a-c) were given for $k_x h =6$, $k_z = 0$, but qualitatively similar results can be obtained for other wavemodes. In fact,
\eqref{eq:rayleigh} depends only on the square of the effective wavenumber, $\kappa^2$, and not on a particular realization of $k_x$, $k_z$. In figure~\ref{fig:Amax_kappa} we show the dependence of $|A|_f$ on $\kappa$, together with the zeros of the function $I_1 (|A|)$. The agreement is again very good, we observe that $|A|_f$ increases linearly with $\kappa^2$. This results in the upward shift of the unstable regions, analogous to the one in figure~\ref{fig:inv_eigs}(d), as $\kappa$ grows and the wavelength becomes smaller.

Further progress could be done if we relate the analytical form of integrals in~\eqref{eq:cr_quad} to the control gain through boundary conditions~\eqref{eq:vtilde_bc}, but that non-trivial task will not be pursued here. From this point on, we will include viscosity in our analysis in order to make it more comparable to the channel flow in DNS. The channel flow is wall-bounded and viscosity becomes important near the channel walls, so one can expect the inviscid instability observed above to be modulated by viscous effects.

\section{The effect of turbulent viscosity}\label{sec:listab_vics}

\subsection{Eigenvalue spectra and similarity with the inviscid flow}\label{sec:visc_invisc_sim}

\begin{figure}
\centering
      \begin{subfigure}[t]{0.5\textwidth}
      \caption{}\label{fig:visc_eig_kx6A1}
  	\centering
  	\includegraphics[width=\textwidth]{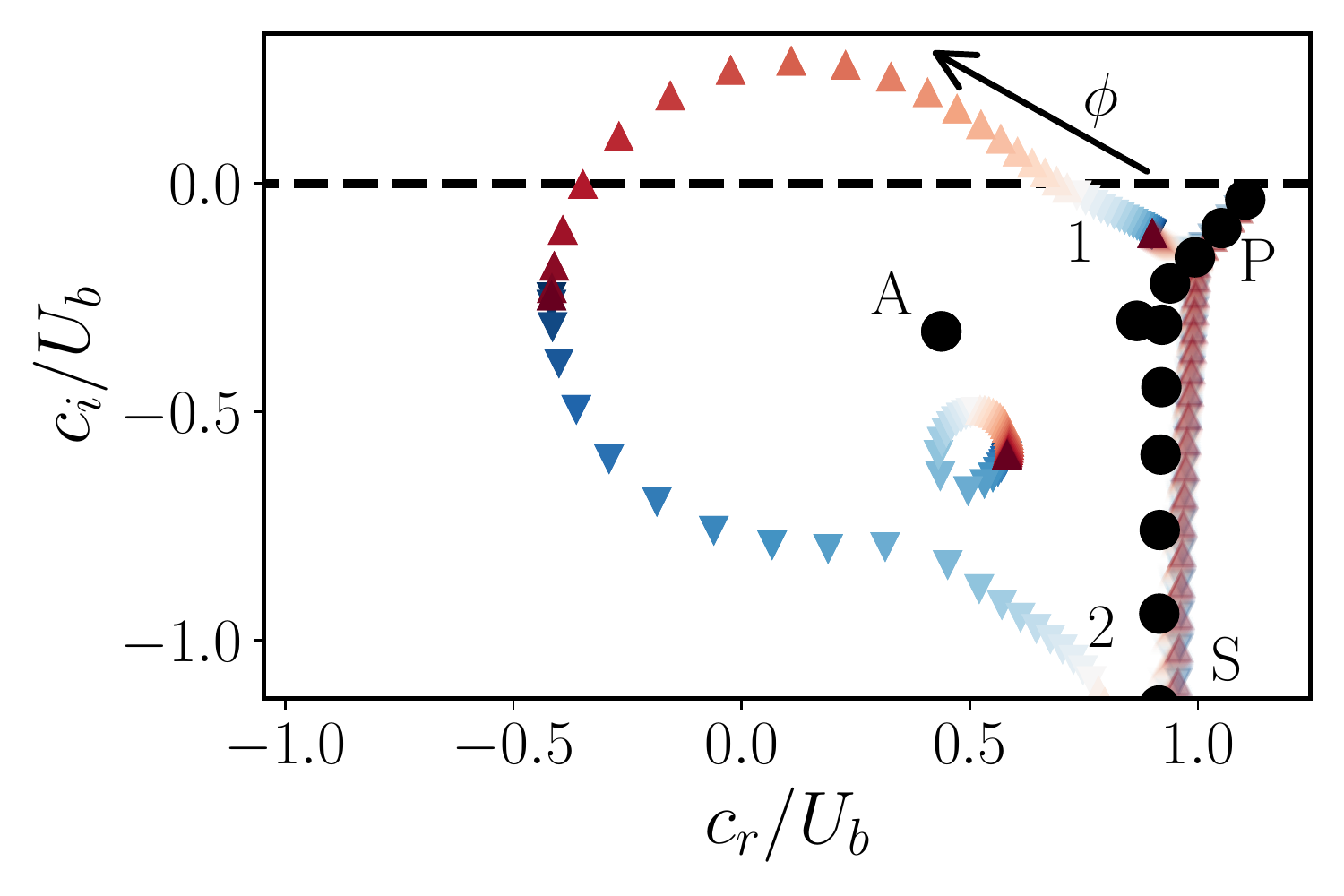} 
  \end{subfigure}
  \begin{subfigure}[t]{0.48\textwidth}
      \caption{}\label{fig:visc_eig_kx6A2}
  	\centering
  	\includegraphics[width=\textwidth]{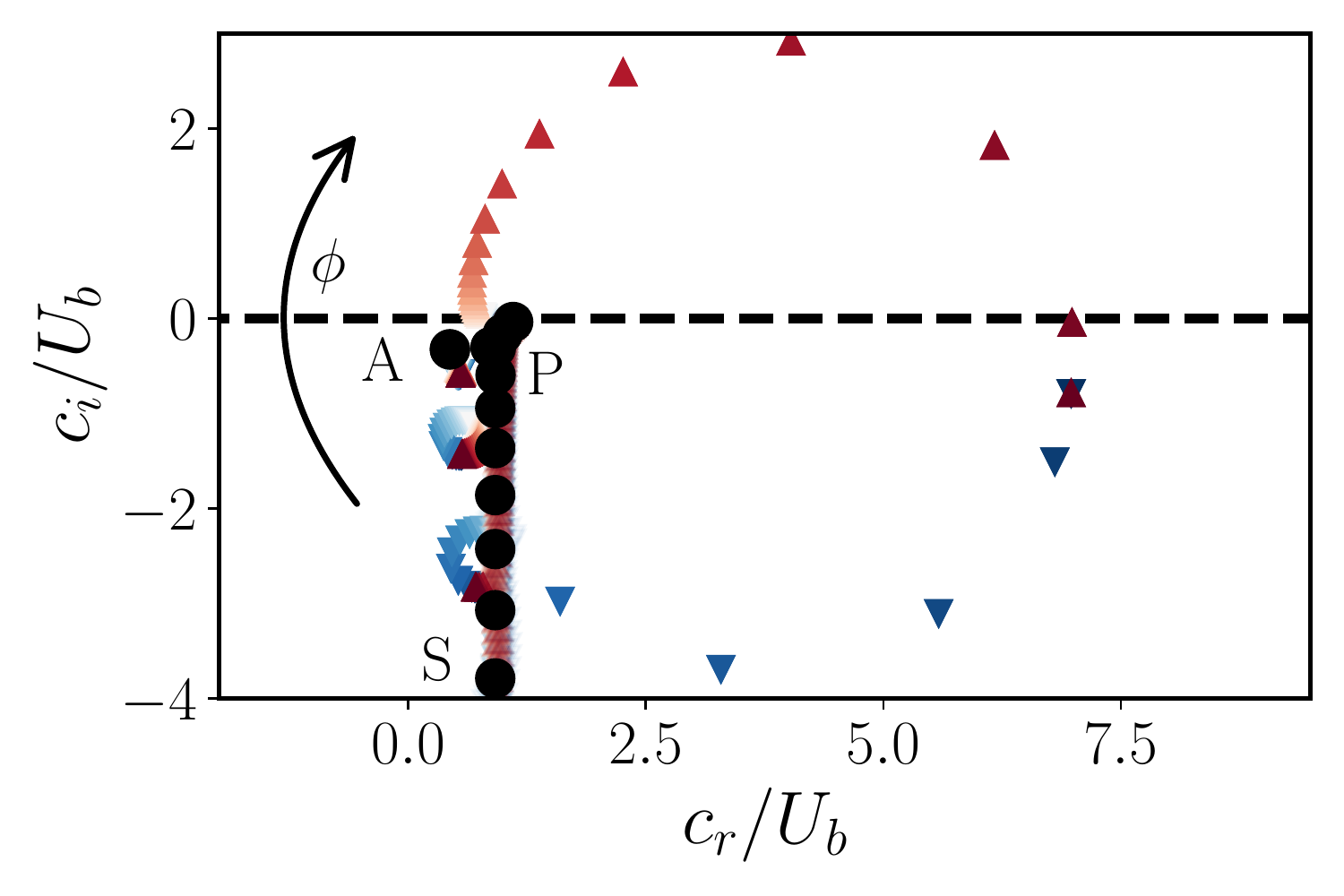} 
  \end{subfigure}\\[1ex]
    \begin{subfigure}[t]{0.48\textwidth}
      \caption{}\label{fig:ci_phi_A1}
  	\centering
  	\includegraphics[width=\textwidth]{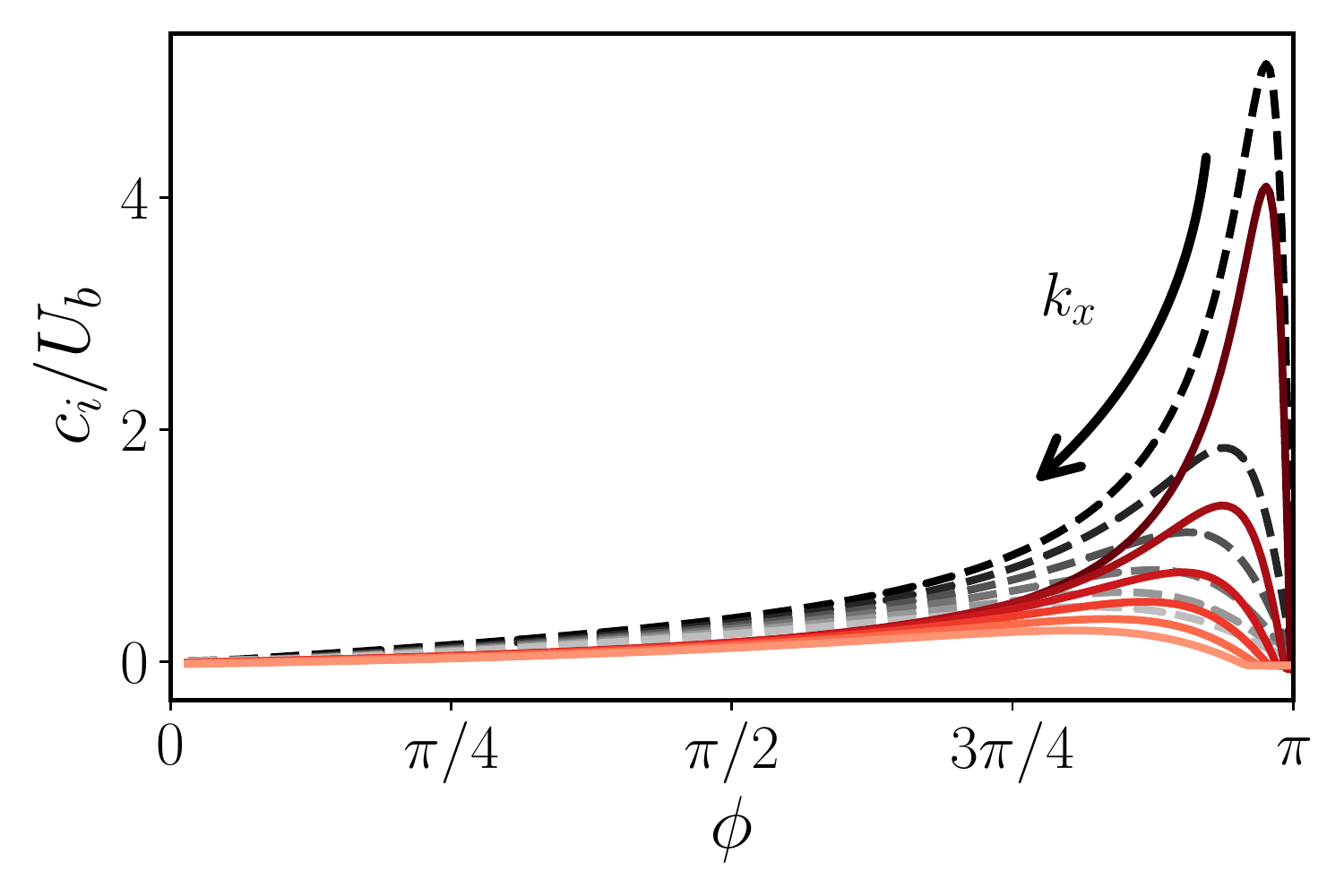} 
  \end{subfigure}
  \begin{subfigure}[t]{0.48\textwidth}
      \caption{}\label{fig:ci_phi_A2}
  	\centering
  	\includegraphics[width=\textwidth]{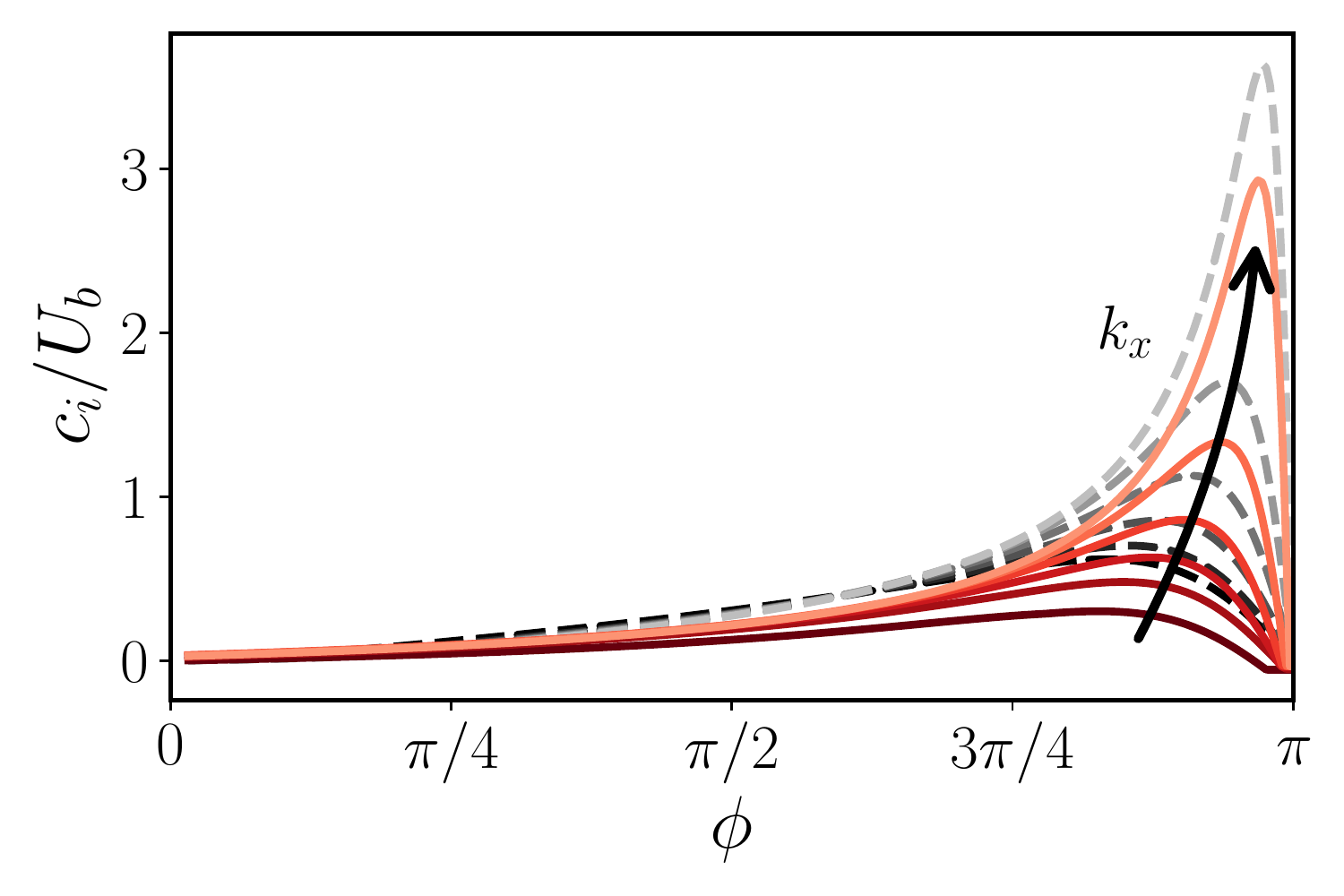} 
  \end{subfigure}\\[1ex]
\caption{ (a,b) Eigenvalue spectrum of the viscous problem as a function of $\phi$. $k_x h =6$, $k_z=0$.  Color and symbols are as in figure~\ref{fig:inv_eigs}(a).  \ding{108}, uncontrolled flow. See text for the description of the branches $A$, $P$ and $S$ and points $1,2$. Some eigenvalues from the branches $P$ and $S$ were removed or made more transparent for clarity. Both axes have the same scale. (a) $|A| = 1$, (b) $|A|=2$. (c,d) Imaginary part, $c_i$, of the eigenvalues as a function of control phase $\phi$. \solid, viscous problem, \dashed, inviscid flow.   Color, from dark to light: $k_x  h\in [1,6]$. (c) $|A| = 1$, (d) $|A|=2$.}
\label{fig:eigs_visc}
\end{figure}

Compared to the inviscid flow, the eigenvalue spectrum of the viscous problem is more complex.  In the conventional linear stability analysis of plane Poiseuille flow close to transition to turbulence, where linearized equations include only molecular viscosity, the eigenvalues are located on three branches: A ($c_r/U_{max} \to 0$), P ($c_r/U_{max} \to 1$), and S ($c_r/U_{max} \approx 2/3$ for $k_x h =1$, $k_z =0$) \citep[p. 64]{mack1976numerical,Schmid2012}. In that case, the unstable eigenmode originates from the A-branch, and is called Tollmien-Schlichting wave. This Y-shaped  eigenvalue spectrum, typical of transitional wall-bounded flows, is preserved under the influence of the eddy viscosity in the flow without control. In figure~\ref{fig:visc_eig_kx6A1}, we use the eigenspectrum of the uncontrolled flow with $k_x h =6$, $k_z =0$ (black dots) to illustrate this. There are also some differences with respect to the flow with only molecular viscosity. Only two eigenvalues remain on branch A, branch P is slightly deformed, and branch S is shifted towards $U_{max}$ (and larger $c_r$). More importantly,  the uncontrolled turbulent mean profile is stable in the presence of turbulent viscosity, as found by \citet{reynolds1967stability}. With control, most of the eigenvalues move slightly from their uncontrolled locations. If the control with $|A| =1$, $\phi = 0$ is applied,  two identical eigenvalues appear in the vicinity of the branch $P$ (see point 1 in figure~\ref{fig:eigs_visc}a). When the phase is positive and increasing, they  move towards the left in figure~\ref{fig:visc_eig_kx6A1}, and their $c_r$ becomes smaller. As their growth rate becomes larger, they cross $c_i = 0$, and the flow becomes unstable. At the same time, another pair of identical stable eigenvalues appears near the branch $S$ (point 2 in figure~\ref{fig:eigs_visc}a). When $\phi<0$ and is decreased further, these eigenvalues also move to the left towards smaller $c_r$, until they join the eigenvalues with $\phi >0$ at $\phi = \pm \pi$, which are stable. The dependence of the eigenvalues on $\phi$ resembles a circular motion, like in the inviscid case (figure~\ref{fig:inv_eigs}b). For $|A|=2$ the circular motion of eigenvalues changes its direction (figure~\ref{fig:eigs_visc}b). Here,  the eigenvalues move towards increasing $c_r$ as $|\phi|$ increases, i.e. the unstable eigenvalues move clockwise and their stable counterparts move counter-clockwise until they meet at $\phi =\pm \pi$. Similarly to the inviscid flow (figure~\ref{fig:inv_eigs}b,c), the eigenvalues follow quasi-circular paths that flip their direction as $|A|$ increases.  The axis of symmetry of these paths, however, is now located below $c_i = 0$, where the flow is stable. We observed a similar behaviour of eigenvalues for all large wave modes of the viscous flow.

Figures~\ref{fig:eigs_visc}(a,b) and~\ref{fig:inv_eigs}(b,c) give a qualitative overview of the eigenvalue behavior under the change of control phase $\phi$, but provide few quantitative details. To further clarify the dependence of the instability on $\phi$, we plot  the imaginary part $c_i$ of the eigenvalue with the largest growth rate as a function of $\phi$ and $k_x$ in figure~\ref{fig:eigs_visc}(c,d). Both viscous (solid) and inviscid (dashed lines) growth rates are presented.  Again, we set the control gain to be $|A|=1$ (figure~\ref{fig:eigs_visc}c), $|A| =2$ (figure~\ref{fig:eigs_visc}d). 
At these values of $|A|$ there are no unstable eigenvalues when $\phi<0$, so the data are plotted only in the half-plane $[0, \pi]$. In the case of $|A| =1$, there is  a well-defined maximum in $c_i$, which moves towards smaller $\phi$ as $k_x$ decreases, and its amplitude decreases with $k_x$. For example, for $k_x h= 1$  the maximum instability growth is attained near $\phi_{max} = 3.1$, while for $k_x h= 6$, it is around $\phi_{max} = 2.5$. In the case of  $|A|=2$, there still exists a pronounced maximum in $c_i$, but the most unstable phase now increases with $k_x$. For example, $\phi_{max} \approx 2.6$ for $k_x h =1$, and $\phi_{max} \approx 3.04$ for $k_x h =6$.  Similar trends can be seen in the real part of the eigenvalue $c_r$ (not shown here), with one important addition: negative values of $c_r$ are observed for $|A| = 1$, while $|A| = 2$ results in positive $c_r$, as expected from the eigenvalue flip in figures~\ref{fig:eigs_visc}(a,b) and~\ref{fig:inv_eigs}(b,c). The similarity of the eigenvalue behaviour in the viscous and inviscid cases in figures~\ref{fig:eigs_visc}(c,d) is remarkable, indicating that we observe the same instability of inviscid origin.

\begin{figure}
\centering
  \begin{subfigure}[t]{0.48\textwidth}
      \caption{}\label{fig:visc_stab_map}
  	\centering
  	\includegraphics[width=\textwidth]{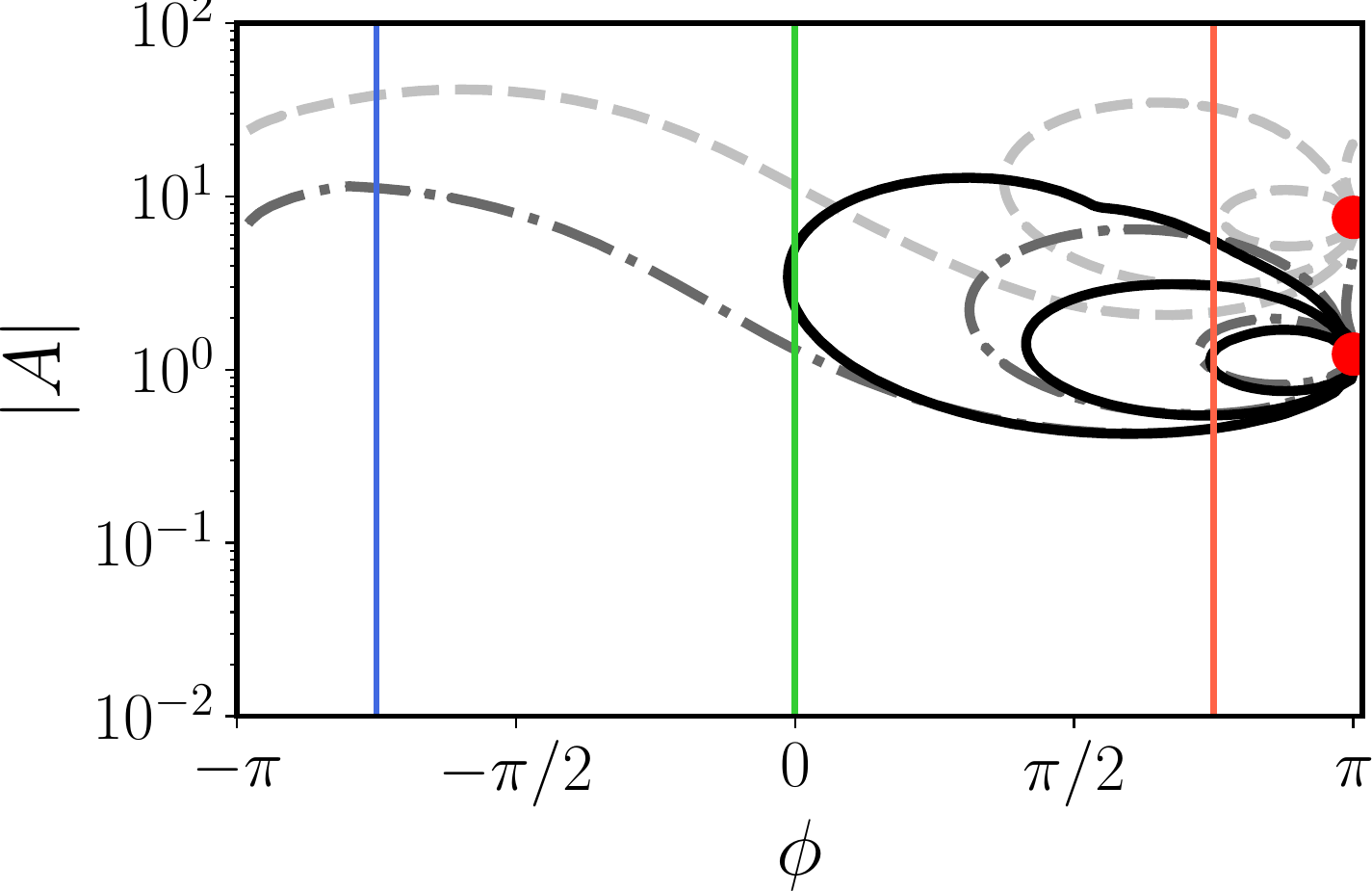}
  \end{subfigure}
    \begin{subfigure}[t]{0.48\textwidth}
      \caption{}\label{fig:cr_ci_kx2_varA}
  	\centering
  	\includegraphics[width=\textwidth]{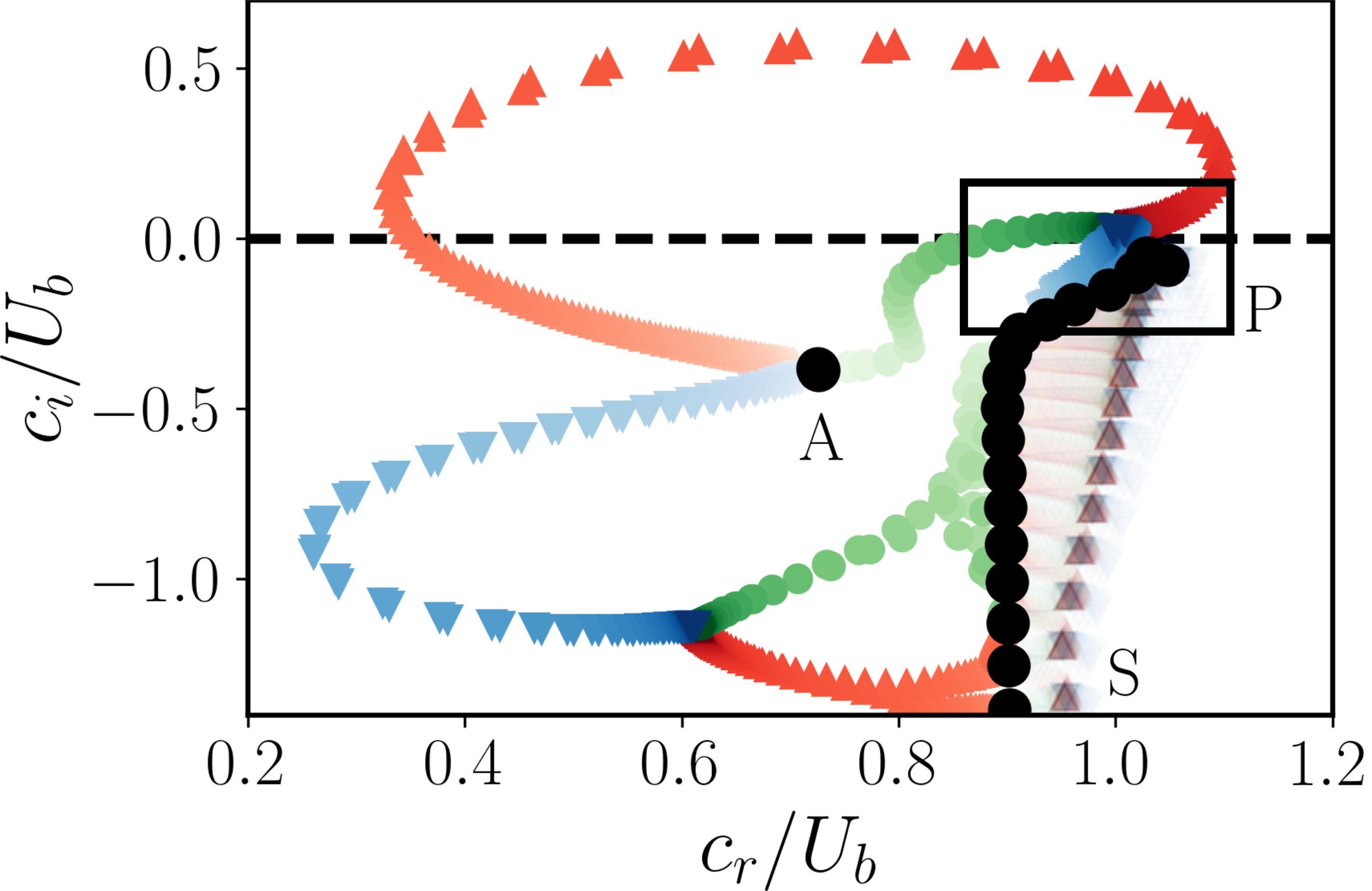}
  \end{subfigure}
  \begin{subfigure}[t]{0.48\textwidth}
      \caption{}\label{fig:cr_ci_kx2_zoom}
  	\centering
  	\includegraphics[width=\textwidth]{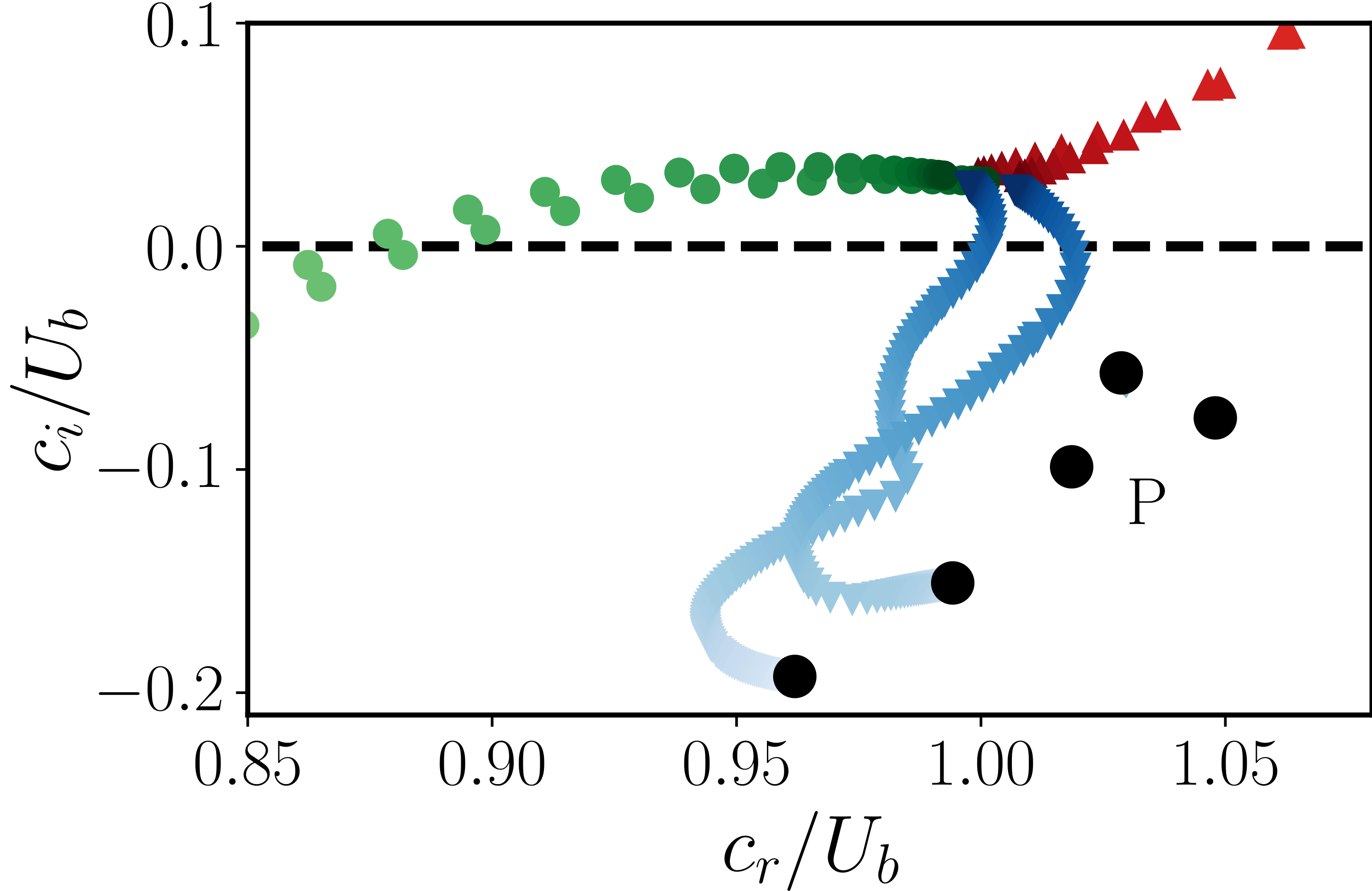}
  \end{subfigure}
    \begin{subfigure}[t]{0.48\textwidth}
      \caption{}\label{fig:ci_amp_sat}
  	\centering
  	\includegraphics[width=\textwidth]{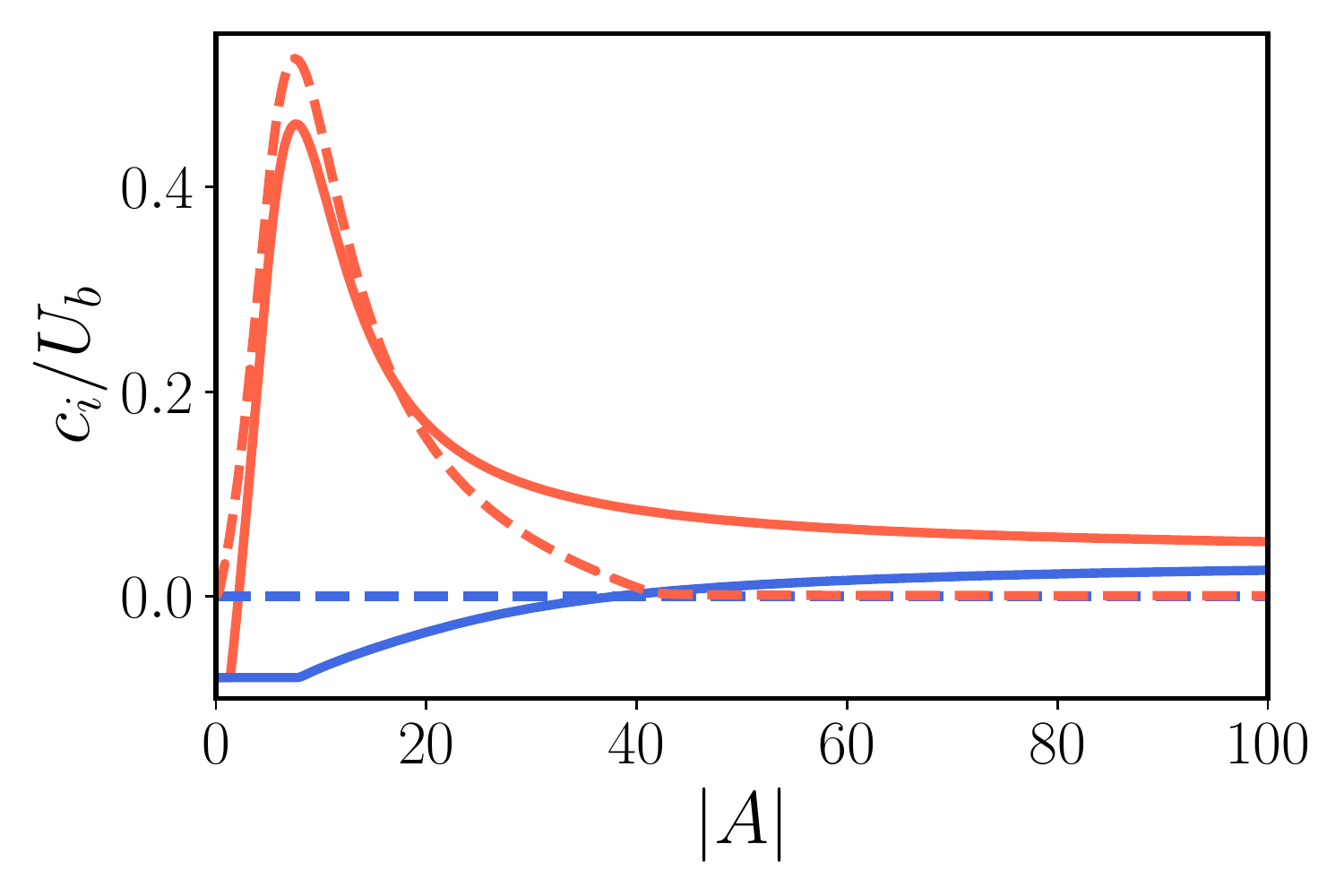}
  \end{subfigure}
\caption{ (a) Stability maps for the viscous problem, analogous to figure~\ref{fig:invisc_stab_map}, $c_i /U_b\in [0, 0.1, 0.5]$. \solid, $k_x =1$, \dashdotted, $k_x = 2$ , \dashed, $k_x = 20$; $k_z = 0$. For each $k_x$, \ding{108} denotes  the most unstable $(|A|_f,\phi_f)$ pair. Vertical coloured lines represent $\phi = [-3\pi/4 , 0, 3\pi/4]$ for the panel (b). (b) Eigenvalue spectrum of the viscous problem as the function of $|A|$ along the vertical lines in (a), for $k_x = 2$, $k_z=0$. Red, $\phi = 3 \pi /4 $, green, $\phi =0$, blue, $\phi = -3 \pi/4$, with color intensity increasing with $|A|$. \ding{108}, the spectrum of the uncontrolled flow with branches $A$, $P$ and $S$. Stable eigenvalues from the branches $P$ and $S$ were made more transparent.  (c) A zoom into panel (b) marked by the black rectangle. Stable eigenvalues of the branch $P$ were removed for clarity. (d) The maximum growth rate $c_i$ as a function of $|A|$, $\phi = -3 \pi/4$ (blue), $\phi = 3 \pi /4$ (red). \solid, viscous flow, \dashed, inviscid. $k_x =20$, $k_z = 0$.}
\label{fig:stab_amp}
\end{figure}

To probe this similarity further, we plot in figure~\ref{fig:visc_stab_map} the isocontours of viscous growth rate for various $k_x$, and $k_z =0$, as a function of $|A|$ and $\phi$. 
With the help of this plot, we will first discuss the common features of the viscous and inviscid stability. Visual inspection shows that the shape of viscous isocontours with large $c_i$ resembles the shape of inviscid ones in figure~\ref{fig:invisc_stab_map}. For each $A$, the growth rate increases with $\phi$.  We marked by circles the pairs  $(|A|_f,\phi_f)$ where the instability reaches its maximum growth. Analogously to the inviscid case, the eigenvalue motion flips its direction at $|A| = |A|_f$, as discussed above. The phase $\phi_f$ is near $\pi$ for all observed wavemodes, the gain $|A|_f$ shifts upwards with increasing $k_x$, at the same time as the unstable region shifts upwards itself.  We search again for $|A|_f$ as a function of the effective wavenumber $\kappa^2 = k_x^2 +k_z^2$, and plot it in figure~\ref{fig:Amax_kappa} together with the inviscid data. Two cases are considered: first, setting $k_z=0$, while varying $k_x h\in [1,20]$, and second, fixing $k_x h =1$, and varying $k_z h \in [1,20]$. One can appreciate that $|A|_f$ again depends linearly on $\kappa^2$ in the viscous flow, with the results being almost identical to those in the inviscid case. Therefore, the shift of the instability region towards higher control gains at large wavenumbers is an inviscid phenomenon, suggesting that equation~\eqref{eq:rayleigh} governs the instability behaviour in a substantial part, even in presence of turbulent viscosity. 

\subsection{Saturation of eigenvalues at large  control gains}\label{sec:visc_invisc_diff}

Now back to the notable differences between the stability maps in figures~\ref{fig:visc_stab_map} and~\ref{fig:invisc_stab_map}. Recall that the instability is confined to the region with $\phi \in [0,\pi]$ for the inviscid flow. In that region, the growth rate $c_i \to 0$ when $|A| \ll 1$, and it increases with $|A|$. After reaching the maximum at $|A|_f$, $c_i$ decays, and the inviscid flow becomes neutrally stable again when $|A| \gg 1$. In contrary to this, viscosity has a damping effect on the instability at low values of $|A|$, and no eigenvalue has a positive growth rate in the bottom-half plane of figure~\ref{fig:visc_stab_map}. The viscous flow becomes unstable when $|A|\sim O(1)$, and the instability is confined to the region with $\phi \in [0,\pi]$ for low $|A|$. As $|A|$ increases further, the instability does not cease to exist for most of the wavemodes,  except for the special case of $k_x h =1$.  Unlike in inviscid flow, unstable eigenvalues also appear  for the negative phases,  $\phi \in [-\pi, 0]$,  where inviscid flow would be stable, as indicated by the neutral isocontours of $c_i = 0$. In other words, some parameter regions in figure~\ref{fig:visc_stab_map}, characterized by large $|A|$, are unstable for all $\phi$. 

Concerned with this issue, we fix three phases $\phi = [-3 \pi /4, 0, 3 \pi /4]$ and compute the eigenvalue spectrum, now as a function of $|A|$.  The results are presented in figure~\ref{fig:cr_ci_kx2_varA} for $k_x h=2$, $k_z=0$. For each amplitude, the eigenvalues again come in pairs. In this case, $k_x h=2$ is large enough and the eigenvalues of these pairs are not identical, but the difference between them is already much smaller than, say, for $k_x h = 1$. It becomes apparent that the  eigenvalues affected by control at $\phi = \pm 3\pi /4$ are connected to the branch $A$ rather than branch $P$ of the uncontrolled flow. At first, the real part of the eigenvalues decreases with the increase of $|A|$, indicating the motion of eigenvalues towards the left half of the complex plane, and the expansion of the eigenvalue circle. As $|A|$ passes through $|A|_f$, this dynamics is reversed and the real part of the eigenvalue begins to increase. The difference between $\phi =  3\pi/4$ and $\phi = -3 \pi /4$ is, obviously, that the growth rates of the eigenvalues exhibiting circular motion become positive in the first case, and even more negative in the second. In the inviscid flow, the case of $\phi = 0$ is neutrally stable and the eigenvalues do not depart from $c_i =0$. In the viscous flow, $\phi = 0$ eventually becomes unstable, although its eigenvalues have increasing $c_r$ for all $|A|$, unlike when $\phi = \pm 3 \pi /4$. Figure~\ref{fig:cr_ci_kx2_zoom} shows a zoom in an area where the unstable eigenvalues for all $\phi$ approach each other asymptotically as $|A| \to 10^2$  (the rest of the eigenvalues has been removed for clarity). As this happens, the eigenvalues for all three values of $\phi$ saturate at a small but positive value. These eigenvalues are not spurious, as our numerical resolution tests  confirm in Appendix C. The case of $\phi = - 3 \pi/ 4$ has a particularly interesting behavior. There, the unstable eigenvalues originate from the branch $P$ rather than the branch $A$  moving upwards as $|A|$ increases. Note that $\phi = 0$ is still connected to the branch $|A|$, and an approximate threshold of the change in the origin from the branch $A$ to the branch $P$ is at about $\phi = - \pi/16$. Finally, in the figure~\ref{fig:ci_amp_sat} the phenomenon of the eigenvalue saturation is addressed quantitatively, with the maximum growth rate $c_i$ as a function of $|A|$. The growth rates are calculated for $k_x= 20$, $k_z = 0$, to highlight that shorter wavelengths have a similar behaviour, and also because the maximum in $c_i$ is weaker, and therefore fits better for visualization purposes. We can think of the solid curves in figure~\ref{fig:ci_amp_sat} with respective colors as following the red upper path in figure~\ref{fig:cr_ci_kx2_varA} and the blue path in figure~\ref{fig:cr_ci_kx2_zoom} (except for the small values of $|A|$, when the growth rates on the branch $P$ are larger). For $\phi = 3\pi /4$, the peak in $c_i$ correlates well in the viscous and inviscid cases, with $c_i$ decaying asymptotically to zero in the latter, and saturating at a small value in the former. In contrast to this, when $\phi = -3 \pi /4$, the inviscid case is always neutrally stable, as expected, and the viscous $c_i$ slowly saturates to a positive level with the increase in $|A|$. This saturation indicates that when magnitude of the input at the walls is very strong, the flow becomes relatively insensitive to the control phase.

\subsection{The effect of control on the shape of the eigenvectors }\label{sec:eigvec}

\begin{figure}
\centering
\setlength{\unitlength}{0.1\textwidth}
\begin{picture}(10,5)
 \put(0,0){\includegraphics[width=0.5\textwidth]{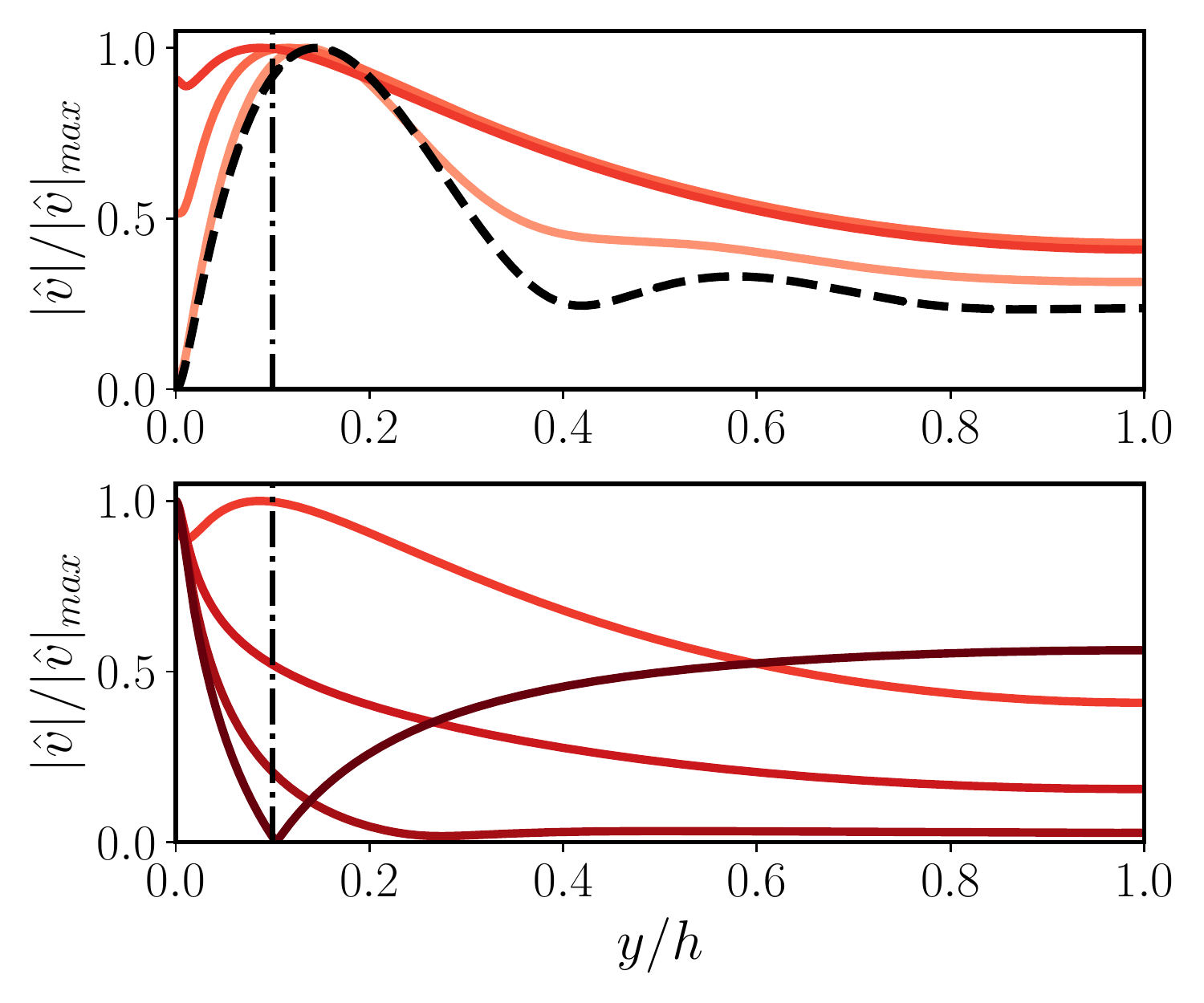}}
 \put(5,0){\includegraphics[width=0.5\textwidth]{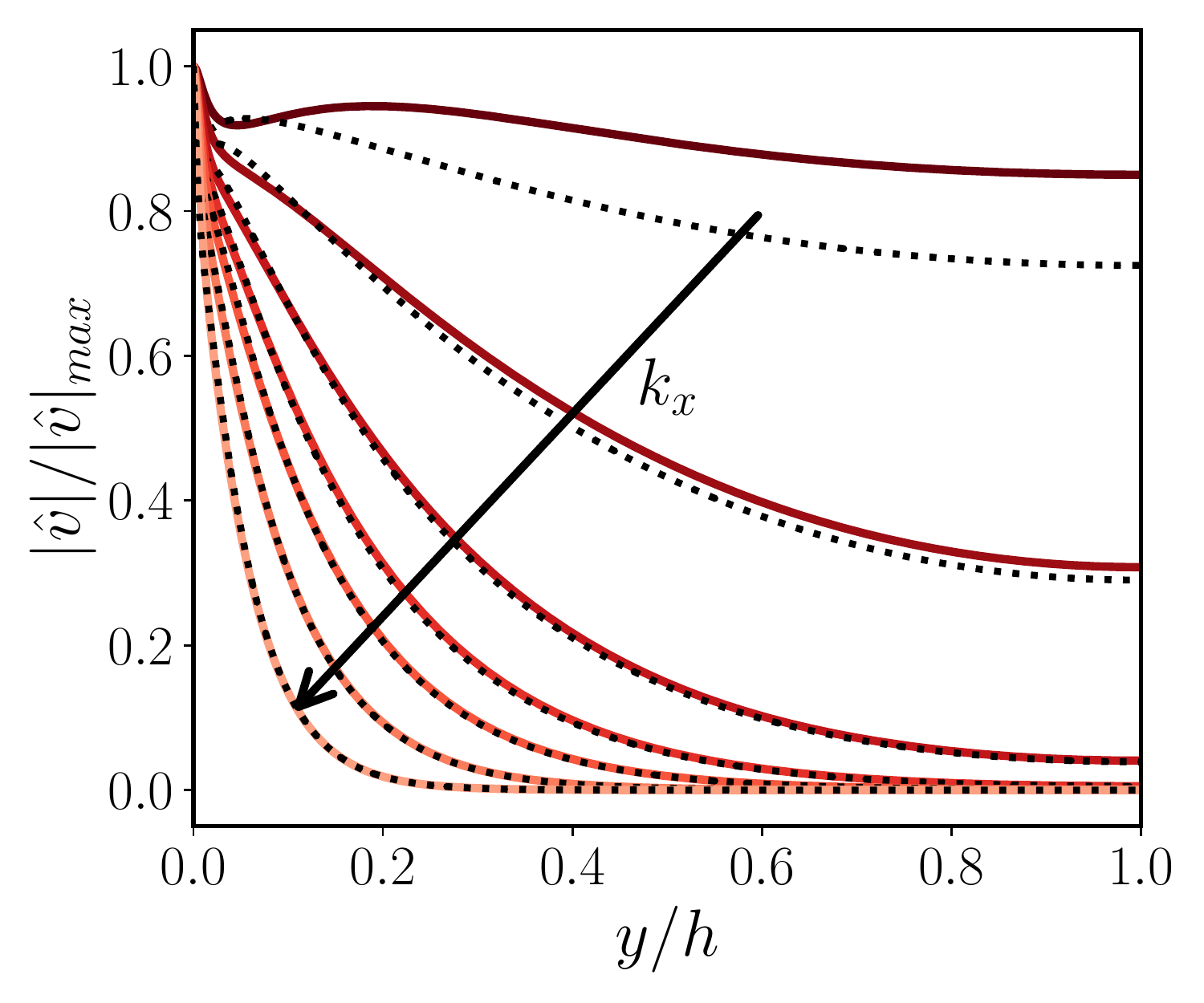}}
\put(4.3,3.7){(a)}
\put(4.3,1.8){(b)}
\put(9.4,3.7){(c)}
 \end{picture}
\caption{ (a)  Normalized eigenvectors of wall-normal velocity as a function of $y$ along the red upper path in figure~\ref{fig:cr_ci_kx2_varA},  $\phi = 3 \pi /4$, $k_x h =2$, $k_z = 0$. Color intensity increases with  $|A| \in [0.05, 0.52, 0.91]$. \dashed, the eigenvector of the uncontrolled flow from the $A$-branch  ($|A| = 0$). (b) As in (a) but for $|A| \in [0.91, 1.9, 5, 100]$.  (c) The unstable eigenvectors of the viscous problem  as a function of wavenumber $k_x h \in [1,2,4,6,8,12,20]$ (decreasing), $k_z = 0$. $\phi = 3 \pi /4$, $|A| = |A|_f (k_x)$ (see figure~\ref{fig:flip}d). \dotted, eigenvectors of the inviscid flow at the same control parameters.  Only symmetric eigenvectors are shown. \dashdotted, detection plane $y_d/h = 0.1$.}
\label{fig:eigvec}
\end{figure}
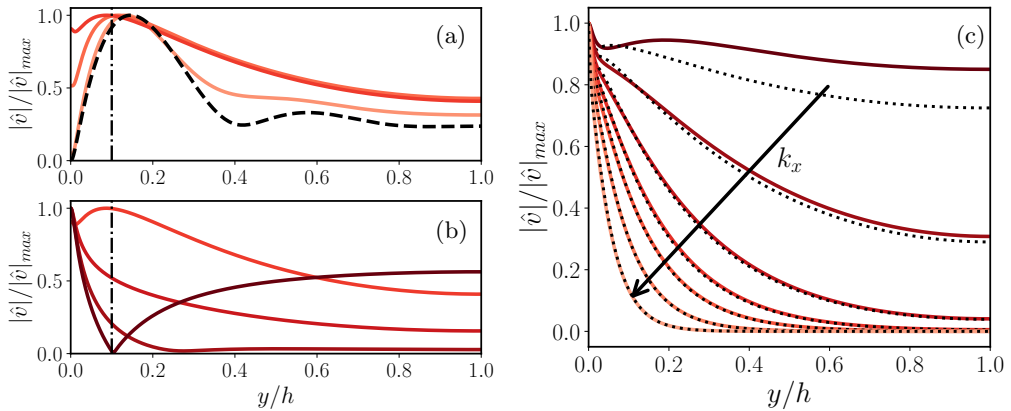

Finally, we show the influence of the control gain on the wall-normal shape of the eigenvectors as $|A|$ increases, and $\phi$ remains fixed. It is reasonable to track the evolution of the eigenvectors with $|A|$ along one of the paths in figure~\ref{fig:stab_amp}(b,c), than by a simple criterion of the maximum growth rate, where eigenvectors can belong to different branches. We consider here the eigenvectors with $k_x h = 2$, $k_z = 0$, associated with the eigenvalues along the red upper path with $\phi = 3\pi /4$ in figure~\ref{fig:cr_ci_kx2_varA}, as it is the most unstable one.  For simplicity, only symmetric eigenmodes are presented, because the near-wall behaviour of the antisymmetric modes is quite similar.  Figure~\ref{fig:eigvec}(a) shows the absolute value of $|\hat{v}|$ close to the wall, normalised with its maximum, for $|A| < 1$. Small values of the gain barely affect the eigenvector, which is very similar to the respective uncontrolled eigenvector from the A-branch. For $|A| = 0.05$ the flow is still stable. 
As $|A|$ increases to $0.5$, the flow becomes unstable, and $|\hat{v}|$ increases at the wall and the bulk of the flow. The shape of the eigenvector at the wall flattens when $|A|$ is increased further. Figure~\ref{fig:eigvec}(b) shows the evolution of eigenvectors for larger $|A|$. Now $\hat{v}$ decreases in the bulk of the flow, and at large enough $|A|$ changes sign with respect to its value at the walls. In the plot for $|\hat{v}|$ it is reflected by a developing minimum between the wall and the middle of the channel, which tends to $y_d/h = 0.1$ when $|A|>>1$. Apparently, for these extreme gains, the linear system adjusts the velocity at $y_d$ to be zero, following condition~\eqref{eq:ctrl_law}.  Effectively, this creates a ``narrower" channel and the instability is weakened as the control gain increases; the respective growth rate saturates at $c_i =0.03$ for  $|A|=100$ (see figure~\ref{fig:stab_amp}c).

  A note of caution must be placed here: the eigenvectors fill the whole height of the channel because the harmonic we show here is quite long ($k_x h = 2$). This does not happen to shorter modes with larger $k_x$, which peak near the walls and decay towards the centre of the channel. We show this effect in figure~\ref{fig:eigvec}(c) comparing the shape of eigenvectors at the control gains $|A|_f (k_x)$ that result in the largest growth, and $\phi=3 \pi /4$, as in the previous plot.  The choice of $|A|_f$ to represent behaviour of eigenvectors with $k_x$ is motivated by the fact that the instability is the strongest there and the instability isocountours for different wavenumbers scale better with $|A|= |A|_f$ (figure~\ref{fig:stab_amp}a). The eigenvector with $k_x h=1$ fills the entire channel, as in the previous case. The eigenvectors with $k_x h = 2,4$ still fill the entire channel, but their absolute value in the bulk of the flow decreases with $k_x$. As $k_x$ is further increased, the eigenvectors no longer occupy the middle of the channel, and are bounded to a region near the wall. The width of this region decreases with $k_x$. The eigenvectors of the viscous flow are in reasonable agreement with the ones of the inviscid flow, and this agreement becomes better with increasing $k_x$, indicating again that the instability is largerly inviscid at its peak.

\subsection{Wavenumbers affected by the instability}\label{sec:listab_wn}
\begin{figure}
\centering
  \begin{subfigure}[t]{0.5\textwidth}
      \caption{}\label{fig:max_ci_kx}
  	\centering
  	\includegraphics[width=\textwidth]{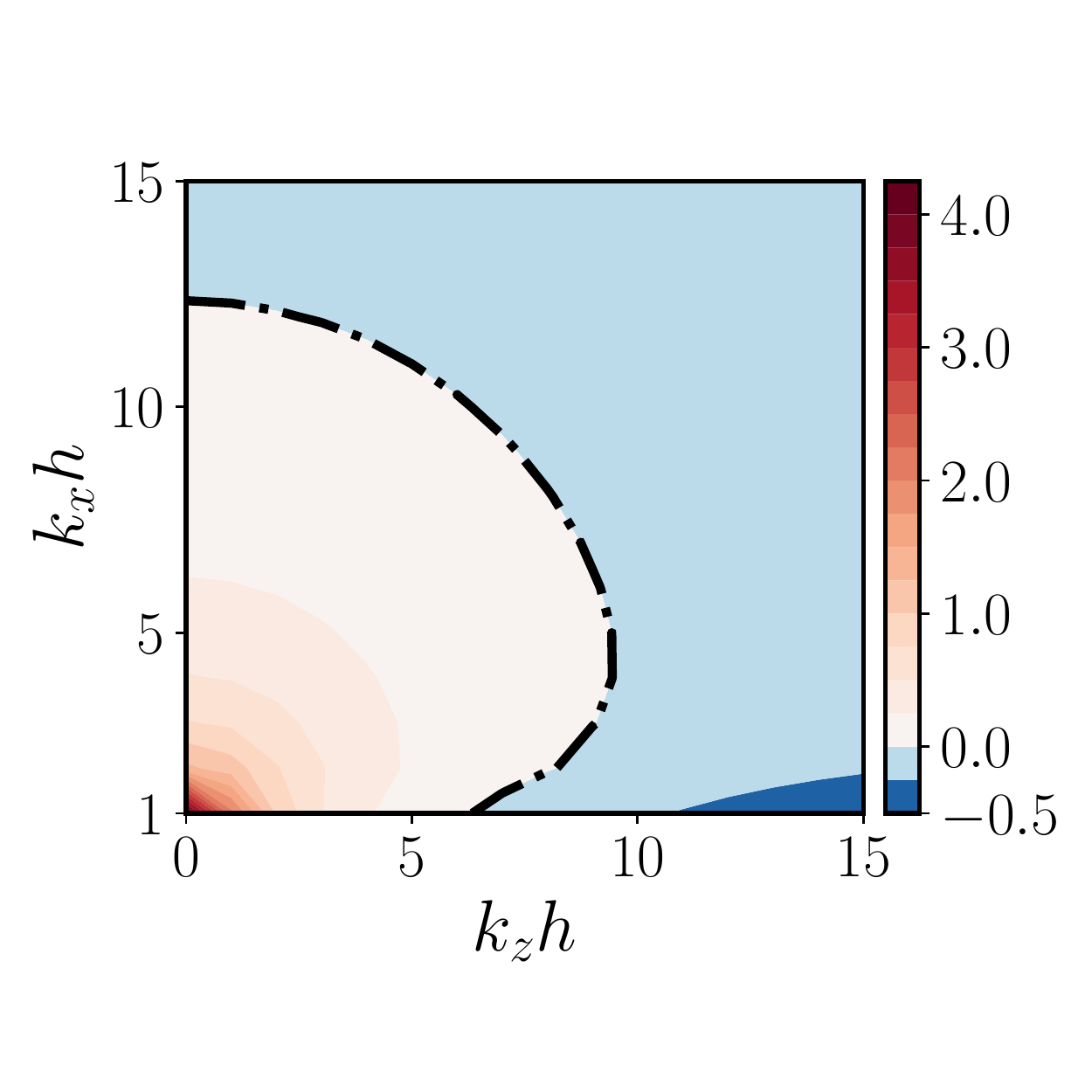} 
  \end{subfigure}
  \begin{subfigure}[t]{0.48\textwidth}
      \caption{}\label{fig:max_cr_kx}
  	\centering
  	\includegraphics[width=\textwidth]{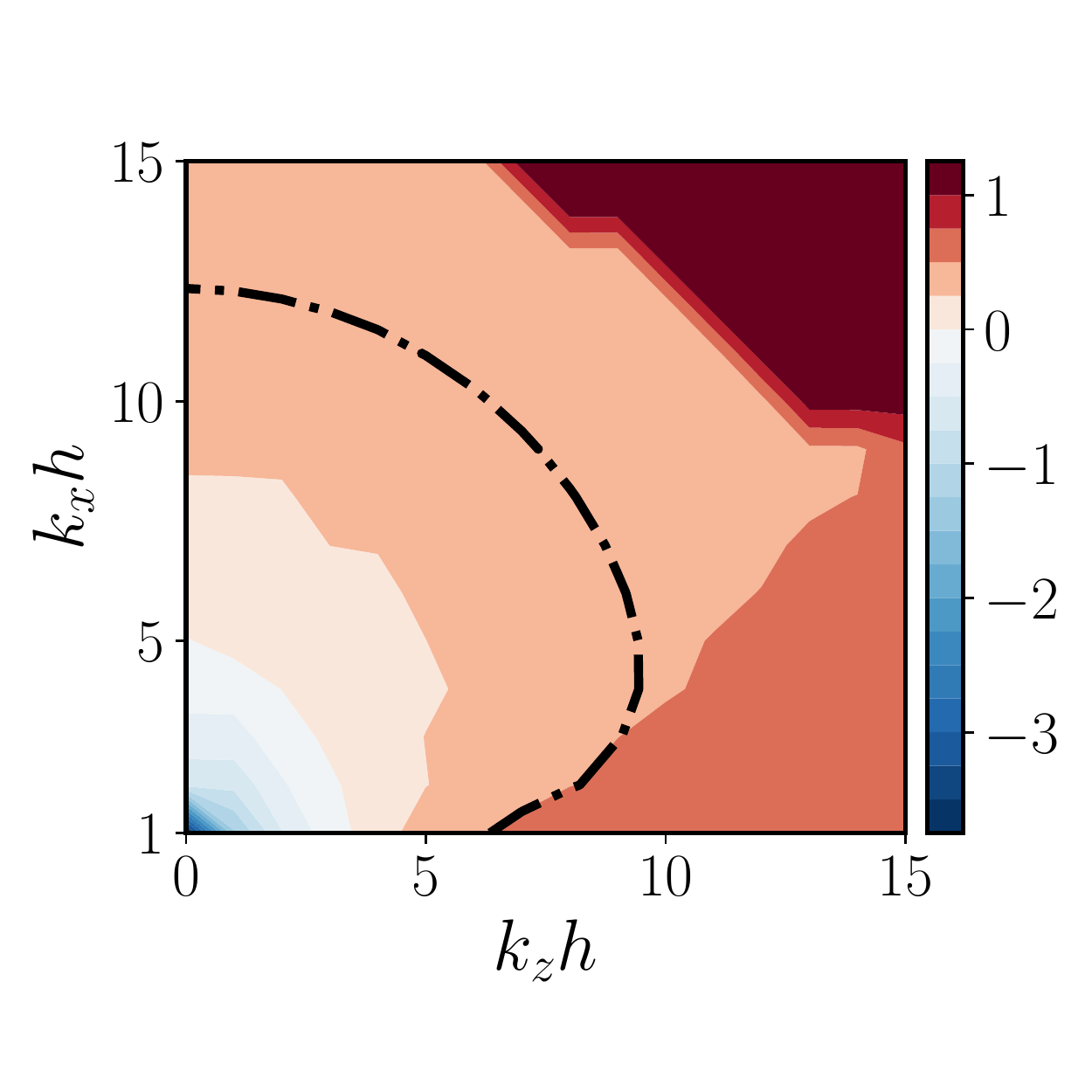}
  \end{subfigure}
  \caption{  (a) Contours of the maximum imaginary part of the eigenvalue $c_i/U_b$ as a function of wavenumbers, $k_x$ and $k_z$. (b) Contours of corresponding $c_r/U_b$.
  For each $k_{x,z}$ pair we search for the maximum growth rate and its respective $c_r$ in a range of amplitudes \anna{$A \in [10^{-2},1]$} and phases $\phi \in [-\pi, \pi]$.  \dashdotted, neutral stability curve with $c_i = 0$.}
  \label{fig:eigvec_stabmap}
\end{figure}
The above discussion emphasizes the effect of large control gains $|A|>1$, but gains larger than $O(1)$ are unlikely to be beneficial in terms of drag reduction. Besides the negative effects of the linear instability discussed above, large gains imply a large energy input at the walls and therefore a large cost of the control. In the following, we perform an optimization that aims to find out which wavenumbers are the most affected by the instability of the viscous flow for reasonably small values of $|A| \in [10^{-2}, 1]$. The importance of this optimization stems from the necessity to exercise caution with these length scales in future numerical and laboratory experiments. We seek for the wavenumbers that have the maximum growth rate in this range of $|A|$, and also maximise $c_i$ over $\phi \in [-\pi, \pi]$.   Figure~\ref{fig:max_ci_kx} shows the resulting stability map as a function of $k_x h$, $k_z h \in [0,15]$, each point of it corresponding to  a pair of $(|A|, \phi)$ that results in maximum $c_i$ for the wave mode of $k_x, k_z$.
The control appears to be more dangerous for longer waves with smaller $k_x$. The contours of $c_i$ are centred around $k_x h = 1$, $k_z=0$, indicating that the instability there grows faster. Large wavenumbers are not affected by the instability when the gain is small enough, and the wavenumbers with $k_x h \geq 13$, $k_z h \geq 7$ remain stable. Figure~\ref{fig:max_cr_kx} shows the phase velocitiy $c_r$ of the modes with the largest $c_i$. As discussed in section~\ref{sec:linstab_invisc}, the unusual effect of this instability is the appearance of  negative phase velocities. For the most unstable modes they are up to four times faster than the maximum of the mean profile (which on the scale of the colorbar is approximately $1$). These upstream-travelling modes, in the form of spanwise rollers, can be observed in the DNS during the linear growth phase in the corresponding parameter regimes. Lastly, we see in figure~\ref{fig:eigvec_stabmap} that high $k_x$-harmonics with infinite spanwise extent ($k_z = 0$) are more affected by the instability than harmonics with higher $k_z$. In fact, the growth rates of wavenumbers with $k_z>0$ are smaller than the ones with $k_z = 0$. We note that the range of unstable wavenumbers in figure~\ref{fig:eigvec_stabmap} partially coincides with the wavenumbers controlled in our DNS ($k_x h\in [0,6]$, $k_z h \in [0,10]$ in table~\ref{tab:sim_param}). 

In the following, we will compare the linearized controlled flow to the DNS results, employing the wavelengths with $k_z = 0$ as a proxy for the linear dynamics of the channel. This comparison will be done for $|A| \leq 1$, as discussed above. But before we move on, we should emphasize the potential importance of our results with $|A|>1$ in sections~\ref{sec:flip_invisc} and~\ref{sec:visc_invisc_diff}. Large control gains, although not very promising for drag reduction, are not unphysical and should be explored for enhancement of friction and mixing.  In addition, the flip of the eigenvalue motion happens for  gains just slightly above $|A| =1$ for very large wave lengths, which are of the same order of magnitude as in the ``classic" opposition control, and therefore may potentially affect its experimental implementations.  More importantly, the flip itself is an exciting physical phenomenon of the linearized controlled flow, which was not explored before.  

\section{Reconciling DNS and linear dynamics.}\label{sec:linstab_and_dns}
\subsection{Instability and the drag increase in the DNS}\label{sec:phase_x0}

\begin{figure}
\centering
  \begin{subfigure}[t]{0.45\textwidth}
      \caption{}\label{fig:phi_dx}
  	\centering
    \includegraphics[width=\textwidth]{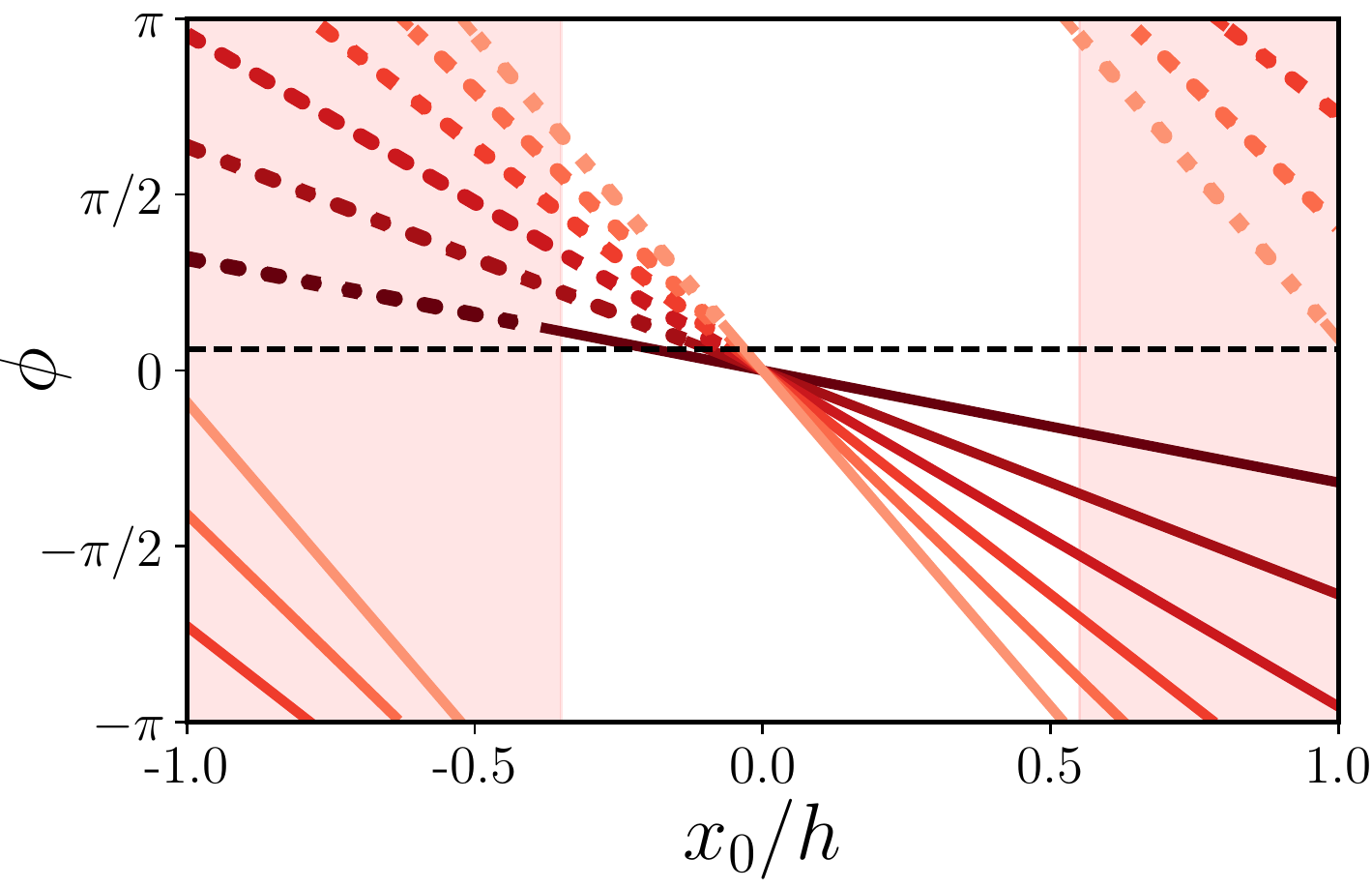}
  \end{subfigure}\hspace{1em}
    \begin{subfigure}[t]{0.45\textwidth}
      \caption{}\label{fig:ci_dx}
  	\centering
  	\includegraphics[width=\textwidth]{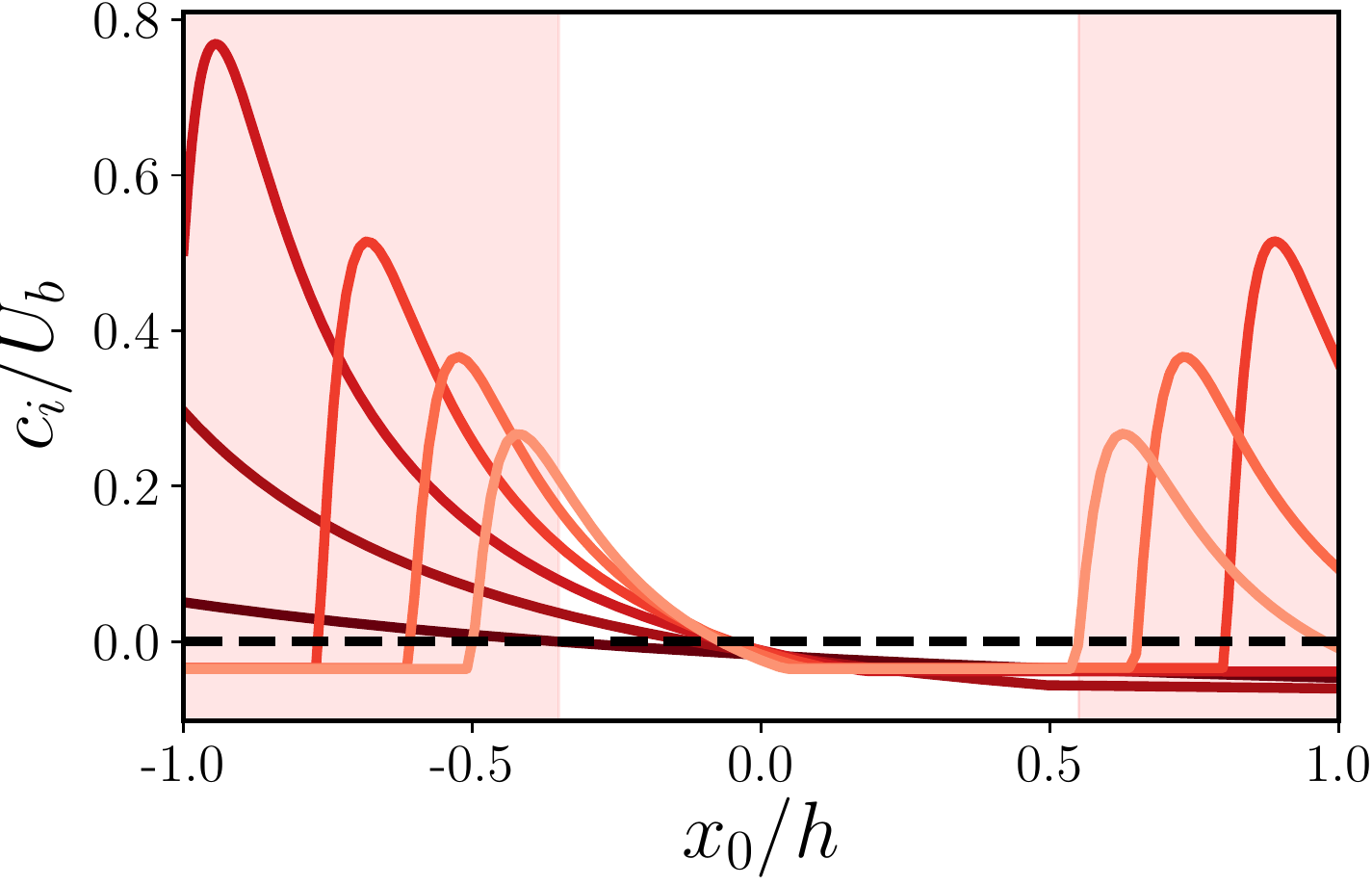}
  \end{subfigure}
  \begin{subfigure}[t]{0.46\textwidth}
      \caption{}\label{fig:cf_and_omega}
  	\centering
  	\includegraphics[width=\textwidth]{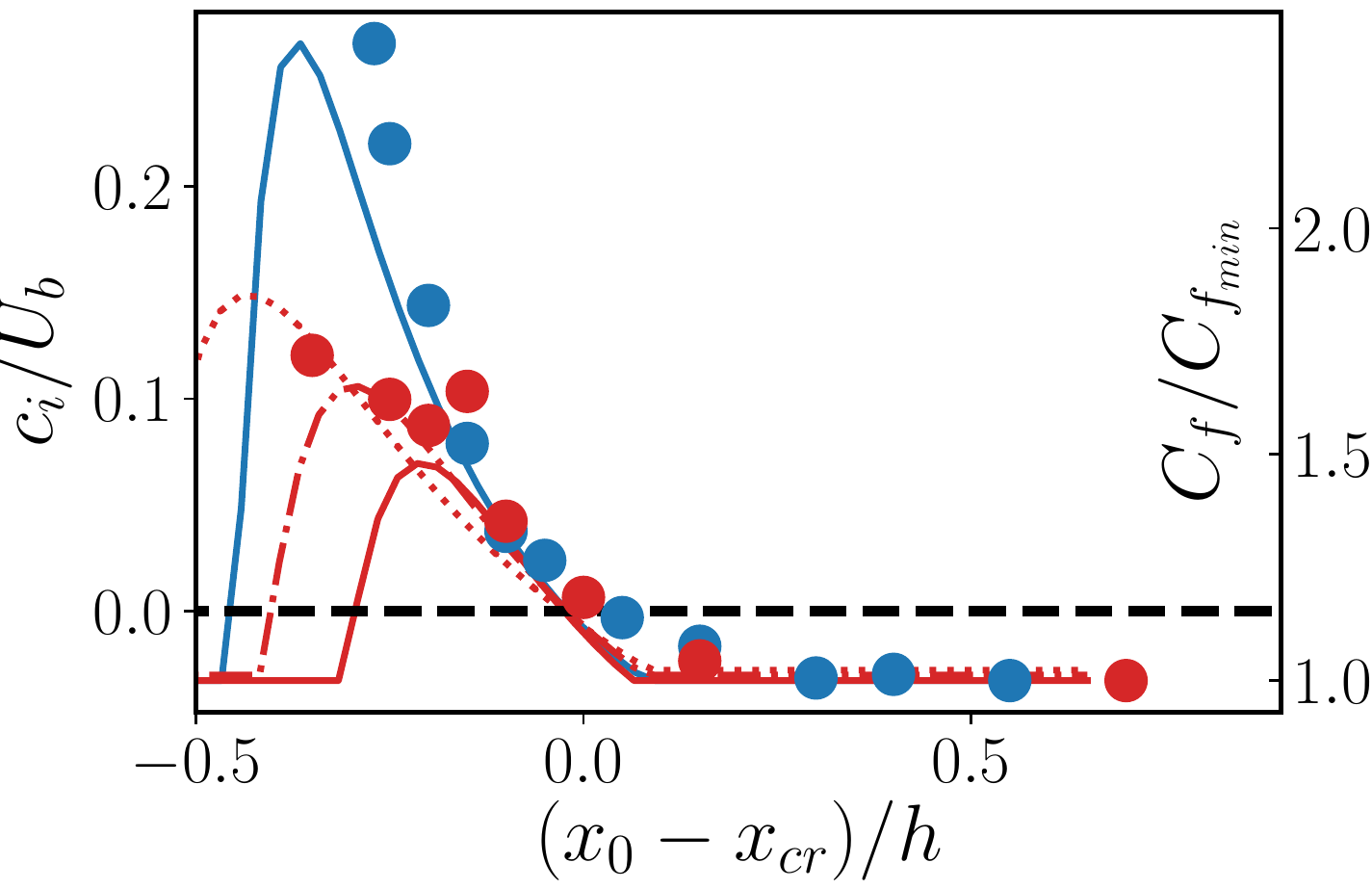}
  \end{subfigure}\hspace{1em}
  \begin{subfigure}[t]{0.44\textwidth}
      \caption{}\label{fig:kx_max_diverge}
  	\centering
  	\includegraphics[width=\textwidth]{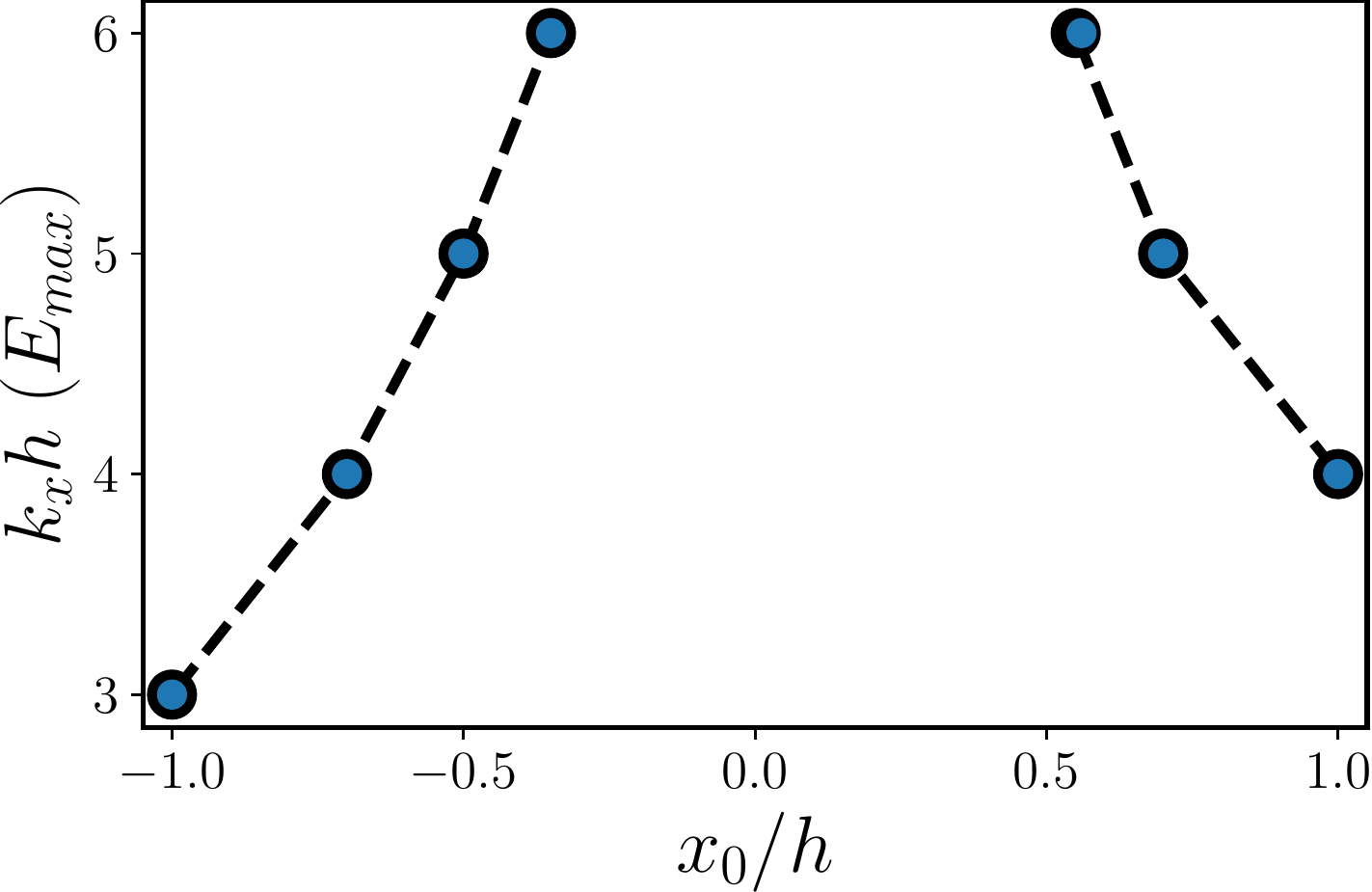}
  \end{subfigure}
\caption{ (a) The control phase $\phi$ as a function of streamwise wavenumber and the shift $x_0/h$. \dotted, highlights the presence of the instability for $|A|=1$. \dashed, the minimum phase for the onset of the instability. Color, from dark to light: $k_x h\in [1, 2,3,4,5,6]$. Shaded regions are the same as in figure~\ref{fig:cf_mean}. (b) $c_i$ as function of $x_0$ and $k_x$.  \dashed, $c_i = 0$, colors as in (a). $|A|=1$. (c)  \protect\marksymbol{*}{black},   friction factor normalized with its minimum (DNS results). Linear instability growth rates: \solid, $k_x h = 6$, \dashdotted, $k_x h= 5$, \dotted, $k_x h = 4$.  Blue color is $|A| = 1$, and red is $|A| = 0.7$. \dashed, neutral stability line $c_i = 0$. The $x$-axis is shifted by the minimum $x_0$ required to trigger the instability, $x_{cr} = (-0.05, -0.15)$ for $|A| = (1, 0.7)$. (d)  \dashed, the most rapidly growing wavenumbers in the diverging DNS cases. The blue circles correspond to the wavenumbers with maximum growth rate in the linear stability analysis. In all cases $k_z = 0$.}
\label{fig:phase_xlag}
\end{figure}

 Equipped with the findings from the previous section, we are in a position to explain some of our DNS results. In the DNS, we focused on relatively small control gains $|A| \leq 1$, where the instability is active for positive phases and does not exist for negative ones (figure~\ref{fig:stab_amp}a). Positive control phases correspond to negative streamwise shifts, and therefore a drag increase at large upstream shifts, $x_0 < 0$, in figure~\ref{fig:cf_mean} is intuitively expected. It is more surprising  that  the simulations also diverge on the right of figure~\ref{fig:cf_mean} where $x_0 >0.55$, without a preceding gradual increase in friction. This seemingly odd result can be explained if  the $2\pi$-periodicity of the control coefficient $A$ is taken into account, namely, a negative phase $-\pi <\phi_1 < 0$ and a positive phase $\phi'_1 = 2\pi + \phi_1$  result in the same value of $A$. To eliminate this redundancy, we project the phases of the streamwise modes controlled in our DNS on the $(-\pi, \pi)$ unit circle and plot them in figure~\ref{fig:phi_dx} as a function of the streamwise shift $x_0$. Let us first focus our attention on the left half of the plot, $x_0<0$.  As the shift decreases from $x_0=0$, the phase applied to the controlled harmonics grows, indicating that they may become unstable.  The unstable interval of $x_0$ for each $k_x$ is drawn dotted in figure~\ref{fig:phi_dx}, computed for $|A| =1$ with linear stability analysis from section~\ref{sec:listab_vics}. The exact threshold depends on $k_x$, but the dashed  horizontal line $\phi = 0.06\pi$ approximately separates the stable and unstable regimes. 
 The first wavenumbers to become unstable are $k_x h =3$, $4$ and $5$ at the shift $x_{cr} \approx -0.05$, immediately followed by  $k_x h = 6$.  Longer modes with $k_x = 1,2$ become unstable one after the other shortly after as $x_0$ decreases further. To simplify the following discussion, let us focus on the wavenumber $k_x h = 6$ with the most rapidly growing phase (larger $k_x$ implies larger $\phi$ for the same $x_0$). With further decrease of $x_0$,  the phase of $k_x h= 6$ reaches $\phi = \pi$ at $x_0/h \approx -0.5$, beyond which the phase of $A$ becomes negative (bottom left corner in figure~\ref{fig:phase_xlag}a), and $k_x h= 6$ becomes stable again. The situation is different for $x_0>0$. As expected, the phase of $k_x h = 6$ is negative at first and further decreases as $x_0$ increases, until it reaches $-\pi$ at $x_0 \approx 0.5$. When the streamwise shift is increased further, the  value of $\phi$ changes from $-\pi$ to $\pi$. Now it is equivalent to a large positive phase shift and the flow becomes unstable again.   The phases of the rest of  streamwise wavenumbers vary in a similar manner.
 
 It is worth mentioning here that approaching the instability on the left side on the figure~\ref{fig:phase_xlag}(a) is different from approaching it on its right-hand side, because the behavior of the two limits of the instability range is different. Figure~\ref{fig:ci_dx} shows how the growth rate $c_i$ of the most unstable eigenvalues varies for each wavenumber $k_x$ along the lines in figure~\ref{fig:phi_dx}. Linearly unstable wavenumbers are those for which $c_i >0$, corresponding to the dotted segments in figure~\ref{fig:phi_dx}. Note that this plot is similar to  figure~\ref{fig:ci_phi_A1} in section~\ref{sec:visc_invisc_sim}, but  with $x_0$ instead of $\phi$ as an argument, so that the $c_i$-curves are shifted with respect to each other. If we again follow  $k_x h =6$ as we make $x_0 $ increasingly negative  (i.e. approaching the instability on the left), the instability growth rate changes slowly, progressively increasing from negative to slightly positive values. A different picture emerges if we begin to increase $x_0$, starting from $x_0 = 0$ (i.e. approaching the instability on the right). Here at the onset of instability, $x_0 = 0.55$,  the $c_i$-curve is almost perpendicular to the $x$-axis and grows sharply. This reflects an abrupt transition of the flow to the instability when the phase of control changes from $-\pi$ to $\pi$. A further small increase in $x_0$ brings the flow to its maximum $c_i$. In fact, the unstable ``bumps" in $c_i$ on the left and on the right of the plot are exactly the same, arising from control with identical $\phi$ (and $A$), and it is their asymmetry that matters for the development of the instability. This asymmetry is also visible in figure~\ref{fig:visc_eig_kx6A1}, where the growth rates increase progressively as the control phase is increased from zero, but would rapidly reach their maximum if the control phase is decreased from $\pi$. Another important observation is that whether we approach instability on the left or on the right, at its onset the growth rate of $k_x = 6$ is larger than of the rest of the controlled wavenumbers. This is either because the control phase of $k_x =6$ grows faster as $x_0$ decays (on the left), or because its phase has a shorter period on $x_0$ (on the right), as seen in figure~\ref{fig:phase_xlag}(a). Besides that, $k_x h = 6$ has the largest effective growth rate, $\omega_i = k_x c_i$, given comparable values of $c_i$ for the rest of unstable wavenumbers in figure~\ref{fig:ci_dx}. Therefore, one would expect the unstable flow to be dominated by this wavelength close to the onset of instability.
 For example, the onset of the instability of $k_x h =6$ on the right correlates with the unexpected increase in friction on the right in figure~\ref{fig:cf_mean}, and the resulting divergence of our DNS. 
 
  In figures~\ref{fig:phase_xlag}(a,b), the onset of the instability on the left correlates with the steep friction increase on the left in figure~\ref{fig:cf_mean}. Thus, we next focus our attention on the relation between the instability growth   and the increase in friction in the DNS at  negative streamwise shifts. Figure~\ref{fig:cf_and_omega} presents both $C_f$ and $c_i$ as functions of streamwise shift, similarly to figure~\ref{fig:cf_mean}. But unlike in figure~\ref{fig:cf_mean}, the $x$-axis is shifted by the minimum $x_0$ leading to the instability. In the following, we will refer to this minimum as the critical value of the streamwise shift $x_{cr}$. Its value is not universal and depends on the most unstable wavenumber and on the control gain. In the linearized flow corresponding to our DNS, $x_{cr}\approx -0.05$ for $|A|=1$, $x_{cr}\approx -0.15$ for $|A|=0.7$.
 Presented in this way, positive values on the $x$-axis are linearly stable, and negative ones are unstable. When the flow is linearly stable, $c_i$ stays at a relatively constant level, which depends more on $k_x$ than on the control gain.  We focus on $k_x h = 6$ again, as it quickly attains the largest growth due to its fastest-growing control phase, and dominates the flow dynamics. Approaching $x_0 - x_{cr}= 0$ from the right, the imaginary part of the eigenvalue begins to increase shortly before the critical point until the flow becomes unstable. After a further growth, $c_i$ reaches its maximum and then quickly decays to the previous stable levels. The DNS behavior is remarkably similar. Although the friction rises slightly before $c_i$ experiences growth itself, i.e. before $x_{cr}$, the pronounced growth in friction factor for $|A| = 1$ on the left of the plot correlates well with the increase in $c_i$. After some point, however, we are unable to advance further our DNS and the behaviour of $C_f$ past the growth rate maximum is unclear. Figure~\ref{fig:ci_dx} suggests that the instability at $k_x =6$ would be subsequently overtaken by $k_x =5$, then by $k_x=4$ and so on as we shift the control further upstream, towards more negative $x_0$. To support this conjecture, we performed additional simulations with $|A| = 0.7$. Here the linear analysis predicts a slower instability growth (compare the red and blue solid lines in figure~\ref{fig:phase_xlag}c), and indeed, we can explore a wider range of  $x_0$ in the DNS. 
 The $C_f$-curves collapse well both in the stable regime and close to the onset of instability for the two values of control gain. Again, the friction factor growth correlates with the onset of instability. The instability at $|A| = 0.7$, $k_x h =6$ is observed in a narrower range of $x_0$, its growth rate reaches its maximum at smaller $|x_0 - x_{cr}|$ that when $|A| =1$, and also decays faster with $x_0$. As the control shifts towards more negative $x_0$, the phase of the next longer mode, $k_xh=5$, becomes 
 positive enough so that eventually its growth rate becomes larger than that of $k_x h=6$. When $c_i(6)$ reaches its maximum and begins to decay, $c_i(5)$ is still growing, and $k_x h=5$ becomes the dominant mode.  This is reflected in the change of slope in $C_f$ at about this location, and also in the spectrum of $v$ (not shown here). At even larger shifts, $k_xh=4$ becomes dominant. 
 
 Finally, to confirm that the linear instability is the cause of significant drag increase and the failure to converge the DNS at larger $x_0$, we show in figure~\ref{fig:kx_max_diverge} the wavenumbers that grow faster as the DNS diverges, and compare them to the controlled wavenumbers with the largest linear growth rate. Those wavenumbers should outgrow the rest when the instability develops, and this is indeed observed in the DNS. As explained above, the wavelength governing the instability become longer for larger $|x_0|$. Note that all these wavemodes have $k_z =0$ and therefore represent a spanwise roller, similar to the one in figure~\ref{fig:u10p_xlag_m03}.


\subsection{The influence of control gain on transition to instability}\label{sec:bifurcation}

\begin{figure}
\centering
  \begin{subfigure}[t]{0.45\textwidth}
      \caption{}\label{fig:ret_time_m03}
  \centering
    \includegraphics[width=\textwidth]{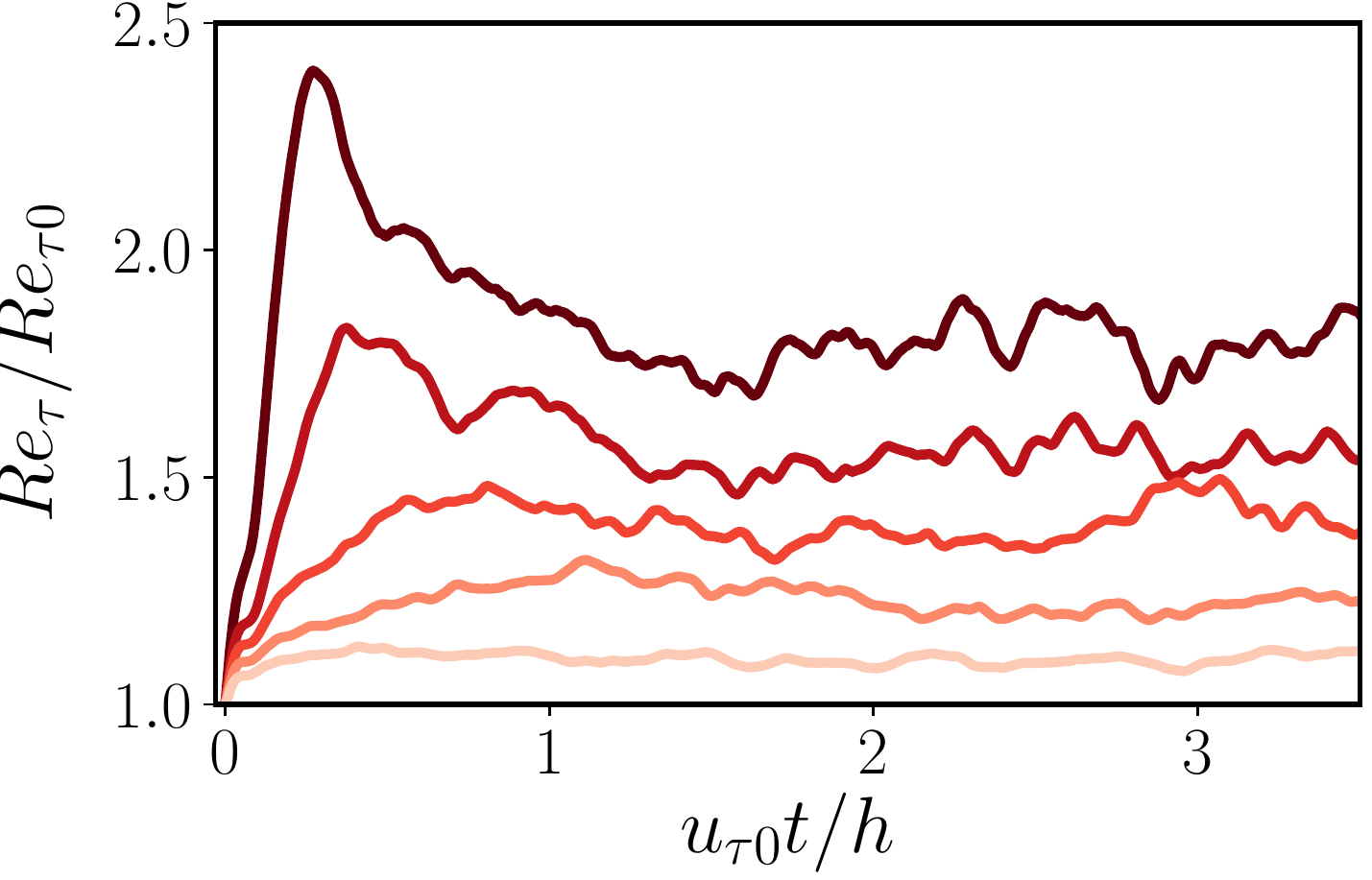}
  \end{subfigure}\hspace{1em}
  \begin{subfigure}[t]{0.45\textwidth}
     \caption{}\label{fig:ret_time_p055}
  \centering
   \includegraphics[width=\textwidth]{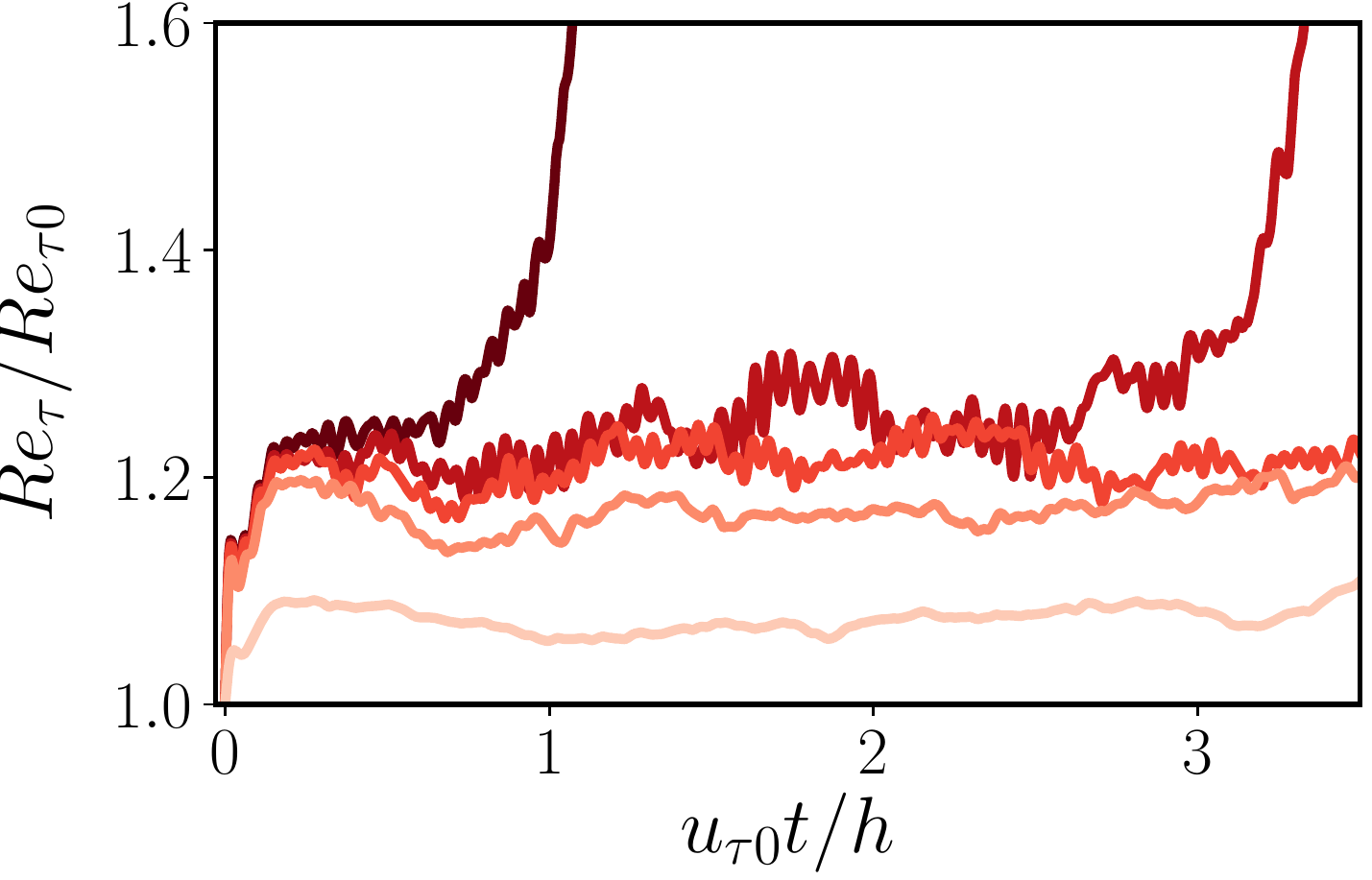} 
  \end{subfigure}
 \begin{subfigure}[t]{0.45\textwidth}
     \caption{}\label{fig:crit_amp_m03}
  	\centering
  	\includegraphics[width=\textwidth]{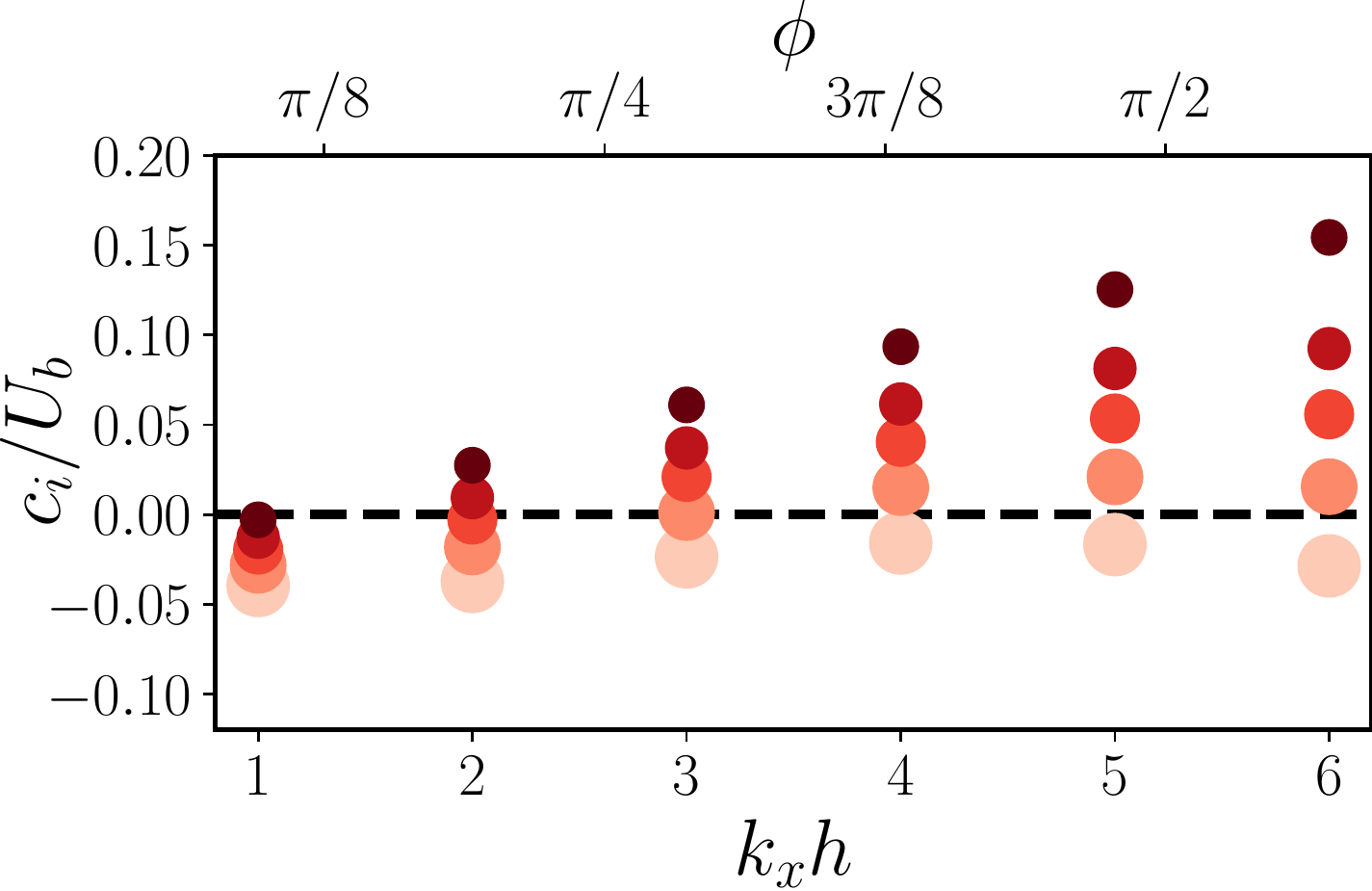} 
  \end{subfigure} \hspace{1em}
  \begin{subfigure}[t]{0.45\textwidth}
     	\caption{}\label{fig:crit_amp_p055}
  	\centering
  	\includegraphics[width=\textwidth]{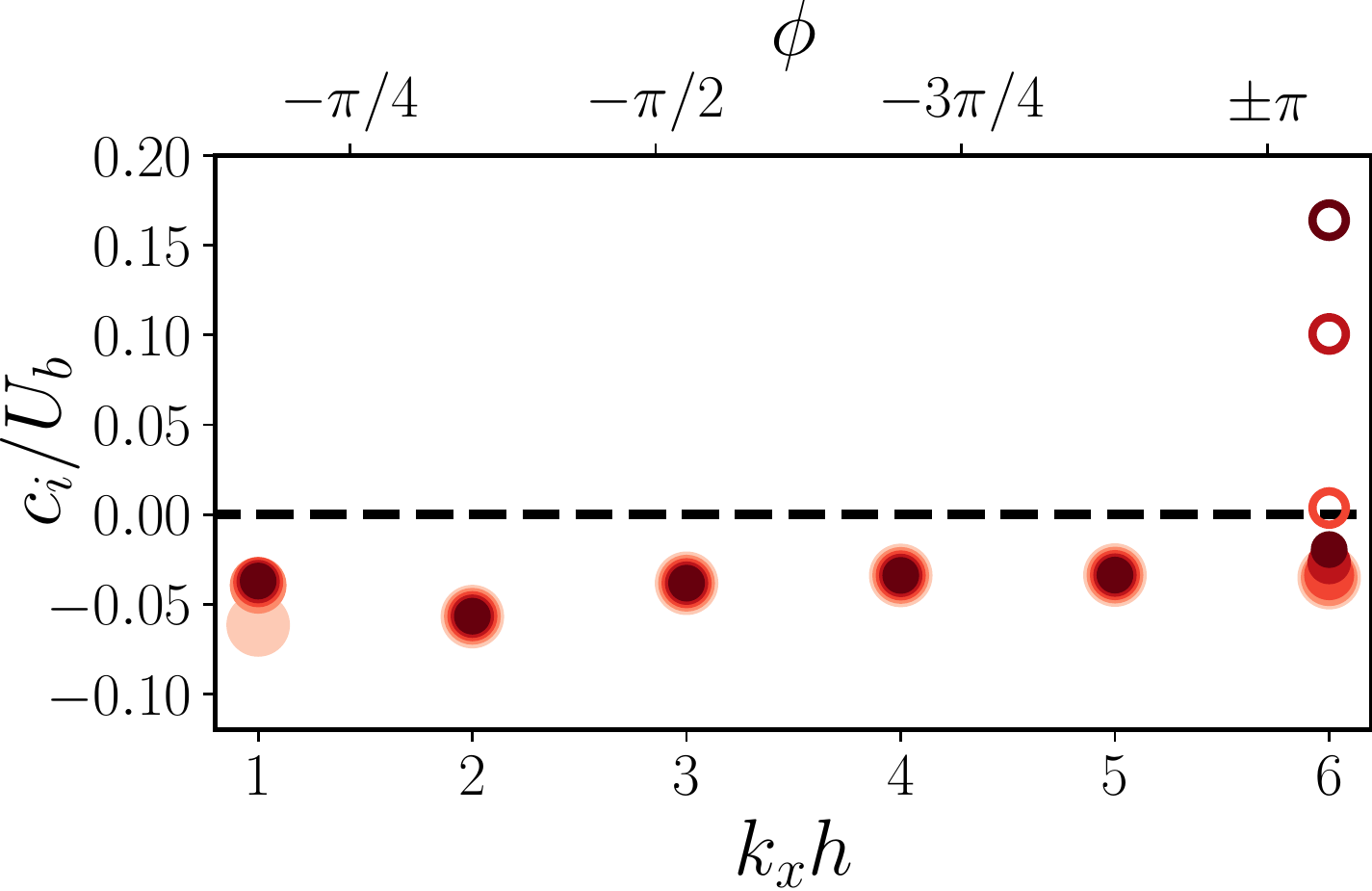}
  \end{subfigure}
\caption{ Top row: friction Reynolds number, normalized with $Re_{\tau 0}$ of the uncontrolled case, as a function of time. Time is normalized with the eddy turnover time $h/u_{\tau 0}$. The control gain $|A|$ grows with color intensity. Bottom row: the largest growth rates of eigenmodes as a function of controlled $k_x$, $\phi$, and $|A|$, $k_z =0$. \dashed, the neutral stability border with $c_i = 0$. The smaller is the symbol size, the larger is $|A|$.  (a), (c) $x_0 = -0.3$, $A \in (0.5, 0.6, 0.7, 0.8, 1)$. (b) $x_0 = 0.55$, $|A| \in (0.5, 0.9, 0.95, 0.96, 0.97)$ (d)  \anna{$x_0 = 0.55$}, \ding{108}, $|A| \in (0.5, 0.9, 0.95, 0.96, 0.97)$, $\bigcirc$, $|A| \in (1.0, 1.1, 1.15)$. }
\label{fig:xlag_transition}
\end{figure}

Further comparison between the linear stability analysis and the DNS gives valuable information about the nature of the flow transition to a new state at both stability edges of figure~\ref{fig:phase_xlag}(a,b). We draw our attention again  to the relation between friction levels in DNS and the linear instability growth rates, but now as a function of $|A|$. Figure~\ref{fig:ret_time_m03} shows the DNS  time history of $Re_\tau$ for $x_0 = -0.3$, close to the onset of the instability on the left-hand side of figure~\ref{fig:ci_dx}. The initial growth of $Re_\tau$ is stronger with increasing $|A|$. It can be clearly discerned for $A= 1.0$ and $0.8$, accompanied by less robust but still noticeable initial increase in $Re_\tau$ at $|A|=0.7$ and $0.6$. After the initial growth phase, the flow saturates to a new  turbulent state with higher $Re_\tau$ than that of the uncontrolled flow. The amplitude of the $Re_\tau$  fluctuations also intensifies, compared to the uncontrolled flow or parameter regimes where the instability is inactive, i.e. $|A| = 0.5$. The saturation of the initial growth takes place on a time scale on the order of the eddy turnover time ($u_{\tau 0}t/h \approx 1 $), which explains the choice of normalization factor in this plot. Now let us compare the behavior of $Re_\tau$ to the growth rates obtained with linear analysis.  In figure~\ref{fig:crit_amp_m03} these growth rates are calculated at $x_0 = -0.3$, for the same values of $|A|$ as in figure~\ref{fig:ret_time_m03}. The growth rates of the longest, most unstable wavenumbers continuously increases with $|A|$ and cross the neutral stability line $c_i = 0$ when $|A| \in [0.5, 0.6]$. This interval correlates with the onset of drag increase in the DNS in figure~\ref{fig:cf_amp}.  The gradual increase in wall friction in figure~\ref{fig:ret_time_m03}, attributed to the presence of instability in figure~\ref{fig:crit_amp_m03}, indicates a supercritical transition occurring at negative streamwise shifts.

In figure~\ref{fig:ret_time_p055}, we show the time history of $Re_\tau$ for $x_0 = 0.55$, the right-hand side limit in figure~\ref{fig:cf_mean}. The initial growth of $Re_\tau$ here is much shorter than in figure~\ref{fig:ret_time_m03}. When $|A|$ is relatively small, the simulations saturate on a low-friction level, where $Re_\tau$ increases only by $10-20$\% with respect to the base flow, as opposed to the $100-200$\% increase in figure~\ref{fig:ret_time_m03}. At $|A| \approx 0.95$, two well discernible frequencies appear and modulate the flow evolution. Further increase in $|A|$ makes the flow wander away from this modulated state. Although  we cannot reach the final high-frequency flow state with our DNS, during the transition we observed that the most rapidly growing wavelengths are roller-type structures (figure~\ref{fig:phase_xlag}d).  This suggests that the final state is also dominated by large-scale rollers, similar to the ones in figure~\ref{fig:u10p_xlag_m03}.  Again, we continue by plotting the growth rates corresponding to $x_0 = 0.55$, as a function of $k_x$ and $|A|$ (figure~\ref{fig:xlag_transition}d). Here the flow remains linearly stable at the amplitudes corresponding to figure~\ref{fig:ret_time_p055}, and there is no noticeable change in $c_i$ until $|A| \approx 1$. This control gain value is  higher than $|A| = 0.96$, 
already resulting in transition to a high-frequency state in the DNS in figure~\ref{fig:ret_time_p055}. If we keep increasing $|A|$, the longest wavenumber $k_x =6$ becomes unstable, as the only wavenumber with positive phase. Its $c_i$ grows more rapidly with $|A|$ than for $x_0 = -0.3$, reaching similar levels with a smaller relative increase in the control gain. The appearance of additional frequencies, modulating the flow, and also the fact that the transition occurs earlier in the DNS, suggests that the transition to the new state for $x_0 = 0.55$ in figure~\ref{fig:phase_xlag}(a,b) is subcritical.

\section{Response to the forcing}\label{sec:res_forced}

\begin{figure}
\centering
  \begin{subfigure}[t]{0.32\textwidth}
      \caption{}\label{fig:DNSvsLin_Eu}
  \centering
    \includegraphics[width=\textwidth]{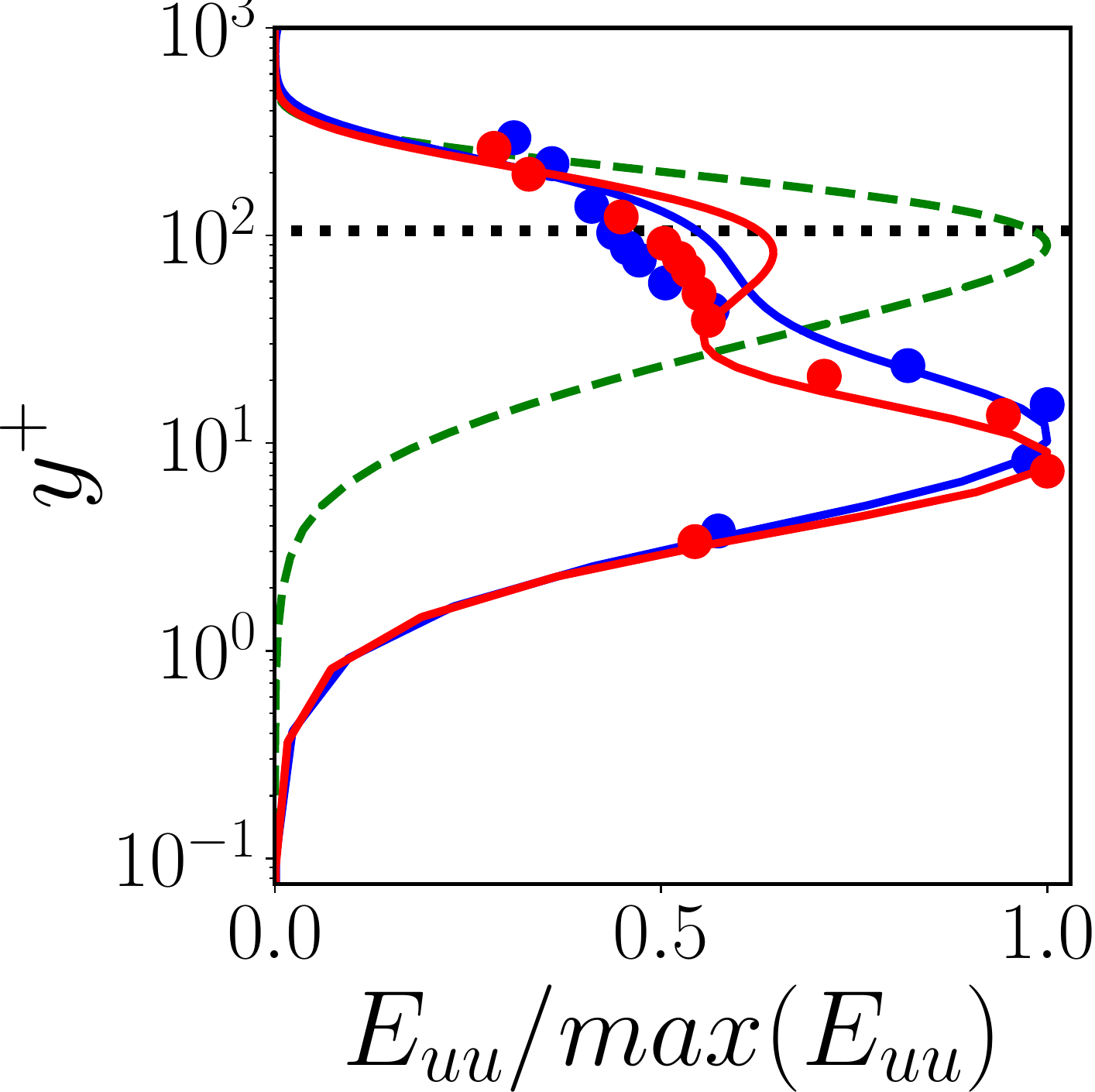}
  \end{subfigure}\hspace{0.1em}
  \begin{subfigure}[t]{0.32\textwidth}
     \caption{}\label{fig:DNSvsLin_Ev}
  \centering
   \includegraphics[width=\textwidth]{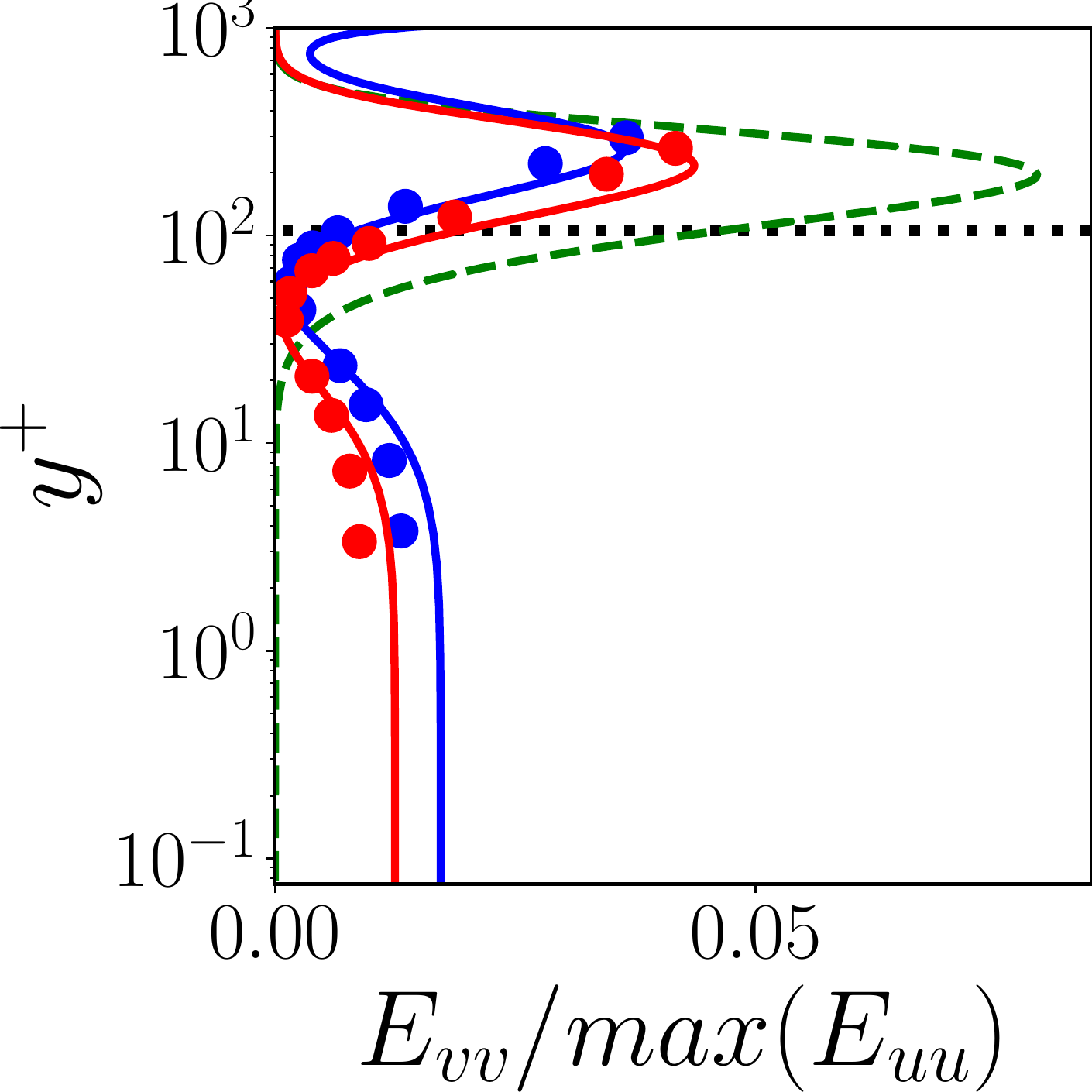}
  \end{subfigure}\hspace{0.1em}
    \begin{subfigure}[t]{0.32\textwidth}
  \centering
     \caption{}\label{fig:DNSvsLin_Euv}
   \includegraphics[width=\textwidth]{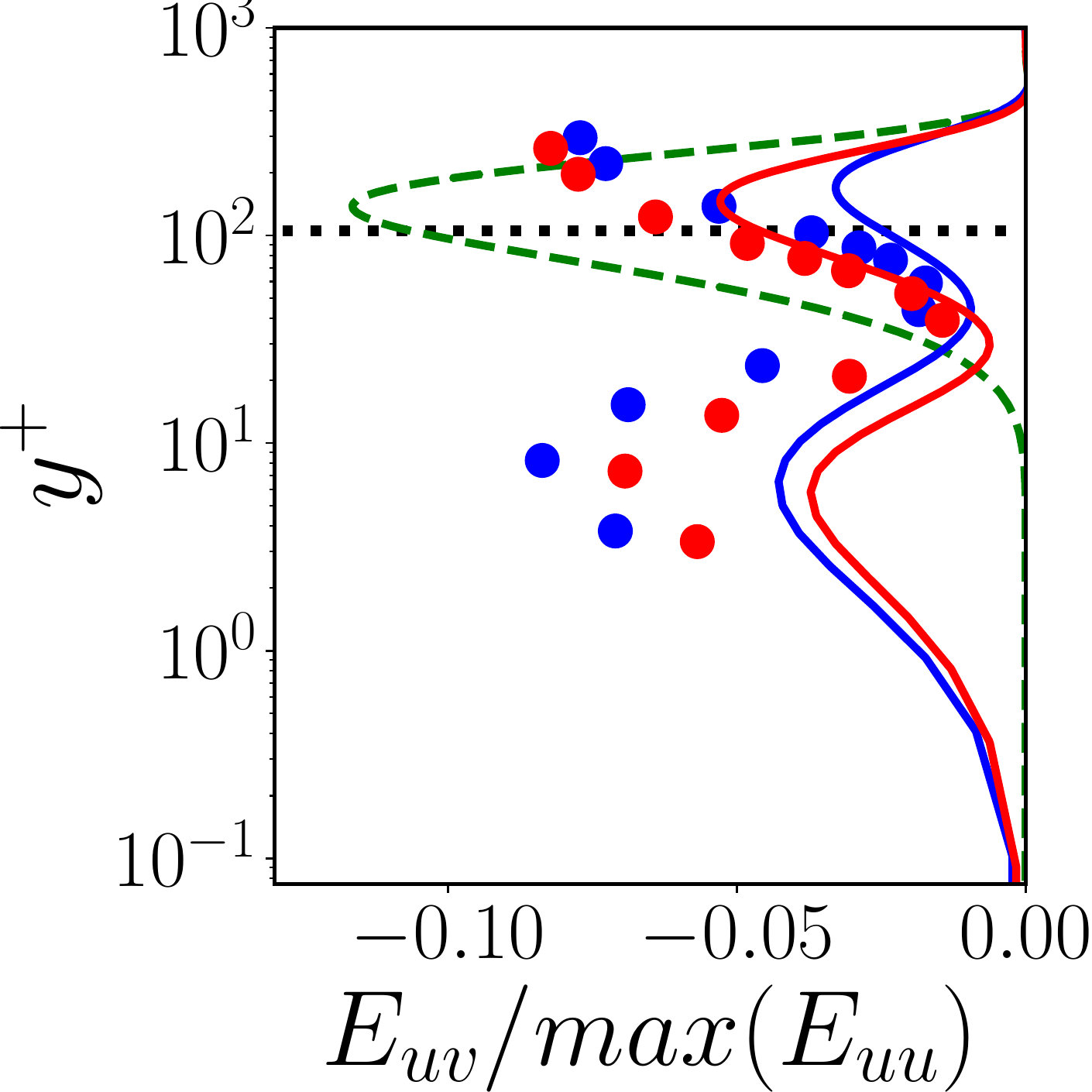}
  \end{subfigure}
  \caption{ Responses to forcing of the linearized Navier-Stokes operator for the mode with $\lambda_x \sim 2h$, $\lambda_z \sim h$, $c_f = U(y_d)$. (a) Energy of the streamwise velocity component.  (b) Energy of the wall-normal velocity component. (c) Tangential Reynolds stresses. \dashed , without control, \solid , with control, \protect\marksymbol{*}{black}, the energy of the corresponding mode in the DNS. $|A|=1$, in blue; $|A| = 0.7$, in red. $x_0 = 0$. \dotted , $y^+_d$, the location of the critical layer for the uncontrolled flow. All energies are normalised with the maximum of $E_{uu}$, and $y^+$ is normalized with the wall units of each case.}
\label{fig:DNSvsLin}
\end{figure}

Figure~\ref{fig:phase_xlag} shows that the flow is linearly stable in the range of $0 \leq x_0<0.55$ for the control parameters corresponding to our DNS. Therefore, the near-wall oblique waves from figure~\ref{fig:u10p_full_kxkz} are caused by  another physical mechanism.
We explore here the possibility of their amplification through a response of the linear system to forcing at a certain frequency. Using the mathematical formalism  outlined in the end of section~\ref{sec:setup_linstab}, we consider the norm~\eqref{eq:res_norm} of the resolvent operator~\eqref{eq:res_op} as the maximum amplification factor for the responses of the linearized controlled flow, and corresponding to it most amplified flow modes.
A necessary modification to the approach in section~\ref{sec:setup_linstab} is needed before we compare it to the DNS results. Before, we took the mean profile and the turbulent viscosity of the uncontrolled flow as the base state for the linear analysis. Now, since we consider time-periodic responses with real frequencies $\omega_f$, neither growing nor decaying, we need to replace the uncontrolled base state with the parameters of the statistically steady controlled flow. This was achieved by extracting the mean velocity $U_c$ and total shear stress profiles from the DNS with control, and replacing $\nu_t$ in \eqref{eq:cess} with the ratio of the total shear stress and the $y-$derivative of $U_c$. Turbulent mean profile and viscosity are different for each $|A|$ and $x_0$. We considered here $x_0 =0$ for brevity, although the conclusions of this section hold for the entire interval of $0 \leq x_0<0.55$.

The operator~\eqref{eq:res_op} requires a frequency $\omega_f$, or phase velocity of the forcing $c_f$, as a parameter. Our control method acts by measuring large scales of $v$ at $y_d/h = 0.1$ and applying them with a corresponding constant factor at the wall. Thus $v$-structures at the wall must have the same advection velocity as the structures of $v$ at $y_d$.   The advection velocity of different flow modes varies with their size, but for large structures in the logarithmic layer it can be  approximated by the mean velocity at their wall-normal location \citep{jimenez2018coherent}. In the following, we adopt $c_f =  U_c(y_d/h = 0.1)$ as the phase speed of resolvent forcing. This translates into  relatively fast advection, $c_f/U_b \approx 0.75 - 0.78$, when scaled with the bulk velocity.  This approach is different from approximating the response velocity field through a sum of the left singular vectors of the operator~\eqref{eq:res_op} with different phase speeds \citep{luhar2014opposition,toedtli2019predicting}. Rather than observing a cumulative effect of the forcing with all possible phase speeds, we want to capture responses to the specific phase speed of the detection plane.

In figure~\ref{fig:DNSvsLin} we compare the wall-normal profile of the most amplified response to the wall-normal distribution of energy in the DNS. The length of the harmonic is set to $\lambda_x/h = 2$, $\lambda_z/h =1$, corresponding to the approximate size of the oblique wave in figure~\ref{fig:u10p_full_kxkz}. The  critical layer for each of the modes is located at $y_d =0.1$ ($y^+_d \approx 100$) for the uncontrolled flow, and is only slightly shifted upwards in wall units when the flow is controlled. In the uncontrolled case, the response modes peak near the critical layer, where advection velocity of forcing is equal to advection velocity of the mean profile. In the controlled case, the linear response to the forcing, besides the expected peak at $y^+ \approx 100$, has a second peak in $E_{uu}$ around $y^+ \approx 10$ (figure~\ref{fig:DNSvsLin}a). Figure~\ref{fig:DNSvsLin_Ev} shows that both the linear response in $v$ and the energy of the selected mode in the DNS exhibit a minimum in $v$ around $y_0^+ = 50$, as already indicated by the minimum in the contribution to the rms of $v$ from the large scales in figure~\ref{fig:vrms_lsonly}. There is also a maximum of $v$ in the logarithmic layer as in the uncontrolled case. Finally, figure~\ref{fig:DNSvsLin_Euv} shows the contribution to the Reynolds stress by this particular mode. Unlike in the uncontrolled case, where Reynolds stress has only one pronounced maximum in the logarithmic layer, in the controlled flow interaction of the new peak in $E_{uu}$ in the buffer layer and non-zero $E_{vv}$ at the wall generates a second peak in $E_{uv}$. This peak is also located in the buffer layer and its magnitude is comparable to the logarithmic-layer maximum (compare to the two-peak distribution in figure~\ref{fig:Euu_spec}d). Responses obtained with our linear amplification model capture the shape of the DNS energies reasonably well both for $|A| = 1$ and $0.7$, indicating that oblique waves observed in the DNS are indeed the amplified linear responses of the flow subject to the forcing with $c_f = U_c(y_d)$. Note that the resolvent analysis will give the same amplification and modal structure for modes with $\pm k_z$, due to the symmetry of the linearized Navier--Stokes operator under transformation $k_z \to -k_z$, and therefore can not predict whether the waves will be travelling in the positive or the negative direction of $z$-axis in the DNS. However, maximum resolvent amplification, quantified by the largest singular values, could explain the energy build-up on the scale of oblique waves,  $\lambda_x/h \approx 2$, $\lambda_z/h \approx 1$.

\begin{figure}
    \centering
    \includegraphics[width=0.65\textwidth]{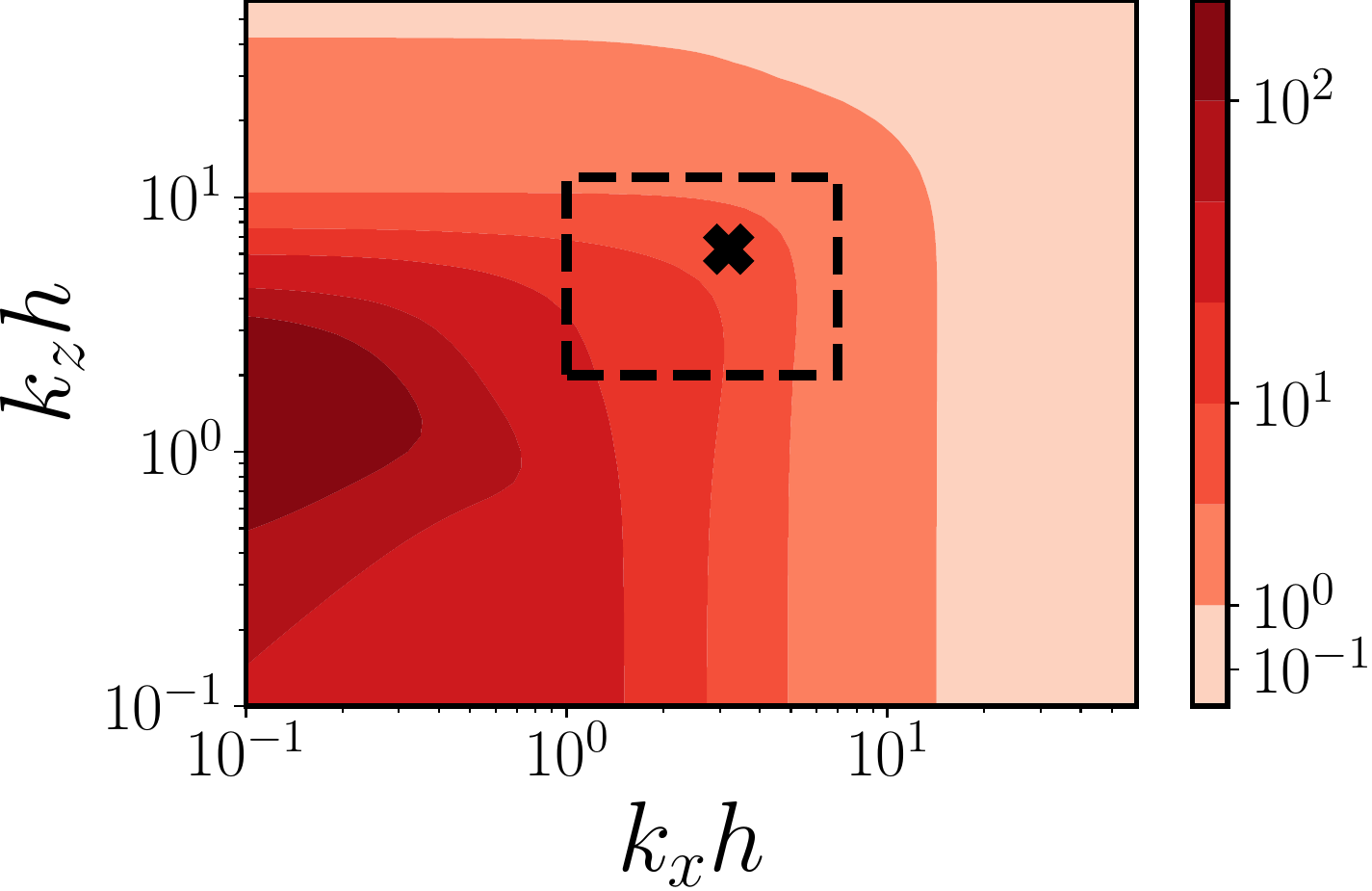}
    \caption{Maximum amplification of the resolvent response $\sigma_0$ as a function of streamwise and spanwise wavenumbers. Controlled in the DNS wave band is $k_x h \in [1,6]$, $k_z h \in [2, 10]$, which is highlighted by the dashed rectangle. \ding{54} marks the length scale of oblique waves from figure~\ref{fig:u10p_full_kxkz} ($\lambda_{x,z}/h \approx 2,1$). Control parameters: $c_f = U(y_d)$, $x_0 = 0$, $|A| = 1$.}
    \label{fig:sigma_kx}
\end{figure} 

Figure~\ref{fig:sigma_kx} shows contours of the first singular values $\sigma_0$ of the operator in \eqref{eq:res_op}, or equivalently, its norm~\eqref{eq:res_norm}, as a function of $k_x$ and $k_z$.  In the linear analysis we can employ much larger wavelengths than permitted by our computational domain in DNS. This allows to capture the general trend in $\sigma_0$ for very large structures. Note that neither infinitely long waves, $k_x = 0$, nor infinitely wide waves, $k_z = 0$, are included in this logarithmically scaled plot. The gradual increase in the response intensity with increase of the length scale indicates that the large scales are more sensitive to the forcing \anna{through the nonlinearity}, which represents an additional challenge for their control. The contours shift towards smaller $k_z$ as $k_x$ decreases, meaning that the most amplified response gets wider as it gets longer. At very long wavelengths, on the order of  $k_x \approx 0.1$ or $\lambda_x/h = 2\pi /0.1 = O(10^2)$, this trend saturates at $k_z \approx 1.5$ ($\lambda_z/h \approx 4.2$), which is larger than the width of the computational box in our DNS. The wavenumbers that are observed and controlled in the DNS are highlighted by the dashed rectangle in figure~\ref{fig:sigma_kx}.  The distribution of $\sigma_0$~contours in this area indicates that the amplification maximum should happen on the largest possible flow scale, especially in $x$. This prediction is in agreement with the mechanism of transient growth, which predicts larger amplification and longer lifetimes of the perturbations with longer streamwise wavelengths \citep{del2006linear}. However, the wavelengths $\lambda_x/h \approx 2$, $\lambda_z/h \approx 1$, marked by the black cross in figure~\ref{fig:sigma_kx}, are smaller than the largest controlled streamwise length scale, in contradiction with the linear prediction. 

\begin{figure}
\centering
  \begin{subfigure}[t]{0.45\textwidth}
     \caption{}\label{fig:sigma_ctrl_ks}
  \centering
   \includegraphics[width=\textwidth]{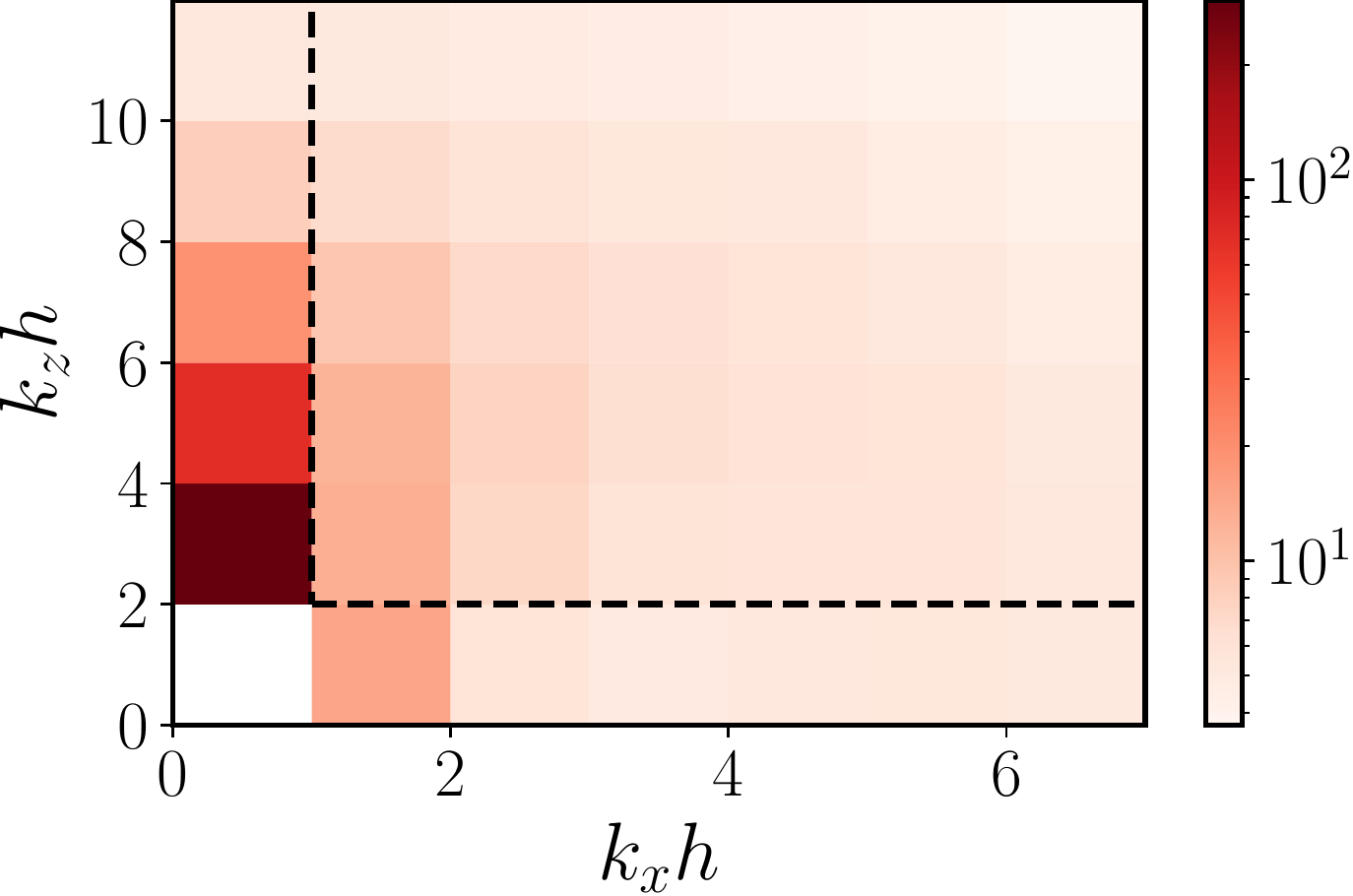} 
  \end{subfigure}\hspace{1em}
  \begin{subfigure}[t]{0.45\textwidth}
     \caption{}\label{fig:Eint}
  \centering
   \includegraphics[width=\textwidth]{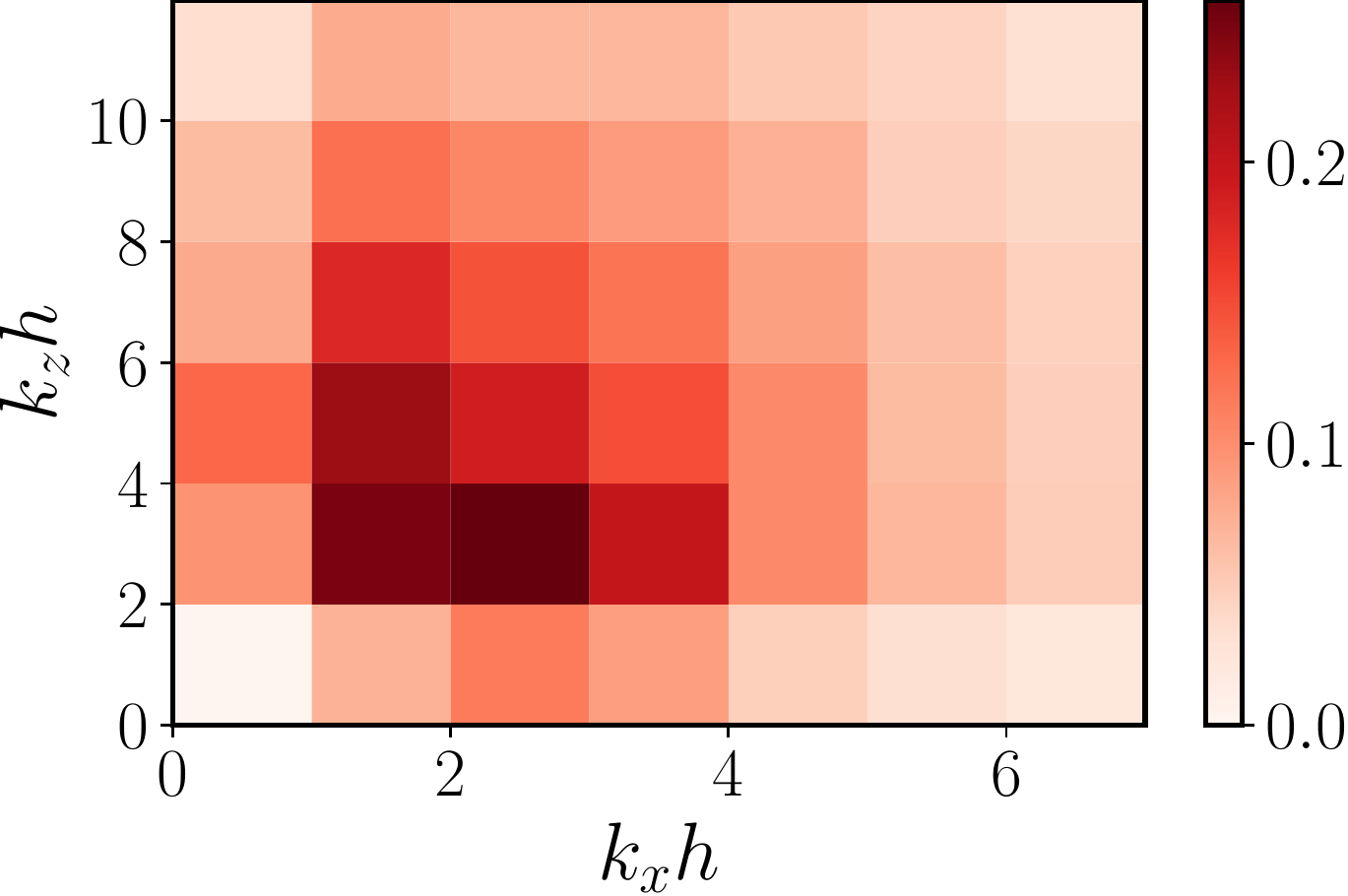} 
  \end{subfigure}
    \begin{subfigure}[t]{0.46\textwidth}
      \caption{}\label{fig:Evwall}
    \centering
    \includegraphics[width=\textwidth]{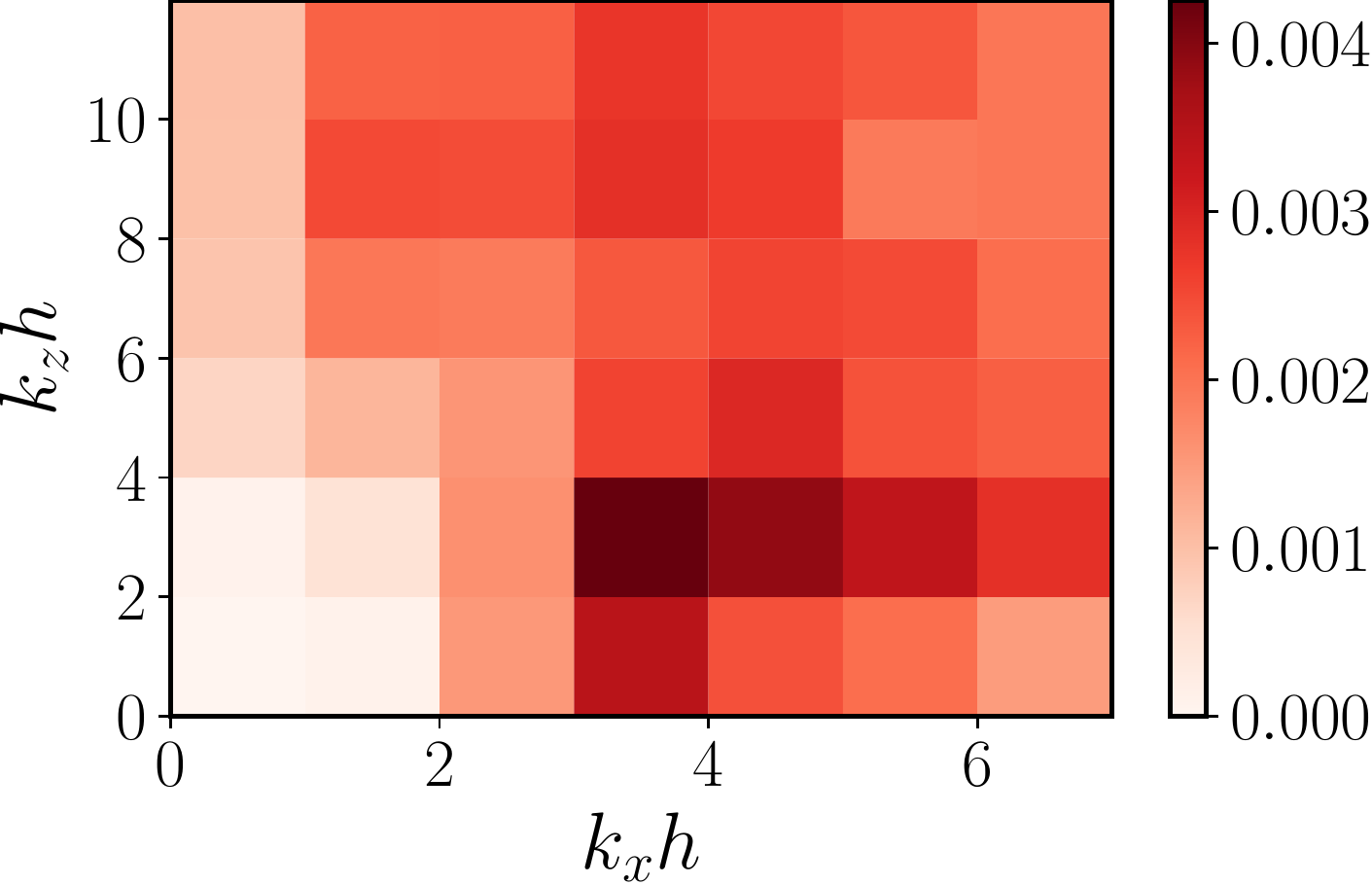}
  \end{subfigure}\hspace{1em}
    \begin{subfigure}[t]{0.45\textwidth}
      \caption{}\label{fig:Eint_Evwall}
    \centering
    \includegraphics[width=\textwidth]{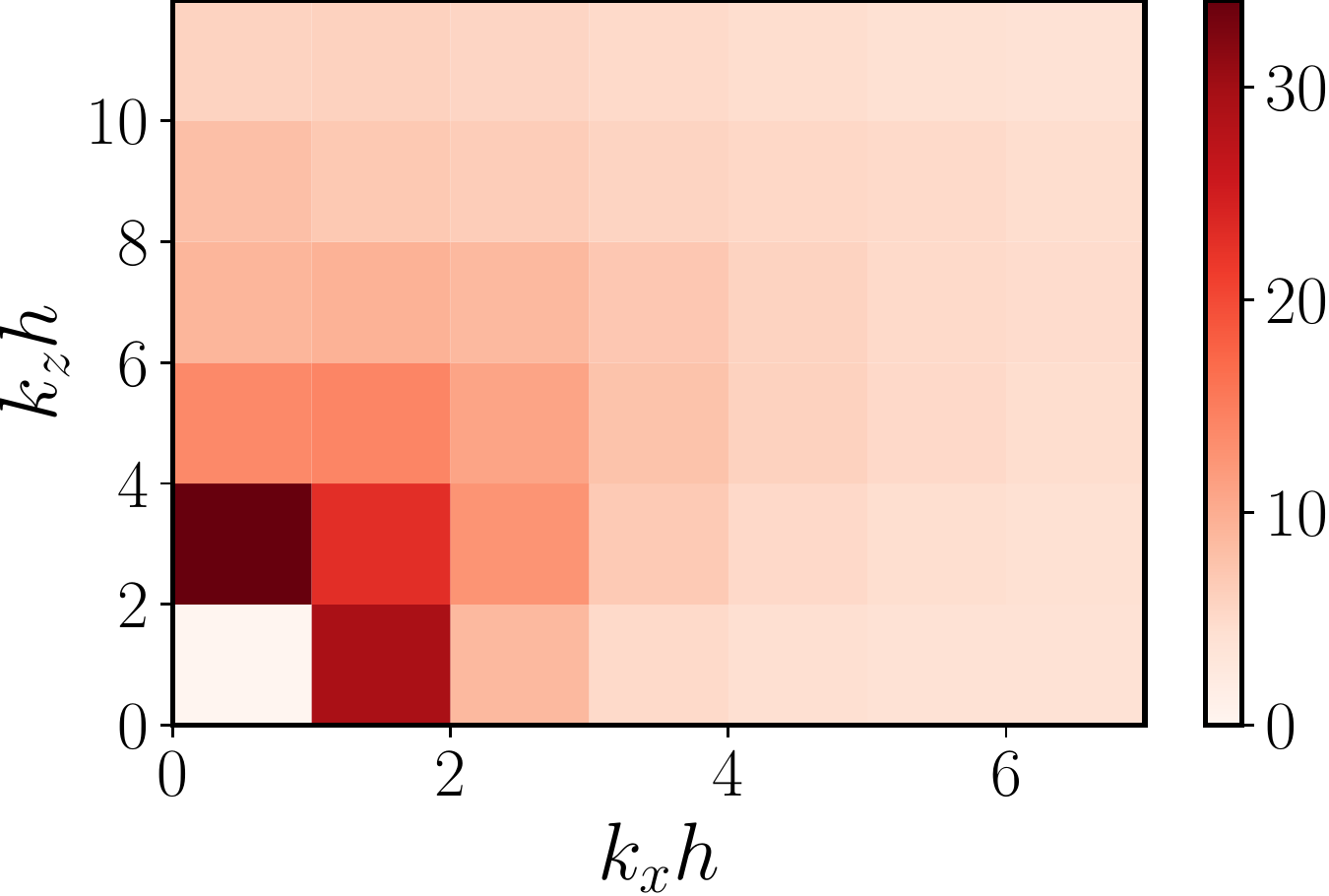}
  \end{subfigure}
\caption{(a) Maximum linear response $\sigma_0$ to forcing at $c_f = U (y_d)$ as function of wavenumbers.  Dashed rectangle contains the same wavenumbers as in figure~\ref{fig:sigma_kx}. (b) Total kinetic energy of the flow $E_{tot}$, integrated over channel height. (c) Energy contribution $E_{vv,wall}$ of wall-normal velocity at the wall.   (d) $\sqrt{E_{tot}/E_{vv,wall}}$, the ratio of (b) to (c).   $k_{x,z}=0$ was not controlled in the DNS and was removed from the plots for clarity. Energies in (b), (c) were normalised with $u^2_\tau$. $x_0 = 0$ for all panels.}
\label{fig:max_resolvent}
\end{figure}

In figure~\ref{fig:sigma_kx}, all scales are affected by control, while only a limited range of scales is controlled in the DNS. To compare the resolvent analysis to the DNS, let us take a closer look at these scales. Figure~\ref{fig:sigma_ctrl_ks} gives the values of $\sigma_0$ for the wavenumbers that are contained in the dashed rectangle in Figure~\ref{fig:sigma_kx}, augmented with $k_x = 0$, $k_z \in [2,10]$ and $k_z =0$, $k_x h\in [1,6]$ which were also controlled. The flow produces a very large linear response at $k_x=0$, which is equivalent to a non-periodic in time forcing with $k_x c_f =0$. For the rest of the controlled scales, the response energy is gradually increasing with scale, which is highlighted by the logarithmically spaced color map. Since the resolvent norm~\eqref{eq:res_norm} was weighted by the kinetic energy of the flow, as in \citet{Schmid2012}, a reasonable quantity to compare with it is the total kinetic energy of the DNS flow. The next panel, figure~\ref{fig:Eint}, presents this quantity, $E_{tot} = E_{uu} + E_{vv} + E_{ww}$, integrated over the channel height. The maximum in $E_{tot}$ does not capture the location of oblique waves in figure~\ref{fig:Euu_spec}, because the energy spectra of $u$ and $v$ in figure~\ref{fig:Euu_spec} were premultiplied and integrated over $\lambda_x$ or $\lambda_z$ instead of $y$. Although figure~\ref{fig:Eint} shows a preference of the flow towards larger scales, the trend is not very clear and infinitely long or wide wavelengths are less energetic than it is suggested by figure~\ref{fig:sigma_kx}.

Alternatively to the resolvent analysis, we can also view the resulting controlled DNS flow in an input-output framework, as a system forced with the wall-normal velocity from the wall. The amplitude of this forcing is unevenly distributed across the scales, since the rms of $v$ at the detection plane increases as the length scale becomes smaller. This way, a particular energy distribution in the flow favouring oblique length scales could stem from the inhomogeneous distribution of control intensity over scales. Figure~\ref{fig:Evwall} shows the distribution of the wall-normal component of the energy at the wall, $E_{vv, wall}$ acting on the flow. It shows a tendency of energy increase towards short scales in $x$, and a peak at $k_{x,z} h =(3,2)$. We take the energy of $v$ at the wall in figure~\ref{fig:Evwall}, and normalise the total energy response $E_{tot}$ with it. The square root of this result, estimating the response amplification, is given in figure~\ref{fig:Eint_Evwall}, and  the tendency of the flow to produce an amplified response at larger scales becomes much more apparent.

Moreover, the DNS responses, normalised this way, also allow for a weighted comparison with our resolvent model. The latter relates $v_w$ and $v(y_d)$ in a way where all wavenumber pairs $(k_x,k_z)$ are treated equally, without taking into account that different scales of wall-normal velocity receive different amounts of energy at the wall through control, as in the DNS. Normalizing the total DNS flow response with the wall-normal velocity distribution at the wall allows to correct for this factor.  The distribution of response magnitudes in figure~\ref{fig:Eint_Evwall} is similar to figure~\ref{fig:sigma_ctrl_ks}, with a preference of the flow response towards the large scales in both cases. There is an intense peak at $k_x = 0$, $k_z h= 2$, which can be a signature of the $\sigma_0$ maximum in figure~\ref{fig:sigma_kx}, but a larger computational domain would be required to prove this point. Overall, figure~\ref{fig:Eint_Evwall} and figure~\ref{fig:sigma_ctrl_ks} correlate reasonably well and support the hypothesis of the proportionality of flow response to the intensity of $v$.

\section{Discussion and conclusions}\label{sec:discussion}

In this work, we performed an analysis of large-scale opposition control targeting structures in the logarithmic layer of the fully turbulent channel flow. The control, applied on the largest Fourier modes of wall-normal velocity  in the DNS, creates a virtual-wall effect on these modes, and simultaneously provokes a large response in streamwise velocity in the buffer layer. This effect is accompanied by the appearance of a near--wall peak in Reynolds stresses and therefore 
a pronounced and undesired increase of friction. Since advancing the detection  upstream with respect to actuation was shown beneficial for drag reduction in opposition flow control by \citet{lee2015opposition}, we introduced a streamwise shift between sensing and control input as a new parameter. Shifting control in the streamwise direction, as depicted in figure~\ref{fig:cf_mean}, divides flow behaviour in two parameter regions. The first one, of approximately $x_0/h \in [0, 0.55)$, features a moderate and roughly constant drag increase, about 50-100\% more than in the uncontrolled flow. It is accompanied by the appearance of oblique waves. In the second parameter region, of approximately $x_0/h \in [-0.35, 0)$, spanwise rollers develop. The drag here increases dramatically, up to 4 times larger than in the uncontrolled flow. Outside these regions, the turbulence was enhanced to such an extent that the simulations diverged. To explain this behaviour, we explored the linear dynamics of the channel flow with boundary conditions adapted for opposition flow control.

We have shown that in the second parameter region, $x_0/h \in [-0.35, 0)$, 
all wavelengths of the channel flow, with or without viscosity, are affected by a linear instability. The instability is manifested as a pair of unstable eigenvalues that migrate along a circular path in the complex plane, as the phase of the complex control coefficient $A$ changes. The direction of the eigenvalue motion in the complex plane depends on the control gain $|A|$. When  $|A|$ is small, the control-related eigenvalues move towards the left, and we observe eigenmodes with unusually large negative phase speeds, $c_r<0$. As $|A|$ increases, the eigenvalues experience a hyperbolic growth, going from $-\infty$ to $\infty$ when $|A| = |A|_f$, $\phi = \pm \pi$. At $|A|_f$, the circle containing \anna{the} eigenvalue motion, flips to the right in the complex plane. We revisited the analysis of the Rayleigh equation for inviscid channel flow, and derived a semi-analytical expression for $c_r (|A|, \phi = \pm \pi)$ explaining this behavior.  In the inviscid case, unstable eigenvalues are encountered for positive phases $\phi = (0, \pi)$ for all $|A|$. The presence of viscosity modulates the instability, dampening it for small control gains (figure~\ref{fig:stab_amp}a). The  instability behavior is also different at large $|A|$ as the instability eventually weakens; the inviscid flow becomes marginally stable, while the growth rates of the viscous problem saturate at small but positive levels.  Physically, the large gains force the eigenvectors to be zero at the detection plane (figure~\ref{fig:eigvec}b), effectively creating a narrower channel.  As the effective wavenumber $\kappa$ increases, and the related structures become smaller, the instability region is displaced towards larger values of $|A|$, as indicated in figure~\ref{fig:stab_amp}(a). Therefore, long and wide wavemodes are more dangerous for control than short and narrow ones, given the same control gain. Considering moderate $|A|\leq 1$, as in our DNS, we find that only the largest flow scales are left under concern and may become unstable (figure~\ref{fig:eigvec_stabmap}).

These findings are easily applied to the DNS results if the linear relation $\phi = - k_x x_0$ between streamwise shift and the phase of control coefficient is considered.  Since the instability region is defined by $\phi$ instead of $x_0$, and the control coefficient $A$ is $2\pi$-periodic, the instability can be approached in two ways in the DNS. The obvious way is to shift control upstream, applying negative $x_0$, until one of the controlled wavenumbers becomes unstable at the ``critical" control shift $x_{cr}$.
In the case of $|A|=1$, several controlled wavenumbers become unstable simultaneously, but $k_x h = 6$ has the largest growth rate among them, and therefore defines the flow evolution near the onset.
We relate its growth rate to the steep increase in friction at $x_0/h \in [-0.35, 0)$ in figure~\ref{fig:phase_xlag}(c). As $x_0$ decreases further, $k_x h = 6$ becomes stable again, but the instability is overtaken by lower $k_x$, so the friction remains high.  On the other hand, it is also possible to approach the instability by shifting control downstream, i.e beyond $x_0/h>0.55$ (figure \ref{fig:phase_xlag}a,b), where the phase for $k_x h = 6$ changes from $-\pi$ to $\pi$. Periodicity of  $A$ also implies that the regions of stability and instability are repeated periodically as the control is moved further upstream or downstream. As the result of the linear instability, the altered steady-state flow takes the form of spanwise rollers while remaining turbulent (figure~\ref{fig:u10p}b). The flow gradually approaches this high-drag state at negative streamwise shifts, indicating a supercritical transition in figure~\ref{fig:xlag_transition}(a). The positive streamwise shift $x_0/h \approx 0.55$  features abrupt transition to the high-drag state, accompanied by two modulating frequencies as $|A|$ increases, and occurs at smaller $|A|$ than predicted by the linear analysis. This suggests a presence of a subcritical bifurcation in figure~\ref{fig:xlag_transition}(b), although more data are needed to confirm this. Our results are in agreement with the recent work of \citet{toedtli2020origin} who links the deterioration of the ``classic" opposition control with the detection plane in the buffer layer, $y_d^+ = 15$, to the formation of spanwise rollers. \citet{toedtli2020origin} finds unstable eigenmodes of the linearized Navier--Stokes equations for control phases $\phi  \in [\pi/4, 3 \pi/4]$, and they are probably of the same physical origin as the ones reported here.

The uniform structure of the high-friction flow state in the form of spanwise rollers in figure~\ref{fig:u10p_xlag_m03} resembles an instability of Kelvin-Helmholtz type.  Such instabilities have a profound effect on wall-bounded flows that allow wall transpiration, for example, channels with porous walls or riblets \citep{jimenez2001turbulent,garcia2011drag}. As in the case of Kelvin-Helmholtz instability, the advection of vorticity with $\hat{v}$ in~\eqref{eq:rayleigh} ($U{''} \hat{v}$) is the necessary ingredient for the instability observed here, and if it is removed from equations, the positive growth rates are no longer observed. However, the control instability has two important differences. First, a Kelvin-Helmholtz instability requires two interacting eigenvalues to form a complex conjugate pair, as illustrated by a simple example of the piecewise linear mixing layer \citep[pp. 26-29]{Schmid2012}. In our case, each of the two control eigenvalues is associated with control at one of the channel walls. The control eigenvalues do not interact, and only one of them appears if the control is applied only to  one wall. At the same time, when the eigenvalues go through a hyperbolic infinity in~\eqref{eq:cr_roots_norm}, they become essentially independent from the only velocity scale of the flow, $U_b$. The simplest interpretation is that the instability observed here is the instability of the control, in the sense of an unstable feedback loop. This is supported by the observation that the growth rates become larger as  $\phi \to  \pi$, which represents reinforcement instead of opposition, since the control at the wall is in phase with the detection plane.

In the region of $x_0/h \in [0, 0.55)$  the linear instability is inactive, but the friction is still increased and oblique waves with a large contribution to Reynolds stress near the wall appear (figure~\ref{fig:u10p}c). We used a linear resolvent model to explore the possibility of these waves to arise  from amplified responses of the linearised Navier-Stokes operator to forcing. Our model captures correctly the shape of oblique waves in the DNS on the length scales  $\lambda_x$, $\lambda_z=2h,h$, if the advection speed of the forcing is set the same as the  advection speed of structures  at the detection plane $y_d /h  = 0.1$,  supporting the suggestion above. On the other hand, resolvent analysis does not predict the largest responses on the scale of oblique waves (figure~\ref{fig:sigma_kx}). We rescaled the total kinetic energy of the flow by the energy of the wall-normal velocity at the wall to obtain a weighted magnitude of the flow response at each flow scale. Normalised this way, the magnitude of response increases with the length scale, and the maximum response is obtained at largest possible scales, in line with the linear prediction of the response norm in figure~\ref{fig:sigma_kx}. Therefore, the visually apparent build-up of energy at this particular scale is an artefact of the distribution of the rms of $v$ at the detection plane. In our work, the focus was on phase velocitites of $c_f /U_b \approx 0.75-0.78$, or $c^+ \in [11,13]$ in wall units, corresponding to ``detached" resolvent modes. \citet{luhar2014opposition} showed that the ``detached'' modes with $12<c^+<16$ will be amplified but create significant blowing and suction at the wall, i.e. are in principle controllable. A comprehensive analysis of the principal forcing frequencies in the controlled DNS, together with a more involved resolvent model with varying $c_f$, treating controlled and uncontrolled length scales separately, could clarify further questions.

Unfortunately, reducing the control gain and therefore inhibiting the possible instabilities does not  reduce drag, as shown by the DNS results in figure~\ref{fig:cf_amp}. A successful control strategy therefore should optimize both control gain and control phase, as suggested by \citet{luhar2014opposition, toedtli2019predicting}. It should also take into account the contribution to friction from all flow length scales, because nonlinearity transfers energy from the large scales affected by control to the uncontrolled small scales. In the current setup, at least two routes for optimization are possible: first, minimizing the total stress of each resolvent response, and second, reducing the maximum of $u$ near the wall. Our preliminary results (not shown here) indicate that the first strategy results in a relatively smaller drag increase in comparison to the second one. Nevertheless, we leave this optimization for a future work, which  should ideally be tailored to approximate an experimental flow.  In a real experimental setup the opposition control is implemented via local blowing and suction \citep{abbassi2017skin}. This requires an adaptation of wave-opposition control, which is global and occupies the whole spatial domain, to spatially local opposition, and, as a result, a different type of optimization problem \citep{pastor2020wall}.

We note here that the most straightforward way of relaxing turbulence is to reduce the non-normality of the linearized Navier-Stokes operator at the wall.  The non-normality is responsible for the near-wall cycle, generating streaks of $u$ from quasi-streamwise vortices. It is known that, if the non-normal coupling between $\omega_y$ an $v$ in the operator is removed or weakened, the turbulence decays \citep{jimenez1999autonomous}.  \citet{kim2000linear} compared the results of this weakening to classic opposition control  of \citet{choi1994active} and suggested that it acts on the flow in a similar manner. 
In the large-scale opposition control implemented here, the forcing is too large to interfere with quasi-streamwise vortices of the near-wall cycle, and therefore this physical mechanism of drag reduction is inactive.  Our results suggest executing caution in applying large-scale control, especially with a lagging delay between sensors and actuators, equivalent to shifting control upstream. Nevertheless, the increase of near-wall activity in the form of oblique structures of streamwise and wall-normal velocities and resulting momentum transport opens the possibility of using this form of control when enhancing of turbulence is beneficial, such as in mixing  or in separation control. \\

This work was supported by the European Research Council under the Coturb grant
ERC-2014.AdG-669505. 

Declaration of Interests. The authors report no conflict of interest.

\appendix
\section{The role of turbulent viscosity}
 In turbulent flows, momentum transport induced by Reynolds stresses modifies velocity profile to adopt a flatter shape, compared to a laminar one.  There have been attempts to explain the specific shape of turbulent mean profile  using the principle of maximum dissipation rate and assumption of neutral stability of the mean profile \citep{malkus1956outline}. Nevertheless, later linear stability analysis  of experimentally observed mean profiles in channels showed that these profiles are far from being neutrally stable and their perturbations decay \citep{reynolds1967stability}, if the turbulent viscosity is taken into account.   \citet{reynolds1972mechanics} demonstrated that including turbulent viscosity gives a much better prediction of experimental results.  In this work,  we also incorporate it into the linear stability analysis. The idea behind is that on a single wave harmonic, the background turbulence acts directly (through Reynolds stresses) and indirectly (through the turbulent mean profile). Turbulent viscosity, introduced into the viscous term, is merely a closure for the mean Reynolds stresses arising in the perturbation equations, and represents the interaction between the wave harmonic and background turbulence. In simple closures, it acts on the wave as an additional damping. Note that in the resolvent analysis of \citet{mckeon2010critical} and many further works  turbulent viscosity is not included in the linear model since nonlinear terms are treated as unknown forcing to the linear part of the Navier-Stokes equations. \citet{morra2019relevance} shows, however, that the resolvent model based on turbulent viscosity performs much better on predicting the velocity spectrum. This suggests that in the absence of additional information about the forcing shape or amplitude it is justified to incorporate the turbulent viscosity in the linear model to get more precise results.

Below we give the Cess analytic approximation for turbulent viscosity, 
\begin{equation}\label{eq:cess}
    \nu_t = \frac{\nu}{2}  \{ 1 + 
    \frac{K^2 Re_\tau}{9} [2 \eta -\eta^2] [3-4 \eta + \eta^2]^2 [1 - \exp{( \frac{- \eta Re_\tau}{A}} )]^2 \}^{0.5} +1/2,
\end{equation}
where $\nu$ is kinematic viscosity, $\eta = y/h$, and $A = 25.4$, $K = 0.426$ are parameters. The turbulent viscosity is later incorporated into \eqref{eq:oss_matrices}, where  Orr-Sommerfeld ($L_{OS}$), Squire ($L_{SQ}$) and wall-normal derivative  ($D$) operators are given by: 
\begin{equation}
 \begin{aligned}\label{eq:los_lsq}
L_{OS} &= \mathrm{i} k_x U (\kappa^2 - D^2) + \mathrm{i} \alpha U^{''} + \nu_t (\kappa^2 - D^2)^2  + 2 \nu_t^{'} (D^3 - \kappa D) + \nu_t^{''} (D^2 + \kappa^2), \\
L_{SQ} &= \mathrm{i} k_x U + \nu_t (k^2 - D^2) +\nu_t^{'} D \\
D &= \frac{\partial}{\partial y}.
\end{aligned}   
\end{equation}

Primes denote derivatives of $U$ and $\nu_t$ with respect to $y$, and $\kappa^2 = k_x^2 + k_z^2$. The the mean profile $U$ can be reconstructed using turbulent viscosity, as well as obtained empirically from the DNS.

\section{Boundary conditions for linearized problem}
Imposing boundary conditions for the eigenvalue problem~\eqref{eq:freq_linsys} is equivalent to implementing additional algebraic constrains on the system. Consider a generalized eigenvalue problem $\omega M v = L v$. Suppose $v \in \mathds{C}^n$, where $n$ corresponds to the spatial discretization of the domain in $y$-direction, and assume finite difference discretization for simplicity. The most general boundary conditions for the Orr-Sommerfeld problem are the clamped boundary conditions 
\begin{equation}\label{eq:clamped_bc}
v_{0,n} = 0, \quad \left( \partial v /\partial y \right)_{0,n} = 0.   
\end{equation}
 Recalling that the left-hand side of the problem $\omega M v$ is related to the time derivative $\partial v/\partial t$, a way to implement conditions~\eqref{eq:clamped_bc} is to set the first and the last row of $M$ to zero. This operation, however, reduces the rank of matrix $M$ and sets the problem ill-conditioned. Moreover, it produces two infinite eigenvalues corresponding to conditions~\eqref{eq:clamped_bc}, since physically these conditions mean that the perturbation on the boundary propagate infinitely fast. To avoid numerical issues, manual filtering of the infinite (not spurious) eigenvalues, or developing manually matrix-reduction algorithms, \citet{goussis1989removal}, \citet{Schmid2012} offer a mapping of the infinite eigenvalues to an arbitrary point in the eigenvalue space. Instead of setting corresponding rows of $M$ to zero, respective rows of the problem are set to satisfy an equation for $v_{0,n}$ with exponentially decaying solutions: 
 \begin{equation}\label{eq:SCH_bc}
     \omega M_{0, ...} v_0 = - |C| \mathrm{i} v_0
 \end{equation}
 The constant $|C|$ is set large enough to map infinite eigenvalues related to boundary conditions into the stable part of the complex plane, way below the line $\omega_i = 0$. For the controlled problem condition~\eqref{eq:SCH_bc} reads as $ \omega M_{0, ...} (v_0 - v_d) = - |C| \mathrm{i} (v_0-v_d)$, which sets the difference between the perturbation at the wall and the detection plane to decay exponentially. The resulting eigenvectors satisfy the control law~\eqref{eq:ctrl_law}.
 
\section{Numerical resolution test for section~\ref{sec:visc_invisc_diff}}
\begin{figure}
\centering
      \begin{subfigure}[t]{0.5\textwidth}
      \caption{}\label{fig:ci_kx6A1}
  	\centering
  	\includegraphics[width=\textwidth]{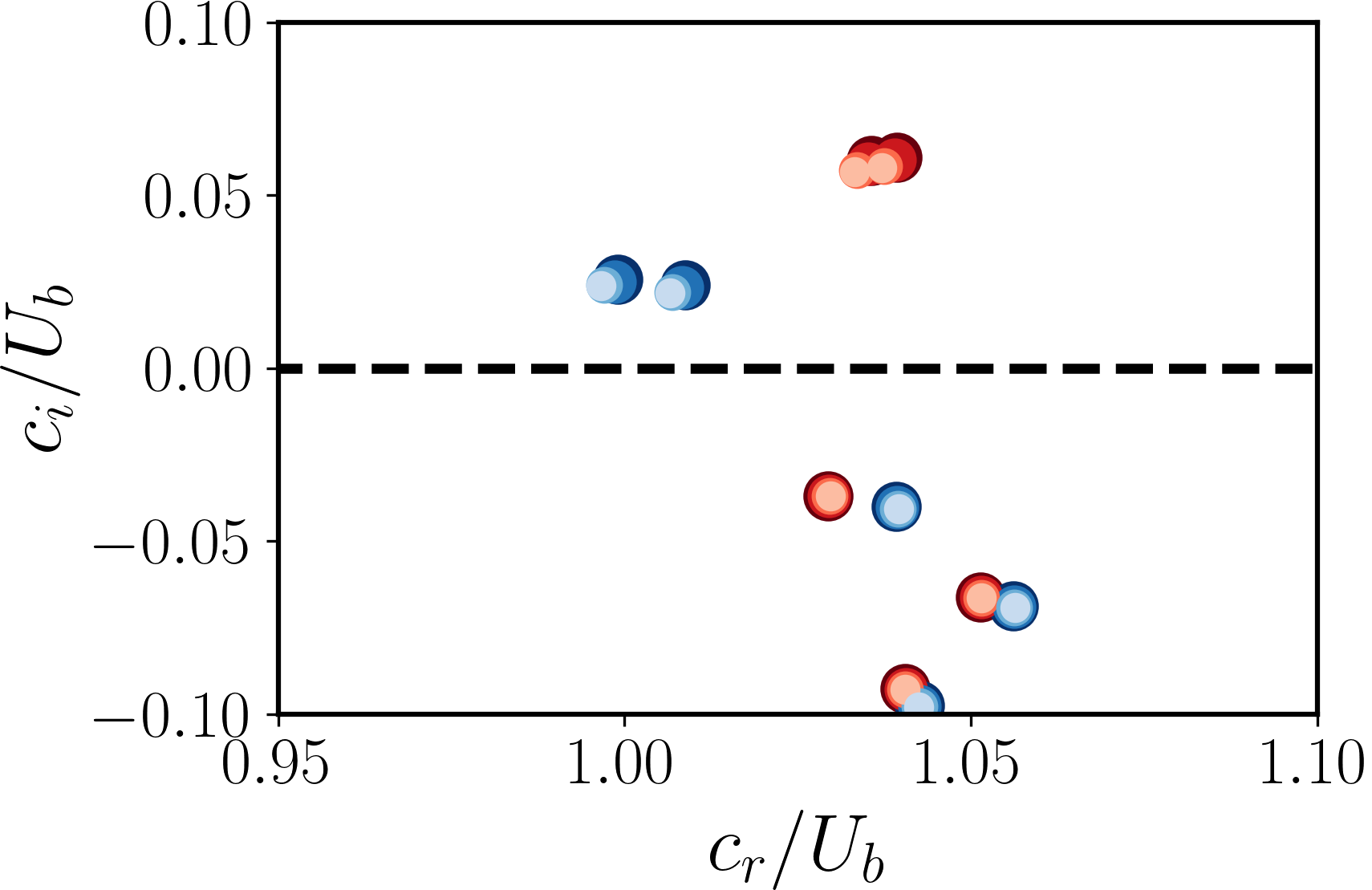} 
  \end{subfigure}
  \begin{subfigure}[t]{0.48\textwidth}
      \caption{}\label{fig:ci_kx6A2}
  	\centering
  	\includegraphics[width=\textwidth]{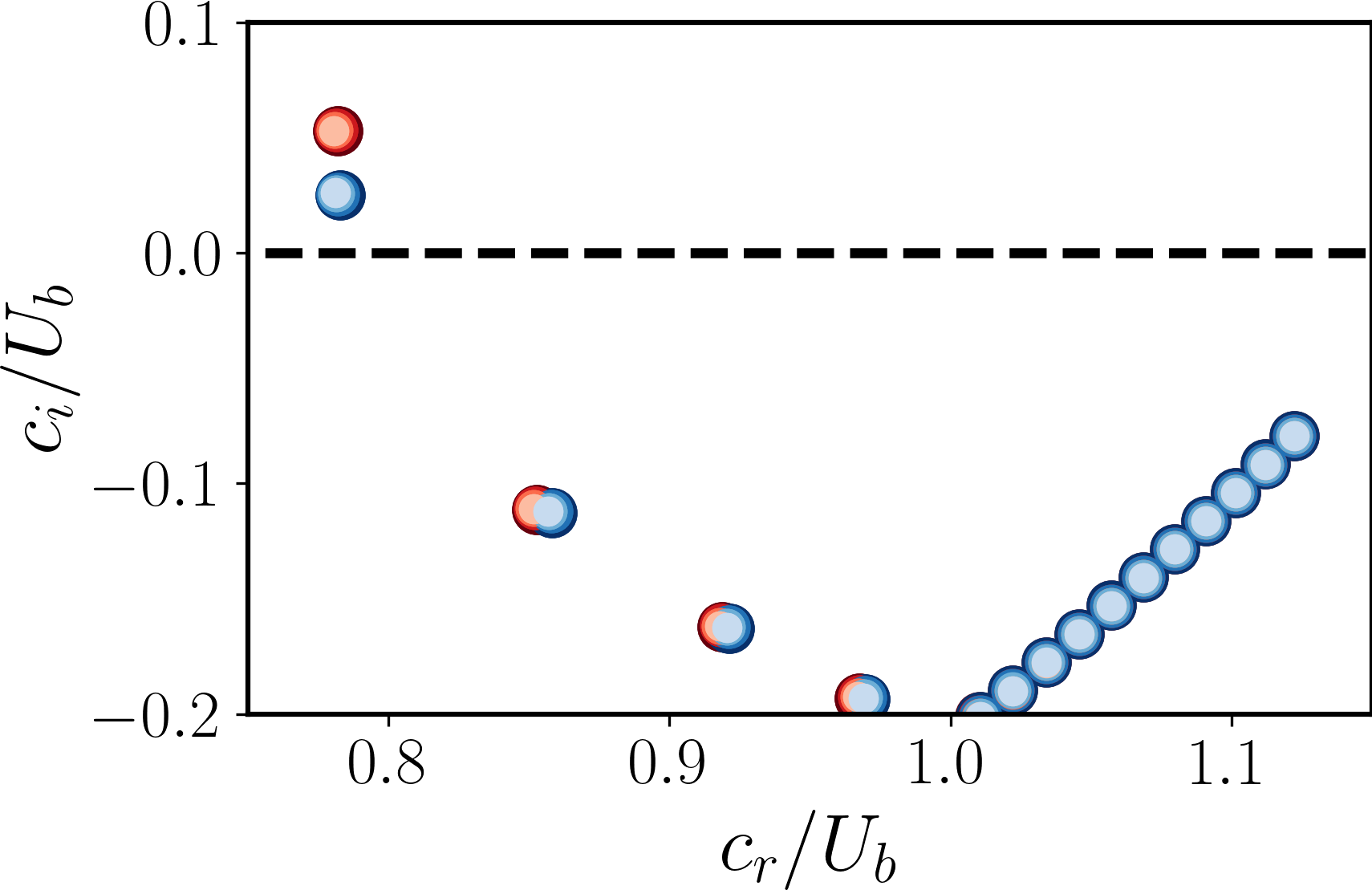} 
  \end{subfigure}
\caption{Numerical resolution test for the linear stability problem with viscosity. Eigenvalue spectrum for (a) $k_x = 2$, $k_z = 0$. Blue, $|A| = 60$ $\phi = -3 \pi/4$. Red, $|A| = 10$, $\phi = 3 \pi/4$. (b) $k_x = 20$, $k_z = 0$. $|A| = 100$. Blue,  $\phi = -3 \pi/4$. Red,  $\phi = 3 \pi/4$. The number of Chebyshev polynomials $N \in [256, 400, 512, 600]$, indicated by decreasing symbol size and color intensity.}
\label{fig:resolution_test}
\end{figure}

It is important to make sure that the numerical resolution of the linear stability analysis is sufficient for the large control gains in section~\ref{sec:visc_invisc_diff}, and that the saturation of eigenvalues is not a spurious effect. We performed resolution tests on various wavenumbers, increasing the number of Chebyshev polynomials $N$ used to discretize the matrices $M$ and $L$ in \eqref{eq:f_linsys}. The tests in figure~\ref{fig:resolution_test} show consistency in eigenvalue positions above $c_i =0$ of the unstable wavenumbers both for small and large $k_x$ with increasing $N$. Since a small variation is nevertheless observed, we increased the numerical resolution in section~\ref{sec:visc_invisc_diff} to $N=512$.

\bibliographystyle{jfm}
\bibliography{controlbib}

\begin{thebibliography}{49}
\expandafter\ifx\csname natexlab\endcsname\relax\def\natexlab#1{#1}\fi
\def\au#1{#1} \def\ed#1{#1} \def\yr#1{#1}\def\at#1{#1}\def\jt#1{\textit{#1}}
  \def\bt#1{#1}\def\bvol#1{\textbf{#1}} \def\vol#1{#1} \def\pg#1{#1}
  \def\publ#1{#1}\def\arxiv#1{#1}\def\org#1{#1}\def\st#1{\textit{#1}}

\bibitem[Abbassi {\em et~al.\/}(2017)Abbassi, Baars, Hutchins \&
  Marusic]{abbassi2017skin}
{\sc \au{Abbassi, M.R.}, \au{Baars, W.J.}, \au{Hutchins, N.} \& \au{Marusic,
  I.}} \yr{2017}  \at{Skin-friction drag reduction in a high-{R}eynolds-number
  turbulent boundary layer via real-time control of large-scale structures}.
  \jt{Int. J. Heat Fluid Fl.}  \bvol{67},  \pg{30--41}.

\bibitem[Bewley \& Liu(1998)]{bewley1998optimal}
{\sc \au{Bewley, T.R.} \& \au{Liu, S.}} \yr{1998}  \at{Optimal and robust
  control and estimation of linear paths to transition}.  \jt{J. Fluid Mech.}
  \bvol{365},  \pg{305--349}.

\bibitem[Bewley {\em et~al.\/}(2001)Bewley, Moin \& Temam]{bewley2001dns}
{\sc \au{Bewley, Thomas~R}, \au{Moin, Parviz} \& \au{Temam, Roger}} \yr{2001}
  \at{Dns-based predictive control of turbulence: an optimal benchmark for
  feedback algorithms}.  \jt{Journal of Fluid Mechanics}  \bvol{447},
  \pg{179--225}.

\bibitem[Butler \& Farrell(1992)]{butler1992three}
{\sc \au{Butler, K.M.} \& \au{Farrell, B.F.}} \yr{1992}  \at{Three-dimensional
  optimal perturbations in viscous shear flow}.  \jt{Phys. Fluids A-Fluid}
  \bvol{4}~(8),  \pg{1637--1650}.

\bibitem[Cess(1958)]{cess1958survey}
{\sc \au{Cess, R.D.}} \yr{1958}  \at{A survey of the literature on heat
  transfer in turbulent tube flow}.  \jt{Res. Rep}  \pg{pp. 8--0529}.

\bibitem[Choi {\em et~al.\/}(1994)Choi, Moin \& Kim]{choi1994active}
{\sc \au{Choi, H.}, \au{Moin, P.} \& \au{Kim, J.}} \yr{1994}  \at{Active
  turbulence control for drag reduction in wall-bounded flows}.  \jt{J. Fluid
  Mech.}  \bvol{262},  \pg{75--110}.

\bibitem[Chung \& Talha(2011)]{chung2011effectiveness}
{\sc \au{Chung, Y.M.} \& \au{Talha, T.}} \yr{2011}  \at{Effectiveness of active
  flow control for turbulent skin friction drag reduction}.  \jt{Phys. Fluids}
  \bvol{23}~(2),  \pg{025102}.

\bibitem[Chung \& Sung(2003)]{chung2003sensitivity}
{\sc \au{Chung, Yongmann~M} \& \au{Sung, Hyung~Jin}} \yr{2003} Sensitivity
  study of turbulence control with wall blowing and suction.  \bt{In {\em Third
  Symposium on Turbulence and Shear Flow Phenomena\/}}. Begel House Inc.

\bibitem[Del~Alamo \& Jim{\'e}nez(2006)]{del2006linear}
{\sc \au{Del~Alamo, J.C.} \& \au{Jim{\'e}nez, J.}} \yr{2006}  \at{Linear energy
  amplification in turbulent channels}.  \jt{J. Fluid Mech.}  \bvol{559},
  \pg{205--213}.

\bibitem[Encinar \& Jim{\'e}nez(2019)]{encinar2019logarithmic}
{\sc \au{Encinar, M.~P.} \& \au{Jim{\'e}nez, J.}} \yr{2019}
  \at{Logarithmic-layer turbulence: A view from the wall}.  \jt{Phys. Rev.
  Fluids}  \bvol{4}~(11),  \pg{114603}.

\bibitem[Flores \& Jim{\'e}nez(2006)]{flores2006effect}
{\sc \au{Flores, O.} \& \au{Jim{\'e}nez, J.}} \yr{2006}  \at{Effect of
  wall-boundary disturbances on turbulent channel flows}.  \jt{J. Fluid Mech.}
  \bvol{566},  \pg{357--376}.

\bibitem[Flores \& Jim{\'e}nez(2010)]{flores2010hierarchy}
{\sc \au{Flores, O.} \& \au{Jim{\'e}nez, J.}} \yr{2010}  \at{Hierarchy of
  minimal flow units in the logarithmic layer}.  \jt{Phys. Fluids}
  \bvol{22}~(7),  \pg{071704}.

\bibitem[Garc{\'\i}a-Mayoral \& Jim{\'e}nez(2011)]{garcia2011drag}
{\sc \au{Garc{\'\i}a-Mayoral, R.} \& \au{Jim{\'e}nez, J.}} \yr{2011}  \at{Drag
  reduction by riblets}.  \jt{Philos. T. Roy. Soc. A}  \bvol{369}~(1940),
  \pg{1412--1427}.

\bibitem[Goussis \& Pearlstein(1989)]{goussis1989removal}
{\sc \au{Goussis, D.~A.} \& \au{Pearlstein, A.~J.}} \yr{1989}  \at{Removal of
  infinite eigenvalues in the generalized matrix eigenvalue problem}.  \jt{J.
  Comput. Phys.}  \bvol{84}~(1),  \pg{242--246}.

\bibitem[Hammond {\em et~al.\/}(1998)Hammond, Bewley \&
  Moin]{hammond1998observed}
{\sc \au{Hammond, EP}, \au{Bewley, TR} \& \au{Moin, P}} \yr{1998}  \at{Observed
  mechanisms for turbulence attenuation and enhancement in
  opposition-controlled wall-bounded flows}.  \jt{Physics of Fluids}
  \bvol{10}~(9),  \pg{2421--2423}.

\bibitem[Ibrahim {\em et~al.\/}(2019)Ibrahim, Guseva \&
  Garcia-Mayoral]{ibrahim2019selective}
{\sc \au{Ibrahim, J.}, \au{Guseva, A.} \& \au{Garcia-Mayoral, R.}} \yr{2019}
  \at{Selective opposition-like control of large-scale structures in
  wall-bounded turbulence}.  \jt{B. Am. Phys. Soc.}  \bvol{64}.

\bibitem[Jim{\'e}nez(1994)]{jimenez1994structure}
{\sc \au{Jim{\'e}nez, J.}} \yr{1994}  \at{On the structure and control of near
  wall turbulence}.  \jt{Phys. Fluids}  \bvol{6}~(2),  \pg{944--953}.

\bibitem[Jim{\'e}nez(2018)]{jimenez2018coherent}
{\sc \au{Jim{\'e}nez, J.}} \yr{2018}  \at{Coherent structures in wall-bounded
  turbulence}.  \jt{J. Fluid Mech.}  \bvol{842}.

\bibitem[Jim{\'e}nez \& Pinelli(1999)]{jimenez1999autonomous}
{\sc \au{Jim{\'e}nez, J.} \& \au{Pinelli, A.}} \yr{1999}  \at{The autonomous
  cycle of near-wall turbulence}.  \jt{J. Fluid Mech.}  \bvol{389},
  \pg{335--359}.

\bibitem[Jim{\'e}nez {\em et~al.\/}(2001)Jim{\'e}nez, Uhlmann, Pinelli \&
  Kawahara]{jimenez2001turbulent}
{\sc \au{Jim{\'e}nez, J.}, \au{Uhlmann, M.}, \au{Pinelli, A.} \& \au{Kawahara,
  G.}} \yr{2001}  \at{Turbulent shear flow over active and passive porous
  surfaces}.  \jt{J. Fluid Mech.}  \bvol{442},  \pg{89--117}.

\bibitem[Kim \& Lim(2000)]{kim2000linear}
{\sc \au{Kim, J.} \& \au{Lim, J.}} \yr{2000}  \at{A linear process in
  wall-bounded turbulent shear flows}.  \jt{Phys. Fluids}  \bvol{12}~(8),
  \pg{1885--1888}.

\bibitem[Kim {\em et~al.\/}(1987)Kim, Moin \& Moser]{kim1987turbulence}
{\sc \au{Kim, J.}, \au{Moin, P.} \& \au{Moser, R.}} \yr{1987}  \at{Turbulence
  statistics in fully developed channel flow at low {R}eynolds number}.  \jt{J.
  Fluid Mech.}  \bvol{177},  \pg{133--166}.

\bibitem[Koumoutsakos(1999)]{koumoutsakos1999vorticity}
{\sc \au{Koumoutsakos, Petros}} \yr{1999}  \at{Vorticity flux control for a
  turbulent channel flow}.  \jt{Physics of Fluids}  \bvol{11}~(2),
  \pg{248--250}.

\bibitem[Lee {\em et~al.\/}(1997)Lee, Kim, Babcock \&
  Goodman]{lee1997application}
{\sc \au{Lee, Changhoon}, \au{Kim, John}, \au{Babcock, David} \& \au{Goodman,
  Rodney}} \yr{1997}  \at{Application of neural networks to turbulence control
  for drag reduction}.  \jt{Physics of Fluids}  \bvol{9}~(6),  \pg{1740--1747}.

\bibitem[Lee(2015)]{lee2015opposition}
{\sc \au{Lee, J.}} \yr{2015}  \at{Opposition control of turbulent wall-bounded
  flow using upstream sensor}.  \jt{J. Mech. Sci. Technol.}  \bvol{29}~(11),
  \pg{4729--4735}.

\bibitem[Lim \& Kim(2004)]{lim2004singular}
{\sc \au{Lim, J.} \& \au{Kim, J.}} \yr{2004}  \at{A singular value analysis of
  boundary layer control}.  \jt{Phys. Fluids}  \bvol{16}~(6),  \pg{1980--1988}.

\bibitem[Lozano-Dur{\'a}n \& Jim{\'e}nez(2014)]{lozano2014effect}
{\sc \au{Lozano-Dur{\'a}n, A.} \& \au{Jim{\'e}nez, J.}} \yr{2014}  \at{Effect
  of the computational domain on direct simulations of turbulent channels up to
  {$Re_\tau= 4200$}}.  \jt{Phys. Fluids}  \bvol{26}~(1),  \pg{011702}.

\bibitem[Luchini {\em et~al.\/}(1991)Luchini, Manzo \&
  Pozzi]{luchini1991resistance}
{\sc \au{Luchini, P.}, \au{Manzo, F.} \& \au{Pozzi, A.}} \yr{1991}
  \at{Resistance of a grooved surface to parallel flow and cross-flow}.  \jt{J.
  Fluid Mech.}  \bvol{228},  \pg{87--109}.

\bibitem[Luhar {\em et~al.\/}(2014)Luhar, Sharma \&
  McKeon]{luhar2014opposition}
{\sc \au{Luhar, M.}, \au{Sharma, A.S.} \& \au{McKeon, B.}} \yr{2014}
  \at{Opposition control within the resolvent analysis framework}.  \jt{J.
  Fluid Mech.}  \bvol{749},  \pg{597--626}.

\bibitem[Mack(1976)]{mack1976numerical}
{\sc \au{Mack, L.~M.}} \yr{1976}  \at{A numerical study of the temporal
  eigenvalue spectrum of the blasius boundary layer}.  \jt{J. Fluid Mech.}
  \bvol{73}~(3),  \pg{497--520}.

\bibitem[Malkus(1956)]{malkus1956outline}
{\sc \au{Malkus, W.V.R.}} \yr{1956}  \at{Outline of a theory of turbulent shear
  flow}.  \jt{J. Fluid Mech.}  \bvol{1}~(5),  \pg{521--539}.

\bibitem[McKeon \& Sharma(2010)]{mckeon2010critical}
{\sc \au{McKeon, B.J.} \& \au{Sharma, A.S.}} \yr{2010}  \at{A critical-layer
  framework for turbulent pipe flow}.  \jt{J. Fluid Mech.}  \bvol{658},
  \pg{336--382}.

\bibitem[Morra {\em et~al.\/}(2019)Morra, Semeraro, Henningson \&
  Cossu]{morra2019relevance}
{\sc \au{Morra, P.}, \au{Semeraro, O.}, \au{Henningson, D.S.} \& \au{Cossu,
  C.}} \yr{2019}  \at{On the relevance of {R}eynolds stresses in resolvent
  analyses of turbulent wall-bounded flows}.  \jt{J. Fluid Mech.}  \bvol{867},
  \pg{969–984}.

\bibitem[Nikuradse(1933)]{nikuradse1933stromungsgesetze}
{\sc \au{Nikuradse, J.}} \yr{1933}  \at{Stromungsgesetze in rauhen rohren}.
  \jt{VDI-Forsch.}  \bvol{361}, (Engl. transl. 1950. Laws of flow in rough
  pipes. NACA TM 1292).

\bibitem[Oehler {\em et~al.\/}(2018)Oehler, Garcia-Guti{\'e}rrez \&
  Illingworth]{oehler2018linear}
{\sc \au{Oehler, S.}, \au{Garcia-Guti{\'e}rrez, A.} \& \au{Illingworth, S.}}
  \yr{2018} Linear estimation of coherent structures in wall-bounded turbulence
  at {$Re_\tau= 2000$}.  \bt{In {\em J. Phys. Conf. Ser.\/}}, ,  \vol{vol.
  1001},  \pg{p. 012006}.

\bibitem[Oehler \& Illingworth(2020)]{oehler2020linear}
{\sc \au{Oehler, S.F.} \& \au{Illingworth, S.J.}} \yr{2020}  \at{Linear control
  of coherent structures in wall-bounded turbulence at {$Re_\tau = 2000$}}.
  \jt{Int. J. Heat Fluid Fl.}  \pg{p. 108735}.

\bibitem[Pastor {\em et~al.\/}(2020)Pastor, Vela-Martin \&
  Flores]{pastor2020wall}
{\sc \au{Pastor, R.}, \au{Vela-Martin, A.} \& \au{Flores, O.}} \yr{2020}
  Wall-bounded turbulence control: statistical characterisation of
  actions/states.  \bt{In {\em J. Phys. Conf. Ser.\/}}, ,  \vol{vol. 1522},
  \pg{p. 012014}.

\bibitem[Pujals {\em et~al.\/}(2009)Pujals, Garc{\'i}a-Villalba, Cossu \&
  Depardon]{pujals2009note}
{\sc \au{Pujals, G.}, \au{Garc{\'i}a-Villalba, M.}, \au{Cossu, C.} \&
  \au{Depardon, S.}} \yr{2009}  \at{A note on optimal transient growth in
  turbulent channel flow}.  \jt{Phys. Fluids.}  \bvol{21},  \pg{015109}.

\bibitem[Quadrio \& Ricco(2004)]{quadrio2004critical}
{\sc \au{Quadrio, M.} \& \au{Ricco, P.}} \yr{2004}  \at{Critical assessment of
  turbulent drag reduction through spanwise wall oscillations}.  \jt{J. Fluid
  Mech.}  \bvol{521},  \pg{251--271}.

\bibitem[Rebbeck \& Choi(2001)]{rebbeck2001opposition}
{\sc \au{Rebbeck, Henry} \& \au{Choi, Kwing-So}} \yr{2001}  \at{Opposition
  control of near-wall turbulence with a piston-type actuator}.  \jt{Physics of
  Fluids}  \bvol{13}~(8),  \pg{2142--2145}.

\bibitem[Rebbeck \& Choi(2006)]{rebbeck2006wind}
{\sc \au{Rebbeck, Henry} \& \au{Choi, Kwing-So}} \yr{2006}  \at{A wind-tunnel
  experiment on real-time opposition control of turbulence}.  \jt{Physics of
  Fluids}  \bvol{18}~(3),  \pg{035103}.

\bibitem[Reynolds \& Hussain(1972)]{reynolds1972mechanics}
{\sc \au{Reynolds, W.C.} \& \au{Hussain, A.K.M.F.}} \yr{1972}  \at{The
  mechanics of an organized wave in turbulent shear flow. {P}art 3.
  {T}heoretical models and comparisons with experiments}.  \jt{J. Fluid Mech.}
  \bvol{54}~(2),  \pg{263--288}.

\bibitem[Reynolds \& Tiederman(1967)]{reynolds1967stability}
{\sc \au{Reynolds, W.C.} \& \au{Tiederman, W.G.}} \yr{1967}  \at{Stability of
  turbulent channel flow, with application to {M}alkus's theory}.  \jt{J. Fluid
  Mech.}  \bvol{27}~(2),  \pg{253--272}.

\bibitem[Schmid \& Henningson(2012)]{Schmid2012}
{\sc \au{Schmid, P.J.} \& \au{Henningson, D.S.}} \yr{2012} {\em Stability and
  Transition in Shear Flows\/}.  \publ{Springer Science {\&} Business Media}.

\bibitem[Spalart \& McLean(2011)]{spalart2011drag}
{\sc \au{Spalart, P.R.} \& \au{McLean, J.D.}} \yr{2011}  \at{Drag reduction:
  enticing turbulence, and then an industry}.  \jt{Philos. T. Roy. Soc. A}
  \bvol{369}~(1940),  \pg{1556--1569}.

\bibitem[Straub {\em et~al.\/}(2017)Straub, Vinuesa, Schlatter, Frohnapfel \&
  Gatti]{straub2017turbulent}
{\sc \au{Straub, S.}, \au{Vinuesa, R.}, \au{Schlatter, P.}, \au{Frohnapfel, B.}
  \& \au{Gatti, D.}} \yr{2017}  \at{Turbulent duct flow controlled with
  spanwise wall oscillations}.  \jt{Flow Turbul. Combust.}  \bvol{99}~(3-4),
  \pg{787--806}.

\bibitem[Toedtli {\em et~al.\/}(2019)Toedtli, Luhar \&
  McKeon]{toedtli2019predicting}
{\sc \au{Toedtli, S.S.}, \au{Luhar, M.} \& \au{McKeon, B.}} \yr{2019}
  \at{Predicting the response of turbulent channel flow to varying-phase
  opposition control: Resolvent analysis as a tool for flow control design}.
  \jt{Phys. Rev. Fluids}  \bvol{4}~(7),  \pg{073905}.

\bibitem[Toedtli {\em et~al.\/}(2020)Toedtli, Yu \& McKeon]{toedtli2020origin}
{\sc \au{Toedtli, S.}, \au{Yu, Ch.} \& \au{McKeon, B.}} \yr{2020}  \at{On the
  origin of drag increase in varying-phase opposition control}.  \jt{Int. J.
  Heat Fluid Fl.}  \bvol{85},  \pg{108651}.

\bibitem[Townsend(1976)]{townsend1980structure}
{\sc \au{Townsend, A.A.}} \yr{1976} {\em The structure of turbulent shear
  flow\/}.  \publ{Cambridge Univ. Press. (2nd ed)}.

\end{thebibliography}

\end{document}